\documentclass[12pt]{article}
\usepackage{amsmath}
\usepackage{mathtools} 
\usepackage[T1]{fontenc}
\usepackage{amssymb}
\usepackage{amsfonts}
\usepackage{amsthm}
\usepackage{mathrsfs}
\usepackage[all]{xy}
\usepackage{tensor}
\usepackage{comment}
\usepackage{stmaryrd}
\usepackage{bm}
\usepackage[normalem]{ulem}
\usepackage[usenames,dvipsnames]{xcolor}
\usepackage[colorlinks=true, citecolor=Blue, linkcolor=Blue]{hyperref}
\usepackage{ragged2e}
\usepackage{tikz}
    \usetikzlibrary{decorations.markings}

\usepackage{diagbox}  
\usepackage{makecell} 
\usepackage{hhline} 
\usepackage{stackengine}

\usepackage[labelfont={small, bf, sf}, textfont={small,sf}]{caption}
\numberwithin{equation}{section} 
\usepackage{authblk}
\usepackage[in]{fullpage}
\usepackage{cite}

\newcommand{\beq}{\begin{equation}}
\newcommand{\eeq}{\end{equation}}
\newcommand{\bes}{\begin{subequations}}
\newcommand{\ees}{\end{subequations}}
\newcommand{\bea}{\begin{eqnarray}}
\newcommand{\eea}{\end{eqnarray}}

\newcommand{\be}{\begin{equation}}
\newcommand{\ee}{\end{equation}}

\newcount\hh
\newcount\mm
\mm=\time
\hh=\time
\divide\hh by 60
\divide\mm by 60
\multiply\mm by 60
\mm=-\mm
\advance\mm by \time
\def\hhmm{\number\hh:\ifnum\mm<10{}0\fi\number\mm}

\def\be{\begin{equation}}
\def\ee{\end{equation}}

\def\ps{\scalebox{.7}{$\scriptscriptstyle +$}}
\def\psps{\scalebox{.55}{$\scriptscriptstyle \stackanchor{+}{+}$}}
\def\psms{\scalebox{.55}{$\scriptscriptstyle \stackanchor{+}{--}$}}
\def\ms{\scalebox{.7}{$\scriptscriptstyle -$}}
\def\msps{\scalebox{.55}{$\scriptscriptstyle \stackanchor{--}{+}$}}
\def\msms{\scalebox{.55}{$\scriptscriptstyle \stackanchor{--}{--}$}}

\newcommand{\tepsilon}{\alpha}
\newcommand{\scrm}{\mathscr{I}^{-}}
\newcommand{\scrp}{\mathscr{I}^{+}}
\newcommand{\Lorentz}{Lorenz }
\newcommand{\scri}{\mathscr{I}}
\def\bftheta{\mbox{\boldmath $\theta$}}
\def\bfbeta{\mbox{\boldmath $\beta$}}
\newcommand{\pb}{\mathcal{P}_{\ast}}

\newcommand{\ve}{\varepsilon}
\newcommand{\bfx}{{\bf x}}
\newcommand{\bfk}{{\bf k}}
\newcommand{\bfl}{{\bf l}}
\newcommand{\bfp}{{\bf p}}
\newcommand{\bfq}{{\bf q}}
\newcommand{\bfn}{{\bf n}}
\newcommand{\bfm}{{\bf m}}
\newcommand{\bfA}{{\bf A}}

\numberwithin{equation}{section}

\usepackage{cleveref}

\crefname{equation}{Eq.}{Eqs.}
\creflabelformat{equation}{#2#1#3}

\crefname{section}{Sec.}{Sec.}
\crefname{appendix}{Appendix}{Appendices}
\crefname{figure}{Fig.}{Figs.}

\crefname{definition}{Def.}{Defs.}
\crefname{prop}{Prop.}{Props.}
\crefname{lemma}{Lemma}{Lemmas}
\crefname{corollary}{Cor.}{Cors.}
\crefname{thm}{Theorem}{Theorems}
\crefname{remark}{Remark}{Remarks}

\begin{document}
\setlength{\parindent}{10pt}

\title{The classical dynamics of gauge theories in the deep infrared}

\author[1]{\'Eanna \'E. Flanagan\thanks{eef3@cornell.edu}}
\author[1]{Ibrahim Shehzad\thanks{is354@cornell.edu}}

\affil[1]{\small \it Department of Physics, Cornell University, Ithaca, NY, 14853, USA}

\maketitle

\begin{abstract}

Gauge and gravitational theories in asymptotically flat settings
possess infinitely many conserved charges associated with large gauge
transformations or diffeomorphisms that are nontrivial at infinity.
To what extent do these charges constrain the scattering in these
theories? It has been claimed in the literature
that the constraints are trivial, due to a decoupling of hard and soft
sectors for which the conserved charges constrain only the dynamics in
the soft sector.  We show that the argument for this decoupling fails due to the failure
in infinite dimensions of a property of symplectic geometry which
holds in finite dimensions.
Specializing to electromagnetism coupled to a massless charged scalar
field in four dimensional Minkowski spacetime, 
we show explicitly that the two
sectors are always coupled using a perturbative classical computation
of the scattering map.
Specifically, while the two sectors are uncoupled at
low orders, they are coupled at quartic order via the electromagnetic
memory effect.  This coupling cannot be removed by adjusting the definitions of the
hard and soft sectors (which includes the classical analog of dressing
the hard degrees of freedom).  We conclude 
that the conserved charges yield nontrivial constraints on the scattering of
hard degrees of freedom.  This conclusion should also apply to
gravitational scattering and to black hole formation and evaporation.

In developing the classical scattering theory, we show that generic
\Lorentz gauge solutions fail to satisfy the matching condition on the vector potential at
spatial infinity proposed by Strominger to define the field configuration space, and we suggest a way to
remedy this.  We also show that when soft degrees of freedom are
present, the order at which nonlinearities
first arise in the scattering map is second 
order in \Lorentz gauge, but can be third order in other gauges.

\end{abstract}

\tableofcontents

\section{Introduction and summary}
\label{sec:intro}

\subsection{Background and motivation}

Recent years have seen significant progress in our understanding of
the dynamics of gauge and gravitational theories in the deep infrared.
In particular, a remarkable web of relations has been discovered between three
seemingly unrelated areas of infrared physics (see the review \cite{Strominger:lectures}):
soft theorems, which govern
universal properties of scattering amplitudes in the limit where
the energy of some external 
massless particles are taken to zero
\cite{Weinberg-soft, CS-soft,Low-1, Low-2,soft-gluons,soft-photinos};
asymptotic symmetries, which are gauge transformations or
diffeomorphisms that act on the boundary of spacetime and that are
associated with conservation laws
\cite{Bondi:1962px,Sachs:1962wk,Sachs:1962zza,Strominger:2013lka,Strominger:lectures,Ashtekar:2018lor};
and memory effects, zero frequency 
components of bursts of radiation that are potentially observable for
example with  
future gravitational wave observations
\cite{BT-memory,Christodoulo-memory,EMmemory-1,EMmemory-2,color-memory}.
New avenues continue to be explored in this rich subject:
aspects of the triangular web of
relations have been extended to enlarged
symmetry groups, to higher dimensional spacetimes, 
to subleading orders, to new spacetime boundaries and to new
theoretical contexts; see, e.g,
Refs.\ \cite{Strominger:2021lvk,Flanagan:2015pxa,higherdim-web-1,higherdim-web-2, 
Compere:2018ylh}.

One interesting aspect of this field has been the discovery of a new kind
of black hole hair in general relativity \cite{HPS-1,HPS-2}.
Asymptotic symmetry transformations at null infinity and at future
horizons are associated with an infinite number of charges called soft
hair which are conserved in scattering processes\footnote{It has been argued that these charges can be used to compute the entropy of Kerr black holes \cite{Haco:2018ske,Chandrasekaran:2020wwn}.}.
It has been argued that during black hole evaporation,
effects involving the heretofore neglected soft hair open 
the door
to possible new avenues for resolving the information loss paradox
\cite{HPS-1,HPS-2,Pasterski:2020xvn,Flanagan:2021ojq}.
One possible mechanism is that the soft hair acts as a kind of
catalyst in its interactions with the Hawking radiation and engenders
correlations between different Hawking quanta.
Another possibility is that the Hawking radiation is purified at late
times by its entanglement with soft hair degrees of freedom
\cite{Strominger:2017aeh}.

A number of arguments have been made against the relevance of the soft
conservation laws to information loss.  First, it has been argued that in some
contexts the number of accessible soft hair degrees of freedom is too
small to be relevant\footnote{See also Ref.\ \cite{Compere:2019rof}
which argues that a maximal entanglement between hard and soft sectors
in black hole evaporation would require involving soft degrees of
freedom beyond the leading and subleading orders.}, for example in anti-deSitter spacetimes \cite{Marolf:2017jkr}.  
Second, it has been argued that the interactions between the 
soft (zero-energy) and hard (non-zero energy) sectors of a
gauge or gravitational theory in any scattering process are too trivial to impact the
information puzzle \cite{Mirbabayi:2016axw,Bousso,Bousso:2017rsx,Donnelly:2018nbv}.

In particular, it has been argued that 
soft charges reside in a sector of the theory
that is decoupled from all the other dynamics \cite{Mirbabayi:2016axw,Bousso,Bousso:2017rsx}.
In other words, the soft conservation laws constrain only a
sector of the theory that is decoupled from the sector that contains all the
hard degrees of freedom, including the Hawking radiation. As a result,
the soft charges do not constrain the
scattering problem in an essential way.  This has been described as
soft hair comprising a ``soft wig'' that is easily removed \cite{Bousso}.

While motivated by deep questions in the quantum theory, the question
of whether the soft and hard degrees of freedom are dynamically
decoupled has a well defined classical limit, and so can be
investigated within the classical theory.
Note that the claimed decoupling is considerably stronger than
properties of the classical theory implied by
the soft factorization theorems in quantum field theory
\cite{Feige:2014wja,Nande:2017dba} (see Sec.\ \ref{sec:softtheorem}).  The claim is that, first, the classical phase
space $\Gamma$ can be expressed as a product
$\Gamma = \Gamma_{\rm soft} \times \Gamma_{\rm hard}$,
for some definitions of soft and hard sectors $\Gamma_{\rm soft}$ and
$\Gamma_{\rm hard}$ (subject to the loose
requirement that $\Gamma_{\rm soft}$ contains only zero energy
degrees of freedom).  Second, the
classical scattering map\footnote{By scattering map we mean the mapping from the phase space of initial data on past null infinity to the phase space of final data on future null infinity.} ${\cal S} : \Gamma \to \Gamma$ is claimed factorize
as
\be
(s,h) \to [{\bar s}(s), {\bar h}(h) ]
\label{eq:factorizes}
\ee
for $s \in  \Gamma_{\rm
  soft}$ and $h \in \Gamma_{\rm hard}$.  We will argue that such factorization cannot occur.

\subsection{Summary of results}

The purpose of this paper is to reexamine the coupling of hard and soft
degrees of freedom in classical gauge theories.
We focus for simplicity on perhaps the simplest context where these
issues arise, electromagnetism coupled to a massless complex scalar
field in four dimensional Minkowski spacetime.
In this context we argue that
for {\it any} choice of definitions of hard and soft sectors, the two sectors
are always dynamically coupled, and that therefore the soft
conservation laws yield nontrivial constraints on the dynamics.
We expect that our results
will generalize to nonabelian gauge theories and to general
relativity, and will constrain the quantum as well as the classical
theories.
This would support the original arguments of Hawking, Perry and Strominger
that these conservation laws constrain in a nontrivial way the
formation and evaporation of black holes \cite{Hawking:2016msc,HPS-2}.

There are two different versions of our analysis.  The two versions
arise because 
the scattering map is not really a map from phase space $\Gamma$ to
itself, as described above.  Instead it is a mapping ${\cal S} : \Gamma_- \to \Gamma_+$ from the
space $\Gamma_-$ of initial data at past null infinity $\scri^-$
to the space $\Gamma_+$ of final data at future null infinity $\scri^+$.
Our analysis allows for adjusting the definitions of hard and soft
sectors,
but imposes that the ``same'' definitions of hard and sectors 
be used at $\scri^-$ and at $\scri^+$
(otherwise it is always trivially possible to find definitions which
decouple the dynamics).  To make this 
requirement precise
requires an identification of $\Gamma_-$ and $\Gamma_+$.
The first version of our analysis (used in most of the paper) adopts 
the identification
\be
\varphi_{0\,*} : \Gamma_- \to \Gamma_+
\label{firstid}
\ee
which is the pullback action of the mapping 
$\varphi_0 : \scri^+ \to \scri^-$ defined by identifying points that are connected by a
null geodesic through Minkowski spacetime\footnote{
In curved spacetimes this construction is problematic since caustics
can occur.  An alternative prescription would be to choose any 
diffeomorphism $\varphi_0 : \scri^- \to \scri^+$ 
that preserves the appropriate asymptotic structures that determine
the BMS group (see, e.g., Eq.\ (E.5) of Ref.\ \cite{Chandrasekaran:2021vyu}).
Such diffeomorphisms are unique up to compositions with BMS transformations.
Our results
in the Minkowski context would be unchanged if we used any of these 
more general identifications.}.
Using this identification effectively means that we are
considering the modified scattering map ${\cal S}_1 : \Gamma_+ \to
\Gamma_+$
given by
\be
   {\cal S}_1 = {\cal S} \circ {\varphi}_{0\,*}^{-1}.
   \label{calS1:def}
\ee
The second version of our analysis instead identifies $\Gamma_-$ and
$\Gamma_+$ using the free field scattering map ${\cal S}_0$, detailed in
Sec.\ \ref{sec:fos} below.  This differs from 
the identification (\ref{firstid}) due to a
nontrivial action of the free field scattering in the soft sector [Eq.\ (\ref{eq:freee-e}) below].
Using the identification ${\cal S}_0$ effectively means that we are
considering the modified scattering map ${\cal S}_2 : \Gamma_+ \to
\Gamma_+$
given by
\be
   {\cal S}_2 = {\cal S} \circ {\cal S}_0^{-1}.
   \label{calS2:def}
\ee
This definition is similar to using the interaction representation in
quantum mechanics.

In the first version of our analysis, we show that the scattering map ${\cal S}_1$
cannot be of the factorized form (\ref{eq:factorizes}),
and we also exclude the more general forms
\be
(s,h) \to [{\bar s}(s,h), {\bar h}(h) ]
\label{eq:factorizes1}
\ee
and 
\be
(s,h) \to [{\bar s}(s), {\bar h}(s,h) ].
\label{eq:factorizes2}
\ee
A scattering map of the form (\ref{eq:factorizes1}) would
support the arguments against the relevance of soft variables to
information loss, since it would 
imply that the hard variables evolve independently of the soft
variables.
Maps of the form (\ref{eq:factorizes1}) and (\ref{eq:factorizes2}) are automatically excluded
if one assumes that the hard and soft variables are symplectically
orthogonal, that is, have vanishing Poisson brackets with one
another, a necessary condition for a factorization of the total Hilbert space into
hard and soft factors (see Appendix \ref{app:so}).  Our analysis
excluding the forms (\ref{eq:factorizes1}) and (\ref{eq:factorizes2})
does not impose this requirement.     

The second version of our analysis yields different results, since the free
field scattering map ${\cal S}_0$ can in general mix the two sectors
together, depending on the definitions of hard and soft sectors
chosen.  It is possible to define versions of the soft degrees of freedom  which are
exactly conserved by the scattering map ${\cal S}_2$, as pointed out
by Bousso and Poratti (BP) \cite{Bousso}. It is natural to use these
variables to define the soft sector, in which case the general form of
${\cal S}_2$ is
\be
(s,h) \to \left[ s, {\bar h}(s,h) \right].
\label{eeee}
\ee
(Note that this form of mapping is disallowed for ${\cal S}_1$.)
However we show that it is
not possible to adjust the definition of the hard sector to eliminate
the dependence of ${\bar h}$ on $s$ in Eq.\ (\ref{eeee}).  That is, we again exclude completely factorized 
scattering maps of the form (\ref{eq:factorizes}) when we use the
conserved soft variables.

\subsection{Overview of paper}

Our analysis will proceed in three main stages: (i) The development of the
classical scattering theory (Secs.\ \ref{U-1} and \ref{sec:scattering map}); (ii) An
analysis of the arguments that have been made for trivial soft
dynamics (Sec.\ \ref{sec:arguments}); and (iii) Explicit computations of the
scattering map in perturbation theory (Secs.\ \ref{sec:secondorder} and \ref{sec:higherorders}).

The classical scattering theory
of electromagnetism in Minkowski spacetime including soft
degrees of freedom has been treated in detail in Refs.\ 
\cite{Ashtekar:1987tt,Strominger:lectures,Ashtekar:1981bq,Satishchandran:2019pyc,Prabhu:2022mcj},
so we focus in the brief overview given here on one or two points
where we deviate from these previous treatments. 
The theory is based on the following well-known
sequence of steps of the covariant phase space approach
\cite{Crnkovic1987,LW,W-noether-entropy,Harlow:2019yfa}:
(i) Define a field configuration space
${\mathscr F}$ by specifying appropriate
  boundary conditions at spatial and null infinity.
(ii) Define a presymplectic form on the on-shell subspace ${\overline {\mathscr F}}$
  of the
  field configuration space in terms of an integral over Cauchy
  slices, and ensure that it is independent of choice of Cauchy slice
  by adjusting the definition and adjusting the choice of ${\mathscr F}$ if necessary.
(iii) The gauge freedom is defined in terms of degeneracy directions
  of the presymplectic form.  By suitably fixing all the asymptotic
  gauge freedom, define a space of initial data $\Gamma_-$ on
  past null infinity $\scri^-$ and a space of final data $\Gamma_+$ on
  future null infinity $\scri^+$.
(iv) Mod out the space ${\overline {\mathscr F}}$ by degeneracy
  directions of the presymplectic form to obtain the phase space
  $\Gamma$, which can be identified 
    with the space of initial data
  $\Gamma_-$ and with the space of final data $\Gamma_+$. These
identifications define the classical scattering map
${\cal S} : \Gamma_- \to \Gamma_+$ which is a symplectomorphism.

A key role in the analysis is played by the leading order piece of the angular
components of the vector potential in an expansion in $1/r$ about null
infinity, which we denote by ${\cal A}_{\ps A}(u,\theta,\varphi) =
{\cal A}_{\ps A}(u,\bftheta)$ on
$\scri^+$ and by ${\cal A}_{\ms A}(v,\bftheta)$ on $\scri^-$.
We write the electric parity pieces of these fields in terms of
potentials as $D_A \Psi^{\rm
  e}_{\ps}(u,\bftheta)$
and $D_A \Psi^{\rm
  e}_{\ms}(v,\bftheta)$.  Finally we define the (assumed finite)
limiting values of the potentials at spatial infinity $i^0$ or at the
timelike infinities $i^\pm$ as
\be
\Psi^{\rm e}_{\psps} = \lim_{u \to \infty} \Psi^{\rm e}_{\ps},
\ \ \ \ \Psi^{\rm e}_{\psms} = \lim_{u \to -\infty} \Psi^{\rm e}_{\ps},
\ \ \ \
\Psi^{\rm e}_{\msps} = \lim_{v \to \infty} \Psi^{\rm e}_{\ms},
\ \ \ \ \Psi^{\rm e}_{\msms} = \lim_{v \to -\infty} \Psi^{\rm e}_{\ms}.
\ee
It is also convenient to use the following linear combinations of
these quantities:
\be
\Delta \Psi^{\rm e}_{\ps} = \Psi^{\rm e}_{\psps} - \Psi^{\rm
  e}_{\psms}, \ \ \ \
{\bar \Psi}^{\rm e}_{\ps} = \frac{\Psi^{\rm e}_{\psps} + \Psi^{\rm
  e}_{\psms}}{2}, \ \ \ \
\Delta \Psi^{\rm e}_{\ms} = \Psi^{\rm e}_{\msps} - \Psi^{\rm
  e}_{\msms}, \ \ \ \
{\bar \Psi}^{\rm e}_{\ms} = \frac{\Psi^{\rm e}_{\msps} + \Psi^{\rm
  e}_{\msms}}{2}.
\ee
These fields are called soft degrees of freedom since they correspond to distributional
components of the Fourier transforms of the fields $\Psi^{\rm
  e}_{\ps}(u,\bftheta)$ and $\Psi^{\rm e}_{\ms}(v,\bftheta)$ at zero
frequency or zero energy.

Some of the issues that arise in our analysis of the classical
scattering theory that are important for our conclusions are as follows
(see Secs.\ \ref{U-1} and \ref{sec:scattering map} for more details):

\begin{itemize}

\item {\it Use independent coordinates on phase space:} We
  decompose the function $\Psi^{\rm e}_{\ms}(v,\bftheta)$ as   
  \be
  \Psi^{\rm e}_{\ms}(v,\bftheta) = {\tilde \Psi}^{\rm e}_{\ms}(v,\bftheta) + g(v) \Delta \Psi^{\rm e}_{\ms}(\bftheta) + {\bar \Psi}^{\rm e}_{\ms}(\bftheta),
  \ee
  where $g(v)$ is a fixed smooth monotonic function with $g(-\infty) =-1/2$ and $g(\infty) = 1/2$.
  Then the field ${\tilde \Psi}^{\rm e}_{\ms}$ obeys the boundary
  conditions ${\tilde \Psi}^{\rm e}_{\ms} \to 0$ as $v \to \pm
  \infty$.  The fields ${\tilde \Psi}^{\rm e}_{\ms}$, $\Delta
  \Psi^{\rm e}_{\ms}$ and ${\bar \Psi}^{\rm e}_{\ms}$ can be varied
  independently and are thus good coordinates on phase space.  Refs.\ \cite{Strominger:lectures,Bousso}
  use instead the coordinates $\Delta \Psi^{\rm e}_{\ms}$, ${\bar
    \Psi}^{\rm e}_{\ms}$ and $\Psi^{\rm e}_{\ms} - {\bar \Psi}^{\rm
    e}_{\ms}$,  which are not independent since
  \be
  \lim_{v \to \infty} \left[ \Psi^{\rm e}_{\ms} - {\bar \Psi}^{\rm 
      e}_{\ms} \right] = \Delta \Psi^{\rm e}_{\ms}/2.
  \ee
This issue underlies why our conclusions on soft-hard coupling differ
from those of Ref.\ \cite{Bousso}.

\item {\it Ensure uniqueness of presymplectic form:} When large gauge
  transformations that are nontrivial at null infinity are allowed,
  the standard expression for the presymplectic form evaluated at past
  null infinity $\scri^-$ does not coincide with the version evaluated
  at $\scri^+$ [see Eq.\ (\ref{pres2}) and Appendix \ref{app:extend}].
Thus it is problematic to choose one of these to define the presymplectic
form as in Refs. \cite{Ashtekar:1987tt,Ashtekar:1981bq,Satishchandran:2019pyc,Prabhu:2022mcj}.
  The usual way of dealing with this difficulty, suggested by Strominger
 \cite{He:2014cra,Strominger:lectures}, is to restrict the field
 configuration space by imposing a matching condition that relates the
 fields $\Psi^{\rm e}_{\msps}$ and $\Psi^{\rm e}_{\psms}$
 [Eq.\ (\ref{matching22})].  While 
 this condition excludes generic \Lorentz gauge solutions (see
 Appendix \ref{app:freesolns}), it is possible to generalize the framework to
 encompass these solutions (see Sec.\ \ref{sec:gauge-specialization}).

\item {\it Include edge modes:} It is well known that when gauge
  theories are formulated in finite regions of spacetime, the theory
  contains physical degrees of freedom on the boundary, so-called edge
  modes, that are necessary to preserve gauge invariance
  \cite{Donnelly:2016auv,Speranza:2017gxd,2017NuPhB.924..312G,Harlow:2016vwg,Freidel:2018fsk}.
  Whether or not it is necessary to include edge modes that arise at asymptotic boundaries
  is less clear, with some treatments in the literature including them \cite{He:2014cra,Strominger:lectures}
  and others excluding them
  \cite{Ashtekar:1987tt,Ashtekar:1981bq,Bousso:2016wwu,Bousso,Prabhu:2022mcj}.
  We shall find it convenient to include  
  edge modes, but our results would be unchanged if they were
  excluded.  See Appendix \ref{app:edge} for further discussion. 

  The way edge modes arise in the present context is as follows \cite{Strominger:lectures}.
  After some preliminary gauge fixing [Eq.\ (\ref{gaugec1})], the
  potentials transform under large gauge transformations as
  $\Psi^{\rm e}_{\ms}(v,\bftheta) \to \Psi^{\rm e}_{\ms}(v,\bftheta) +
  \varepsilon_{\ms}(\bftheta)$ and
    $\Psi^{\rm e}_{\ps}(u,\bftheta) \to \Psi^{\rm e}_{\ps}(u,\bftheta)
  + \varepsilon_{\ps}(\bftheta)$ for some free functions
  $\varepsilon_{\ms}$ and $\varepsilon_{\ps}$.
  One linear combination of these two functions is
fixed by the matching condition referred to above.
The other linear combination is not a degeneracy direction
of the presymplectic form and thus is not a true gauge degree of freedom.
It is the physical symmetry that underlies the soft conservation laws
\cite{He:2014cra}. The quantity that transforms under this symmetry, a
corrected version of ${\bar \Psi}^{\rm e}_{\ms}$ [Eq. (\ref{hatPsi})], is
essentially the edge mode (see Secs. \ref{sec:gauge-specialization}
and \ref{sec:Dirac} and Appendix \ref{app:edge}).  Note that excluding
this edge mode as in
Refs.\ \cite{Ashtekar:1987tt,Ashtekar:1981bq,Bousso:2016wwu,Bousso,Prabhu:2022mcj}
does not exclude memory effects.

\end{itemize}

We now turn to a consideration of the arguments that have been made
for the decoupling of hard and soft degrees of freedom
\cite{Mirbabayi:2016axw,Bousso,Bousso:2017rsx,Donnelly:2018nbv}
(Sec.\ \ref{sec:arguments}).
There is a theorem in symplectic geometry in finite dimensions
which we review in Sec.\ \ref{sec:theorem} and Appendix
\ref{app:theorem}. It 
says that if a set of phase space functions are
exactly conserved by a symplectomorphism, and their Poisson brackets
with each other are constants, then the symplectomorphism factorizes
as in Eq.\ (\ref{eq:factorizes}).
All of the conditions of this theorem, except for the restriction to
finite dimensions, are satisfied in the context of the scattering map ${\cal S}_2$ defined in
Eq.\ (\ref{calS2:def}) above \cite{Bousso}.
As noted above, it is possible to
define versions of the soft degrees of freedom  which are 
conserved by ${\cal S}_2$.
This arises
because of the soft conservation laws and because of the matching
condition\cite{Bousso}.
Also the Poisson brackets of these soft degrees of freedom
are constants on phase space (Sec.\ \ref{sec:theorem}).
Thus one expects the factorization result to apply; this is
essentially the factorization argument of Bousso and Poratti (BP) \cite{Bousso}\footnote{More precisely, instead of
explicitly invoking the theorem BP attempt to compute
the phase space coordinates that give rise to the
decoupled dynamics (\ref{eq:factorizes}).  The resulting coordinates
are not independent which invalidates the decoupling conclusion.}.
In Sec.\ \ref{sec:theorem} we show explicitly that this argument fails because
the theorem does not apply in our infinite dimensional context.  This
indicates that the soft and hard degrees of freedom are coupled in a manner which
has no analog in finite dimensions.  

Another argument for decoupled soft and hard sectors was given in 
Ref.\ \cite{Mirbabayi:2016axw}, based on the soft factorization
theorems for S-matrix elements in quantum field theory\cite{Weinberg-soft}.
However, this argument is based on an assumption on how the
factorization theorems constrain the quantum theory which we argue is invalid
in Sec.\ \ref{sec:softtheorem}.

The last portion of the paper is an explicit perturbative computation of the
scattering map that confirms that the hard and soft sectors are always
coupled.  The scattering is trivial at second order with the gauge
fixing we use in this paper, as we show in Sec.\ \ref{sec:secondorder}
and Appendix \ref{app:secondorder} (although there is a nonlinear
hard-soft coupling in \Lorentz gauge).
Section \ref{sec:higherorders} considers third and fourth order
scattering.  We first show in Sec.\ \ref{sec:fourth-order} and
Appendix \ref{app:memory} that there are two particular pieces of the
fourth order scattering which are nonzero in general, one of which is
the change in electromagnetic memory
in the scattering process.
The change in memory constitutes a transformation between an incoming purely hard
scalar field and an outgoing electromagnetic field with a nontrivial
soft component, so it is a soft-hard coupling.  We then show in
Sec.\ \ref{sec:firstversioncoupled} that these pieces of the fourth
order scattering cannot be removed by
any adjustment of the definitions of soft and hard sectors, using the
first version ${\cal S}_1$ of the scattering map, ruling out the possibilities
(\ref{eq:factorizes}), (\ref{eq:factorizes1}) and (\ref{eq:factorizes2}).
Section \ref{sec:secondversioncoupled} considers the second version
${\cal S}_2$ of the scattering map, and again shows that the hard and
soft sectors are always coupled, when the soft variable definitions are restricted
to quantities which are conserved by the scattering.

\section{Electromagnetism coupled to a charged scalar field: phase space of asymptotic data}
\label{U-1}

In this section we review the phase space and symplectic form of
electromagnetism coupled to a charged scalar field in Minkowski spacetime, expressed in terms
of data at past null infinity or at future null infinity.
This subject has been treated in detail by Strominger and collaborators
\cite{He:2014cra,Strominger:lectures}, by Ashtekar \cite{Ashtekar:1981bq}, and
by Prabhu, Satishchandran and Wald \cite{Satishchandran:2019pyc,Prabhu:2022mcj}.
We mostly follow these references, but we also highlight below
some points where we deviate from them, as discussed in the Introduction.

To construct the phase space we follow the strategy outlined in the Introduction.
In particular,
we would like to fix the asymptotic gauge degrees of freedom in order
to obtain good coordinates on phase space and a nondegenerate symplectic form that can be inverted to obtain Poisson brackets.
Gauge transformations can be divided into two categories, those which
correspond to degeneracy directions of the presymplectic form (true
gauge transformations) and those which do not.  The latter category
includes the so-called ``large gauge transformations'' that have non-trivial
behavior at infinity and that correspond to the infinity of conserved
charges in gauge theories discovered in the past few years \cite{Strominger:2013jfa,Strominger:2013lka,He:2014cra}.
Our strategy here will be to fix all of the true gauge degrees of
freedom, but avoid fixing any of the gauge degrees of freedom that
correspond to nondegenerate directions of the presymplectic form.

\subsection{Foundations}
\label{sec:found}

The action of the theory is
\be
\label{action}
S = - \frac{1}{4 e^2} \int d^4x \sqrt{-g} F_{ab} F^{ab} - \int d^4x
\sqrt{-g} \, (D^a \Phi)^* D_a \Phi,
\ee
where $A_a$ is the vector potential, $\Phi$ is a complex scalar field,
$e$ is the electric charge, $D_a = \nabla_a - i A_a$, and
$F_{ab} = \nabla_a A_b - \nabla_b A_a$.
The equations of motion
are
\begin{subequations}
  \label{eom00}
  \begin{eqnarray}
\label{currentdef}
    \Box A_a - \nabla_b \nabla_a A^b &=& e^2 j_a = -i e^2 (\Phi \nabla_a \Phi^* - \Phi^* \nabla_a \Phi)
    + 2 e^2 A_a \Phi^* \Phi, \\
    \Box \Phi &=& 2 i A^a \nabla_a \Phi + A^a A_a \Phi + i \Phi
    \nabla_a A^a.
    \label{kleingordon}
  \end{eqnarray}
  \end{subequations}
The theory is invariant under the local gauge transformations
\be
\label{gauge1}
A_a \to A_a + \nabla_a \varepsilon, \ \ \ \ \Phi \to e^{i \varepsilon}
\Phi,
\ee
which may or may not be true gauge transformations.

We now consider asymptotic conditions near future null infinity $\scri^+$.  We
use retarded coordinates $(u,r,\theta,\varphi) = (u,r,\theta^{A})$ in terms of which
the metric is
\be
ds^2 = - du^2 - 2 du dr + r^2 h_{AB} d\theta^A d\theta^B,
\ee
where the unit metric on the two-sphere
is $h_{AB} d\theta^A d\theta^B = d\theta^2 + \sin^2 \theta d\varphi^2$.
We assume the
following asymptotic behavior of the components of the Maxwell tensor
in the limit to $\scri^{+}$, that is, $r \to \infty$ at fixed $u$:
\begin{subequations}
  \label{Fexpand}
  \begin{eqnarray}
    F_{ur} &=& \frac{1}{r^2} {\cal F}_{\ps ur} + \frac{1}{r^3} {\hat {\cal
    F}}_{\ps ur} + O \left( \frac{1}{r^4} \right), \\
F_{uA} &=& {\cal F}_{\ps uA} + \frac{1}{r} {\hat {\cal
    F}}_{\ps uA} + O \left( \frac{1}{r^2} \right), \\
F_{rA} &=& \frac{1}{r^2} {\cal F}_{\ps rA} + \frac{1}{r^3} {\hat {\cal
    F}}_{\ps rA} + O \left( \frac{1}{r^4} \right), \\
F_{AB} &=& {\cal F}_{\ps AB} + \frac{1}{r} {\hat {\cal
    F}}_{\ps AB} + O \left( \frac{1}{r^2} \right).
\end{eqnarray}
\end{subequations}
We are using a notational convention where caligraphic quantities are
used to represent the pieces of fields that appear at leading order in
an expansion in $1/r$, the $+$ subscripts denote quantities on $\scri^+$,
and hatted caligraphic quantities are subleading.
The scalings of the leading terms can be deduced from the physical
arguments given by Strominger \cite{Strominger:lectures}, or from
demanding
smoothness of the solution on the conformal completion of the
spacetime \cite{Ashtekar:2017wgq}.

We assume the following asymptotic behavior of the vector potential
and scalar field as
$r \to \infty$ at fixed $u$, slightly more general than that of
\cite{Strominger:lectures}:
\begin{subequations}
  \label{Aexpand}
  \begin{eqnarray}
    A_A &=& {\cal A}_{\ps A} + \frac{1}{r} {\hat {\cal A}}_{\ps A} + O
    \left( \frac{1}{r^2} \right), \\
    A_u &=& {\cal A}_{\ps u} + \frac{1}{r} {\hat {\cal A}}_{\ps u}
    + \frac{1}{r^2} {\hat {\hat  {\cal A}}}_{\ps u}
    + O\left( \frac{1}{r^3} \right), \\
A_r &=& \frac{1}{r^2} {\cal A}_{\ps r} + \frac{1}{r^3} {\hat {\cal A}}_{\ps r} + O
\left( \frac{1}{r^4} \right), \\
\Phi &=& \frac{1}{r} \chi_{\ps} + \frac{1}{r^2} {\hat \chi}_{\ps} + O\left( \frac{1}{r^3} \right).
\label{eq:twopsix}
      \end{eqnarray}
  \end{subequations}
The expansion coefficients in the expansions (\ref{Fexpand}) and (\ref{Aexpand}) are
then related by
\begin{subequations}
  \label{related}
  \begin{eqnarray}
    \label{Furf}
    {\cal F}_{\ps ur} &=& \partial_u {\cal A}_{\ps r} + {\hat {\cal
        A}}_{\ps u}, \ \ \ \     {\hat {\cal F}}_{\ps ur} = \partial_u
    {\hat {\cal A}}_{\ps r} + 2 {\hat {\hat {\cal
          A}}}_{\ps u},     \\
    \label{FaAf}
    {\cal F}_{\ps uA} &=& \partial_u {\cal A}_{\ps A} - D_A {\cal
        A}_{\ps u}, \ \ \ \         {\hat {\cal F}}_{\ps uA} =
    \partial_u {\hat {\cal A}}_{\ps A} - D_A {\hat {\cal
        A}}_{\ps u},           \\
        \label{FrAf}
    {\cal F}_{\ps rA} &=& -D_A {\cal A}_{\ps r} - {\hat {\cal A}}_{\ps
      A},     \\
\label{FABformula}
    {\cal F}_{\ps AB} &=& D_A {\cal A}_{\ps B} - D_B {\cal A}_{\ps A},
      \end{eqnarray}
  \end{subequations}
where $D_A$ is a covariant derivative with respect to the two-sphere
metric $h_{AB}$.

The components of the current (\ref{currentdef}) are
\be
\label{currentexpand}
j_u = {\cal  J}_{\ps u}/r^2 + O(r^{-3}), \ \ \ \ j_r = {\cal J}_{\ps r}/r^4 +
O(r^{-5}), \ \ \ \ j_{\ps A} =
{\cal J}_A/r^2 + O(r^{-3}),
\ee
where
\begin{subequations}
  \begin{eqnarray}
    \label{calJu}
    {\cal J}_{\ps u} &=& - i (\chi_{\ps} \partial_u \chi_{\ps}^* - \chi_{\ps}^* \partial_u \chi_{\ps})
    + 2 {\cal A}_{\ps u} \chi_{\ps}^* \chi_{\ps}, \label{eq:current-leading} \\
    {\cal J}_{\ps r} &=&  i (\chi_{\ps} {\hat \chi}_{\ps}^* - \chi_{\ps}^* {\hat \chi}_{\ps})
    + 2 {\cal A}_{\ps r} \chi_{\ps}^* \chi_{\ps}, \label{eq:2pteighta} \\
    {\cal J}_{\ps A} &=& - i (\chi_{\ps} \partial_A \chi_{\ps}^* - \chi_{\ps}^* \partial_A \chi_{\ps})
    + 2 {\cal A}_{\ps A} \chi_{\ps}^* \chi_{\ps}.
  \end{eqnarray}
\end{subequations}
The leading order pieces of Maxwell's equations are
\cite{Strominger:lectures}
\begin{subequations}
  \label{maxwellexpand}
  \begin{eqnarray}
    \label{maxwellexpand1}
  \partial_u {\cal F}_{\ps ur} + D^A {\cal F}_{\ps Au}    &=&  e^2
          {\cal J}_{\ps u} , \\          
    {\hat {\cal F}}_{\ps ur} + D^A {\cal F}_{\ps Ar}  &=& e^2 {\cal
      J}_{\ps r}  , \\
    \label{maxwellexpand3}
    - \partial_u {\cal F}_{\ps rA} + {\hat {\cal F}}_{\ps uA} + D^C
    {\cal F}_{\ps CA}  &=& e^2 {\cal J}_{\ps A},
  \end{eqnarray}
\end{subequations}
where $D^A = h^{AB} D_B$.  

The expansion coefficients of the various fields transform under gauge
transformations as follows.  Under the gauge transformation
(\ref{gauge1}) with
\be
\label{gaugeexpand}
\varepsilon = \varepsilon_{\ps} + \frac{1}{r} {\hat \varepsilon}_{\ps} + \frac{1}{r^2} {\hat
  {\hat \varepsilon}}_{\ps} + O\left( \frac{1}{r^3} \right),
\ee
we have
\begin{subequations}
\label{gggs}
  \begin{eqnarray}
\label{gg1}
    {\cal A}_{\ps u} &\to& {\cal A}_{\ps u} + \partial_u \varepsilon_{\ps},
    \ \ \ \ {\hat {\cal A}}_{\ps u} \to {\hat {\cal A}}_{\ps u} +
    \partial_u {\hat \varepsilon}_{\ps}, \ \ \ \ {\hat {\hat {\cal A}}}_{\ps u}
    \to {\hat {\hat {\cal A}}}_{\ps u} +
    \partial_u {\hat {\hat \varepsilon}}_{\ps}, \\ 
        {\cal A}_{\ps A} &\to& {\cal A}_{\ps A} + D_A \varepsilon_{\ps},
    \ \ \ \ {\hat {\cal A}}_{\ps A} \to {\hat {\cal A}}_{\ps A} +
    D_A {\hat \varepsilon}_{\ps}, \\
    \label{gg3}
{\cal A}_{\ps r} &\to& {\cal A}_{\ps r} - {\hat \varepsilon}_{\ps},
\ \ \ \ {\hat {\cal A}}_{\ps r} \to {\hat {\cal A}}_{\ps r} - 2
        {\hat {\hat \varepsilon}}_{\ps}, \\
        \chi_{\ps} & \to& e^{i \varepsilon_{\ps}} \chi_{\ps}, \ \ \ \
        {\hat \chi}_{\ps}  \to e^{i \varepsilon_{\ps}}( {\hat  \chi}_{\ps}
        + i {\hat \varepsilon}_{\ps} \chi_{\ps} ).
  \end{eqnarray}
  \end{subequations}

A similar analysis can be carried out for the limiting behavior of the fields near 
past null infinity $\scri^-$.  We use advanced coordinates $v$, $r$, $\theta^A$ given by $v = u + 2 r$.
The expansion of the Maxwell tensor at $\scri^{-}$, as $r \to \infty$ at fixed $v$, is similar to the expansion (\ref{Fexpand}) and is given by
\begin{subequations}
  \label{Fexpandminus}
  \begin{eqnarray}
    F_{vr} &=& \frac{1}{r^2} {\cal F}_{\ms vr} + \frac{1}{r^3} {\hat {\cal
    F}}_{\ms vr} + O \left( \frac{1}{r^4} \right), \\
F_{vA} &=& {\cal F}_{\ms vA} + \frac{1}{r} {\hat {\cal
    F}}_{\ms vA} + O \left( \frac{1}{r^2} \right), \\
F_{rA} &=& \frac{1}{r^2} {\cal F}_{\ms rA} + \frac{1}{r^3} {\hat {\cal
    F}}_{\ms rA} + O \left( \frac{1}{r^4} \right), \\
F_{AB} &=& {\cal F}_{\ms AB} + \frac{1}{r} {\hat {\cal
    F}}_{\ms AB} + O \left( \frac{1}{r^2} \right).
\end{eqnarray}
\end{subequations}
Here the subscripts $-$ denote expansion coefficients of an expansion near $\scri^-$.
The corresponding expansions of the vector potential and scalar field are 
\begin{subequations}
  \label{Aexpandminus}
  \begin{eqnarray}
    A_A &=& {\cal A}_{\ms A} + \frac{1}{r} {\hat {\cal A}}_{\ms A} + O
    \left( \frac{1}{r^2} \right), \\
    A_v &=& {\cal A}_{\ms v} + \frac{1}{r} {\hat {\cal A}}_{\ms v}
    + \frac{1}{r^2} {\hat {\hat  {\cal A}}}_{\ms v}
    + O\left( \frac{1}{r^3} \right), \\
A_r &=& \frac{1}{r^2} {\cal A}_{\ms r} + \frac{1}{r^3} {\hat {\cal A}}_{\ms r} + O
\left( \frac{1}{r^4} \right), \\
\Phi &=& \frac{1}{r} \chi_{\ms} + \frac{1}{r^2} {\hat \chi}_{\ms} + O\left( \frac{1}{r^3} \right),
      \end{eqnarray}
  \end{subequations}
and the gauge transformation parameter can be expanded as
\be
\label{gaugeexpand1}
\varepsilon = \varepsilon_{\ms} + \frac{1}{r} {\hat \varepsilon}_{\ms} + \frac{1}{r^2} {\hat
  {\hat \varepsilon}}_{\ms} + O\left( \frac{1}{r^3} \right).
\ee

\subsection{Space of solutions of the field equations}
\label{sec:ss}

We will restrict attention to field configurations on $\scri^-$ for
which satisfy three conditions:
\begin{enumerate}
\item The
    limits
    \be
    \lim_{v \to \pm \infty} {\cal A}_{\ms A}, \ \ \ \
    \lim_{v \to \pm \infty} {\cal A}_{\ms r}, \ \ \ \
    \lim_{v \to \pm \infty} {\cal A}_{\ms v},
\label{assumedlimits}
    \ee
    exist, as functions on the two-sphere.
\item The field ${\cal A}_{\ms A}$ satisfies the fall off conditions
  near timelike infinity and spatial infinity
\be
\partial_v {\cal A}_{\ms A} \sim \frac{1}{|v|^{1+\epsilon}},
\ \ \ \ \ v \to \pm \infty,
\label{assumedfalloff0}
\ee
for some $\epsilon > 0$.
This condition is sufficient to ensure the convergence of the
symplectic form (\ref{eq:sympform}) below.

\item The initial data $\chi_{\ms}$ for the scalar field 
  falls off like\footnote{Our assumed fall off for the scalar field is
  stronger than that for the gauge field.  We are excluding for
  simplicity any nontrivial soft behavior on $\scri^-$ in the scalar sector.}
  \be
  \chi_{\ms} \sim \frac{1}{|v|^{1+\epsilon}}, \ \ \ \ v \to \pm \infty,
  \label{assumedfalloff}
  \ee
\end{enumerate}
for some $\epsilon>0$.
We will assume that the conditions (1) and (2) are preserved by the scattering
process, so that solutions which obey these conditions at $\scri^-$
also satisfy analogous conditions at $\scri^+$.

We introduce the following notations for the limiting values of fields on $\scri^-$
at past timelike infinity $i^-$ and at spatial infinity $i^0$.  For
any function $f_{\ms} = f_{\ms}(v,\bftheta)$ defined on $\scri^-$,
where $\bftheta = (\theta^1,\theta^2)$,
we define
\be
\label{conv1}
f_{\msms}(\bftheta) = \lim_{v \to -\infty} f_{\ms}(v,\bftheta), \ \ \ \ f_{\msps}(\bftheta) = \lim_{v \to \infty} f_{\ms}(v,\bftheta).
\ee
Similarly for functions $f_{\ps}$ defined on $\scri^+$ we denote the limiting functions at $i^+$ and at $i^0$ by
\be
\label{conv2}
f_{\psps}(\bftheta) = \lim_{u \to \infty} f_{\ps}(u,\bftheta), \ \ \ \ f_{\psms}(\bftheta) = \lim_{u \to -\infty} f_{\ps}(u,\bftheta).
\ee

We now assume that for solutions to the field equations we
have the following behavior near past timelike infinity \cite{Strominger:lectures}:
\begin{subequations}
  \label{falloffiminus}
  \begin{eqnarray}
    \label{c1}
          {\cal F}_{\msms vr}(\bftheta) &=& 0, \\
     \label{c2}
           {\cal F}_{\msms rA}(\bftheta) &=& 0, \\
     \label{c4}
           {\cal F}_{\msms AB}(\bftheta) &=& 0, \\
     \label{c3}
     {\hat \chi}_{\msms}(\bftheta)  &=& 0.
\end{eqnarray}
\end{subequations}
These conditions can be derived if the initial data for
$\chi_{\ms}$ is of compact support on $\scri^-$.  In this case there exists
a neighborhood of $i^-$ in which $\Phi$ vanishes,
and so in that neighborhood $A_a$ satisfies the homogeneous Maxwell equations with no sources.  Solutions of these equations
satisfy the conditions (\ref{c1}) -- (\ref{c4}) (see Appendix \ref{app:freesolns}).
Similar arguments can be given for the condition (\ref{c3}).  We will
assume that the conditions (\ref{falloffiminus}) continue to be valid
under the weaker assumption (\ref{assumedfalloff}) on the behavior of
$\chi_{\ms}$; this should follow from continuity of the dependence of the solutions of the equations of motion on the initial data on $\scri^-$.
Conditions analogous to (\ref{falloffiminus}) at future timelike infinity $i^+$, namely
\begin{subequations}
  \label{falloffiplus}
  \begin{eqnarray}
    \label{fc1}
          {\cal F}_{\psps ur}(\bftheta) &=& 0, \\
     \label{fc2}
           {\cal F}_{\psps rA}(\bftheta) &=& 0, \\
     \label{fc4}
           {\cal F}_{\psps AB}(\bftheta) &=& 0, \\
     \label{fc3}
     {\hat \chi}_{\psps}(\bftheta)  &=& 0,
\end{eqnarray}
\end{subequations}
should similarly follow
from the assumptions on the final data on
$\scri^+$ that we have discussed.

Another key property of the solutions is the validity of the matching conditions
\begin{subequations}
\label{matching00}
\begin{eqnarray}
  \label{matching}
{\cal F}_{\psms ur}(\bftheta) &=& {\cal P}_* {\cal F}_{\msps
  vr}(\bftheta), \\
\label{Bmatch}
{\cal F}_{\pm AB}(\bftheta) &=& -{\cal P}_* {\cal F}_{\msps AB}(\bftheta).
\end{eqnarray}
\end{subequations}
Here ${\cal P} : S^2 \to S^2$ is the antipodal inversion mapping given by
$(\theta,\varphi) \to (\pi - \theta, \pi+ \varphi)$, and ${\cal P}_*$ is the pullback 
$({\cal P}_*f) (\theta,\varphi) = f(\pi - \theta, \pi+\varphi)$ for functions $f$.
These identities are related to the existence of charges related to large
gauge transformations and were discovered by Strominger
\cite{Strominger:lectures,Strominger:2015bla}.
They were proven rigorously in Minkowski spacetime by Campiglia and Eyheralde
\cite{Campiglia:2017mua}, and generalized to all asymptotically flat
spacetimes by Prabhu \cite{Prabhu:2018gzs}.

We next claim that solutions are determined up to gauge by specifying on $\scri^-$ the initial data
\be
\label{leading}
   {\cal A}_{\ms A}, \ {\cal A}_{\ms v}, \ {\cal A}_{\ms r}, \chi_{\ms}.
   \ee
 Subleading fields can be obtained from these fields from the analog
 on $\scri^-$ of the asymptotic expansion (\ref{maxwellexpand}) of
 Maxwell equations.  
 For example, the subleading field  ${\hat {\cal A}}_{\ms v}$  can be obtained from the leading fields
 (\ref{leading}) and from ${\cal F}_{\ms vr}$, from the analog of Eq.\ (\ref{Furf}) on $\scri^-$.
 In turn, the field ${\cal F}_{\ms vr}$ can be obtained from its
 evolution equation, the analog on $\scri^-$ of
 Eq.\ (\ref{maxwellexpand1}), together with the initial condition
 (\ref{c1}) at $v=-\infty$.  Similarly ${\hat {\cal A}}_{\ms A}$ is
 obtained from ${\cal F}_{\ms rA}$ from the analogs of Eqs.\ (\ref{FrAf}), the evolution equation 
 (\ref{maxwellexpand3}) and the initial condition (\ref{c2}).
Similar arguments apply to the subleading scalar field ${\hat \chi}_{\ms}$
using an expansion of the scalar field equation (\ref{kleingordon}). 
 
Therefore we can take the four fields (\ref{leading}) on $\scri^-$ to parameterize
the phase space (up to gauge transformations which we discuss below).
Similarly the fields 
\be
\label{leading1}
   {\cal A}_{\ps A}, \ {\cal A}_{\ps u}, \ {\cal A}_{\ps r},
   \chi_{\ps}
      \ee
on $\scri^+$ also uniquely determine the solution, up to gauge transformations.

\subsection{Presymplectic form}

The presymplectic form obtained from the action (\ref{action}) depends on
a pair of linearized perturbations $(\delta_1 A_a, \delta_1 \Phi)$ and
$(\delta_2 A_a, \delta_2 \Phi)$ about a background solution
$A_a,\Phi$ \cite{WZ}.  It is given by the integral over any Cauchy
surface $\Sigma$,
\be
\label{presym}
\Omega_\Sigma(A_a,\Phi; \delta_1 A_a, \delta_1 \Phi, \delta_2 A_a,
\delta_2 \Phi) = \int_\Sigma \omega_{abc}
\ee
of the $3$-form
\be
\omega_{abc} = \frac{1}{e^2} \epsilon_{abcd} \delta_1 F^{df} \delta_2
A_{ f}
+ \epsilon_{abcd} ( D^d \delta_1 \Phi^* \delta_2 \Phi + D^d \delta_1
\Phi \delta_2 \Phi^*) - (1 \leftrightarrow 2).
\ee
Here the orientation of $\Sigma$ is that determined by $t^a \epsilon_{abcd}$ where $t^a$ is any future-pointing timelike vector field.
We define
\be
\label{Omegapm}
\Omega_{\scri^-} =  \int_{\scri^-} \omega_{abc}, \ \ \ \ \Omega_{\scri^+} = \int_{\scri^+} \omega_{abc},
\ee
the limiting integrals over past and future null infinity.  The limits to $\scri^+$ and $\scri^-$ of $\omega_{abc}$ exist
by virtue of our assumed expansions (\ref{Fexpand}), (\ref{Aexpand}), (\ref{Fexpandminus}) and
(\ref{Aexpandminus}).  We will discuss in Sec.\ \ref{sec:Dirac} below the conditions necessary for the integrals
(\ref{Omegapm}) to converge and be finite.

Before considering the presymplectic forms (\ref{Omegapm}) for general perturbations, it will be useful to
first specialize to the case where the second perturbation is a pure
gauge transformation,
\be
\delta_1 A_a = \delta A_a, \ \ \ \ \ \delta_1 \Phi = \delta \Phi, \ \ \ \ \ \delta_2 A_a = \delta_\varepsilon A_a =  \nabla_a \varepsilon,
\ \ \ \ \ \delta_2 \Phi = \delta_\varepsilon \Phi = i \varepsilon \Phi,
\label{puregauge}
\ee
In this case $\omega_{abc}$ is always exact \cite{WZ}, and here we have 
$\omega = - d \delta Q$,
where
\be
\label{exact}
Q_{ab} = \frac{1}{2 e^2} \varepsilon \, \epsilon_{abcd} \, F^{cd}.
\ee
We assume expansions of the form (\ref{gaugeexpand}) for the gauge transformation
function $\varepsilon$ near $\scri^+$ and $\scri^-$, and we further
assume that at each angle $\bftheta$ the leading order coefficients $\varepsilon_{\ps}$,
$\varepsilon_{\ms}$ asymptote to constants as $|u|$ or $|v|$ go to infinity,
to be consistent the assumed asymptotic behavior of the fields at $i^\pm$ and $i^0$ discussed above Eq.\ (\ref{conv2}).
Following the
notational conventions (\ref{conv1}) and (\ref{conv2}) these limiting values
will be denoted $\varepsilon_{\psps}$, $\varepsilon_{\psms}$, $\varepsilon_{\msps}$
and $\varepsilon_{\msms}$.
Converting the integrals over
$\scri^+$ and $\scri^-$ to integrals over their boundaries using Eq.\ (\ref{exact}) we obtain
\begin{subequations}
  \label{omegas}
  \begin{eqnarray}
    \label{Omegaplus0}
\Omega_{\scri^+} &=& - \frac{1}{e^2} \int d^2 \Omega 
\left[ -\delta {\cal F}_{\psps ru} \varepsilon_{\psps} + \delta {\cal
    F}_{\psms ru} \varepsilon_{\psms} \right], \\
\Omega_{\scri^-} &=& - \frac{1}{e^2} \int d^2 \Omega 
\left[ -\delta {\cal F}_{\msms rv}
  \varepsilon_{\msms}
  + \delta {\cal F}_{\msps rv} \varepsilon_{\msps}  \right].
\label{pres1}
  \end{eqnarray}
  \end{subequations}
We now simplify using the asymptotic conditions (\ref{c1}) and
(\ref{fc1}), and the matching condition (\ref{matching}).  This gives
\be
\Omega_{\scri^+} - \Omega_{\scri^-} = - \frac{1}{e^2} \int d^2 \Omega 
\, \delta {\cal F}_{\psms ru} \left[  \varepsilon_{\psms} - {\cal P}_* \varepsilon_{\msps}
    \right].
\label{pres2}
\ee

Now we would like to specialize our definition of the field
configuration space to make
\be
\Omega_{\scri^+} = \Omega_{\scri^-}
\label{unitarity}
\ee
for general on-shell perturbations, that is, to
make the presymplectic forms on past null infinity 
and future null infinity coincide. In other words, the scattering map
from data at past null infinity to data at future null infinity should
be a symplectomorphism (the classical version of unitarity in the quantum
theory).  From Eq.\ (\ref{pres2}) we see that the condition (\ref{unitarity}) cannot
be preserved under general transformations of the form $A_a \to A_a + \nabla_a \varepsilon$.
In the next section we will discuss 
a specialization
of the definition of the field configuration space suggested by
Strominger \cite{He:2014cra} 
which makes
the quantity (\ref{pres2}) vanish, and thus removes the obstruction to
achieving (\ref{unitarity}) for general perturbations.

\subsection{Gauge specialization and scattering map}
\label{sec:gauge-specialization}

We now fix some of the gauge degrees of freedom that 
that correspond to degeneracy directions of the
presymplectic form (\ref{presym}), and so correspond to true gauge degrees of freedom. 
First, we use the free function $\varepsilon_{\ps}$ to set ${\cal A}_{\ps
  u}$
to zero, using Eq.\ (\ref{gg1}).  In doing so we specialize to
$\varepsilon_{\psms} = \varepsilon_{\ps}(u=-\infty)=0$, in order to correspond to a degeneracy
direction of (\ref{Omegaplus0}), from Eq.\ (\ref{fc1}).
Next, we use the free function ${\hat
  \varepsilon}_{\ps}$ to set ${\cal A}_{\ps r}$ to zero, from Eq.\ (\ref{gg3}).
We perform similar specializations at $\scri^-$. Summarizing, we have
fixed the gauge so that
\be
\label{gaugec1}
{\cal A}_{\ps u} = {\cal A}_{\ps r} = {\cal A}_{\ms v} = {\cal A}_{\ms
  r} = 0.
\ee
The remaining gauge freedom that acts on the data on $\scri^+$ and $\scri^-$ consists of functions $\varepsilon_{\ps} = 
\varepsilon_{\ps}(\bftheta)$ and $\varepsilon_{\ms} = \varepsilon_{\ms}(\bftheta)$
that are functions of angle only.  In particular we have
$\varepsilon_{\psms} = \varepsilon_{\psps} = \varepsilon_{\ps}$ and
$\varepsilon_{\msps} = \varepsilon_{\msms} = \varepsilon_{\ms}$.
We define the even and odd linear
combinations of these gauge transformations via
\begin{subequations}
  \label{sigmaeo}
  \begin{eqnarray}
    \label{sigmaeven}
\varepsilon_{\rm e} &=& \frac{1}{2} (\varepsilon_{\ps} + {\cal P}_*
\varepsilon_{\ms}), \\
\label{sigmaodd}
\varepsilon_{\rm o} &=& \frac{1}{2} (\varepsilon_{\ps} - {\cal P}_* \varepsilon_{\ms}).
  \end{eqnarray}
  \end{subequations}

We next decompose the fields ${\cal A}_{\ps A}$ and ${\cal A}_{\ms A}$ into electric and magnetic parity pieces on the two sphere,
as
\begin{subequations}
  \label{AAdecompose}
\begin{eqnarray}
  {\cal A}_{\ps A} &=& D_A \Psi_{\ps}^{\rm e} + \varepsilon_{AB} h^{BC} D_C \Psi_{\ps} ^{\rm m}, \\
  {\cal A}_{\ms A} &=& D_A \Psi_{\ms}^{\rm e} + \varepsilon_{AB} h^{BC} D_C \Psi_{\ms} ^{\rm m},
  \label{AAdecompose1}
\end{eqnarray}
\end{subequations}
which determines the potentials $\Psi_{\ps}^{\rm e}$ etc. up to their $l=0$ parts which we take to vanish.
Under the gauge transformation given by Eqs.\ (\ref{gaugeexpand}) and (\ref{gaugeexpand1}) we have
\begin{subequations}
  \label{ggg0}
  \begin{eqnarray}
    \label{ggg1}
        \Psi^{\rm e}_{\ms} &\to& \Psi^{\rm e}_{\ms} + \varepsilon_{\ms}, \ \ \ \     \Psi^{\rm e}_{\ps} \to \Psi^{\rm e}_{\ps} + \varepsilon_{\ps}, \\
    \Psi^{\rm m}_{\ms} &\to& \Psi^{\rm m}_{\ms}, \ \ \ \     \Psi^{\rm m}_{\ps} \to \Psi^{\rm m}_{\ps},
  \end{eqnarray}
\end{subequations}
assuming that $\varepsilon_{\ms}$ and $\varepsilon_{\ps}$ have no $l=0$ pieces.

Finally, following Strominger \cite{He:2014cra}, we specialize our definition of the field configuration space
by imposing the
condition\footnote{Note that the corresponding relation
 for the magnetic potentials $\Psi^{\rm m}$,
 \be
 \label{matching22}
 \Psi^{\rm m}_{\psms} =  {\cal P}_* \Psi^{\rm m}_{\msps},
\ee
is a consequence of the matching condition (\ref{Bmatch}) together with Eqs.\ (\ref{FABformula}) and (\ref{AAdecompose}).}
\be
\label{matching2}
\Psi^{\rm e}_{\psms} =  {\cal P}_* \Psi^{\rm e}_{\msps},
 \ee
which can be achieved by using the odd gauge transformation (\ref{sigmaodd}), from Eqs.\ (\ref{ggg1}) and (\ref{AAdecompose}). 
The condition
(\ref{matching2}) then eliminates
the odd gauge transformation freedom, and consequently the quantity
(\ref{pres2}) vanishes as desired, removing the obstruction to the
scattering symplectomorphism property (\ref{unitarity}) discussed in the last
section.  In Appendix \ref{app:extend} we generalize this result and show that the symplectic forms on $\scri^-$ and $\scri^+$ coincide
in general, and not just for field configurations of the form (\ref{puregauge}),
using results of Campiglia and Eyheralde\cite{Campiglia:2017mua}.
We also show there that the condition (\ref{matching2}) follows from
imposing a Lorenz-like gauge condition on the fields to leading order
in an expansion around spatial infinity. 
We note that the condition (\ref{matching2}) is generally incompatible with
\Lorentz gauge (see Appendix \ref{app:freesolns}); this point is
discussed further after Eq.\ (\ref{scatter0}) below.

The condition (\ref{matching2}) is
  called
  a ``matching condition'' in Refs.\ \cite{He:2014cra,Bousso}.
  However its status is very different from that of the matching
  conditions (\ref{matching00}), being a specialization
of the definition of the configuration space
rather than a
property of generic physical solutions.  Note also that it is not accurate to call it 
a gauge specialization, even though it is enforced by making use of the transformations (\ref{sigmaodd}),
since true gauge transformations are defined in terms of degeneracy directions of the presymplectic
form, and the specialization (\ref{matching2}) is needed before a unique presymplectic form can be defined.

The condition (\ref{matching2}) together with the gauge fixing (\ref{gaugec1})
determines a unique gauge for initial data on $\scri^-$ and final data
on $\scri^+$ which we will call the {\it preferred asymptotic
  gauge}.  From Eq.\ (\ref{leading}), the space $\Gamma_-$ of initial
data in this gauge consists of the pairs $({\cal A}_{\ms A},
\chi_{\ms})$ on $\scri^-$.
Similarly the space $\Gamma_+$ of final data consists of the pairs 
$({\cal A}_{\ps A},
\chi_{\ps})$ on $\scri^+$.

We do not fix the even transformation freedom (\ref{sigmaeven}), as it
corresponds to a nondegenerate  
direction of the presymplectic form, so it is a physical symmetry
transformation rather than a gauge freedom.  The corresponding charges
are the new conserved charges of
\cite{Strominger:lectures}.  Specifically, the variation in the charge
is given by \footnote{This is Eq.\ (8)  of Ref.\ \cite{WZ} but with
the sign flipped.
We use the sign convention of Ref.\ \cite{Strominger:lectures}
for associating charges with symmetries, in agreement
with Sec.\ 3.3 of \cite{Harlow:2019yfa}, but opposite to that of
\cite{WZ}. This choice 
gives the conventional sign for the charge (\ref{Echarge}).}
\be
\delta Q_\varepsilon = -\Omega(A_a,\Phi; \delta A_a, \delta \Phi, \delta_\varepsilon A_a,
\delta_\varepsilon \Phi),
\ee
and using Eqs.\ (\ref{omegas}) and (\ref{sigmaeven}) and integrating in phase space gives \cite{Strominger:lectures}
\be
Q_\varepsilon = 
\frac{1}{e^2} \int d^2 \Omega \,
     {\cal F}_{\psms ru} \varepsilon_{\psms}=
     \frac{1}{e^2} \int d^2 \Omega \, 
 {\cal F}_{\msps rv} \varepsilon_{\msps}.
\label{Echarge}
\ee
There are also magnetic conserved charges
analogous to the electric charges (\ref{Echarge})
\cite{Strominger:2015bla}, which we review in 
Appendix \ref{app:fieldconfig-magsector}.
For simplicity will restrict attention to the sector of the theory
where all the magnetic charges (\ref{magcharges}) vanish, 
which by Eqs.\ (\ref{FABformula}), (\ref{AAdecompose}), (\ref{kkk1}) and (\ref{kkk2}) is equivalent to the requirement that
\be
\Psi_{\psms}^{\rm m} = \Psi_{\msps}^{\rm m} = 0.
\label{kkk3}
\ee

To summarize, the field configuration space ${\mathscr F}$ of the theory
is given by 
the set of fields
that obey the asymptotic conditions (\ref{Aexpand}) and
(\ref{Aexpandminus}), the matching condition (\ref{matching2}), the
vanishing magnetic charges condition (\ref{kkk3}), and
for which the initial data on $\scri^-$ obeys the conditions
(\ref{assumedlimits}) -- (\ref{assumedfalloff}) and the final data
satisfies analogous conditions on $\scri^+$.
The phase space $\Gamma$
of the theory is the on-shell subspace ${\overline {\mathscr F}}$ of the field configuration space,
modded out by degeneracy directions of the presymplectic form
\cite{Harlow:2019yfa}, which correspond to gauge transformations that
act trivially on the boundaries $\scri^-$ and $\scri^+$.
This phase space is in one-to-one correspondence with the space
$\Gamma_-$ of initial data $({\cal A}_{\ms A}, \chi_{\ms})$ on
$\scri^-$ in the preferred asymptotic gauge.  It is similarly in one-to-one correspondence with the space $\Gamma_+$
of final data $({\cal A}_{\ps A}, \chi_{\ps})$ on $\scri^+$ in the preferred asymptotic gauge.
In the remainder of the paper we will be concerned with properties of
the scattering map
\be
   {\cal S}: \Gamma_- \to \Gamma_+
   \label{scatter0}
\ee
which is a symplectomorphism.

While our on-shell field configuration space ${\overline {\mathscr F}}$ is well defined, it is also useful to
consider the larger space ${\overline {\mathscr F}}_{\rm ext}$ of solutions that is obtained by
relaxing the matching condition (\ref{matching2}).  This larger
space is necessary, for example, to accommodate \Lorentz gauge
solutions, as we show in Appendix \ref{app:freesolns}.  We extend the
definition of the presymplectic form $\Omega$ from ${\overline {\mathscr F}}$ to ${\overline {\mathscr
    F}}_{\rm ext}$ as follows.  Given a pair of solutions in ${\overline {\mathscr F}}_{\rm
  ext}$, we transform that pair to preferred asymptotic gauge
using the transformations (\ref{sigmaodd}),
and
then use the  
prescription for $\Omega$ discussed above.
Modding out by degeneracy directions of $\Omega$ then yields the same
phase space $\Gamma$ as before.
From the point of view of
the extended space ${\overline {\mathscr F}}_{\rm ext}$, the role of the matching condition
(\ref{matching2}) is to modify the definition of the presymplectic
form rather than to restrict the field configuration space.
In the remainder of the paper we will work mostly within the extended 
space ${\overline {\mathscr F}}_{\rm ext}$, transforming back and forth between \Lorentz
gauge and preferred asymptotic gauge as needed.

\subsection{Soft and hard variables and Poisson brackets}
\label{sec:Dirac}

The presymplectic form (\ref{presym}) when evaluated on the space
$\Gamma_-$ of initial data $({\cal A}_{\ms A}, \chi_{\ms})$ is now 
nondegenerate, so from now on we will call it the symplectic form. It
can be written as
\begin{eqnarray}
  \label{eq:sympform}
&&\Omega_{\scri^-}(A_a,\Phi; \delta_1 A_a, \delta_1 \Phi, \delta_2 A_a,
\delta_2 \Phi) =  \int dv \int d^2 \Omega  \left[
  \frac{1}{e^2} h^{AB} \partial_v \delta_1 {\cal A}_{\ms A} \delta_2
       {\cal A}_{\ms B} - ( 1 \leftrightarrow 2) \right] \nonumber \\
&&  + \int dv \int d^2 \Omega  \left[
  \partial_v \delta_1
  \chi_{\ms}^* \delta_2 \chi_{\ms}
+    \partial_v \delta_1
  \chi_{\ms} \delta_2 \chi_{\ms}^*
- (1 \leftrightarrow 2)  \right].
\end{eqnarray}
In terms of the potentials 
$\Psi^{\rm e}$ and $\Psi^{\rm m}$ defined in Eq.\ (\ref{AAdecompose})
the symplectic form is
\begin{eqnarray}
  \label{presymfinal1}
&&\Omega_{\scri^-} =  -\frac{1}{e^2} \int dv \int d^2 \Omega  \left[
  \partial_v \Psi^{\rm e}_{1\,\ms} D^2 \Psi^{\rm
    e}_{2\,\ms}
+  \partial_v \Psi^{\rm m}_{1\,\ms} D^2 \Psi^{\rm
    m}_{2\,\ms}
         - ( 1 \leftrightarrow 2) \right] \nonumber \\
&&  + \int dv \int d^2 \Omega  \left[
  \partial_v \delta_1
  \chi_{\ms}^* \delta_2 \chi_{\ms}
+    \partial_v \delta_1
  \chi_{\ms} \delta_2 \chi_{\ms}^*
- (1 \leftrightarrow 2)  \right].
\end{eqnarray}
Here for simplicity we have written $\Psi^{\rm e}_1$ instead of
$\delta_1 \Psi^{\rm e}$ etc.

We next make a change of coordinates on $\Gamma_-$, from
the set
$
\left[ \Psi^{\rm e}_{\ms}(v,\bftheta), \Psi^{\rm
    m}_{\ms}(v,\bftheta), \chi_{\ms}(v,\bftheta) \right]
$
to a new set
\be
\left[
  {\tilde \Psi}^{\rm e}_{\ms}(v,\bftheta),
  \Psi^{\rm m}_{\ms}(v,\bftheta),
  \chi_{\ms}(v,\bftheta),
  \Delta \Psi^{\rm e}_{\ms}(\bftheta),
  {\bar \Psi}^{\rm e}_{\ms}(\bftheta)
  \right]
\label{newset}
\ee
defined as follows.  We pick a smooth function $g(v)$ which increases
monotonically with boundary values
\be
g(-\infty) = -1/2, \ \ \ \ g(\infty) = 1/2.
\label{gbvals}
\ee
We
define
\begin{subequations}
  \label{transform}
  \begin{eqnarray}
    \label{DeltaPsidef}
    \Delta \Psi^{\rm e}_{\ms} &=& \Psi^{\rm e}_{\msps} - \Psi^{\rm      e}_{\msms}, \\
\label{barPsidef}
    {\bar \Psi}^{\rm e}_{\ms} &=& \frac{1}{2} \left( \Psi^{\rm e}_{\msps} + \Psi^{\rm e}_{\msms} \right),\\
{\tilde \Psi}^{\rm e}_{\ms}(v,\bftheta) &=& \Psi^{\rm
  e}_{\ms}(v,\bftheta) - {\bar \Psi}^{\rm e}_{\ms}(\bftheta) - g(v)
{\Delta \Psi}^{\rm e}_{\ms}(\bftheta).
\label{third}
  \end{eqnarray}
\end{subequations}
Note that it follows from these definitions that
\be
   {\tilde \Psi}^{\rm e}_{\msps} = {\tilde \Psi}^{\rm e}_{\msms} = 0.
\label{bc33}
   \ee
We will use the terminology ``hard variables'' for the 
the quantities ${\tilde \Psi}^{\rm e}_{\ms}$, $\Psi^{\rm m}_{\ms}$, and $\chi_{\ms}$ which depend on $u$, 
and ``soft variables'' for the
the quantities $\Delta \Psi^{\rm e}_{\ms}$, ${\bar \Psi}^{\rm e}_{\ms}$
which do not and which correspond to zero energy degrees of freedom, following 
Ref.\ \cite{Strominger:lectures}.
We similarly define the phase space variables 
\be \label{eq:phasespace-variables}
\left[
  {\tilde \Psi}^{\rm e}_{\ps}(u,\bftheta),
  \Psi^{\rm m}_{\ps}(u,\bftheta),
  \chi_{\ps}(u,\bftheta),
  \Delta \Psi^{\rm e}_{\ps}(\bftheta),
  {\bar \Psi}^{\rm e}_{\ps}(\bftheta)
  \right]
\ee
at future null infinity $\scri^+$, using the same function $g$.

It may seem strange that we have to introduce an arbitrary function
$g(v)$ in order to to separate out
the soft variables from the hard variables.  However, it is necessary
to make such a choice in order to get a complete separation.  Of
course, the choice of $g(v)$ is arbitrary and no observable quantities
will depend on this choice.  Any two phase space coordinate systems
corresponding to 
two different choices of $g(v)$ are related by a symplectomorphism
[see Eqs.\ (\ref{pbs}) below, which are independent of $g$].

We note that a transformation of phase space variables similar to our
transformation (\ref{transform})
was used in
Refs.\ \cite{Strominger:lectures,Bousso}, except that those authors
did not include the third term on the right hand side of Eq.\ (\ref{third}).
As a consequence, their variables are not independent, obeying the constraint (in our notation) of
$\Delta \Psi^{\rm e}_{\ms} = {\tilde \Psi}^{\rm e}_{\ms}(u=\infty) - {\tilde \Psi}^{\rm e}_{\ms}(u=-\infty)$.
This lack of independence of phase space coordinates is the key reason why our results
on the coupling of hard and soft degrees of freedom differ from those of Ref.\ \cite{Bousso},
as discussed in more detail in Sec.\ \ref{sec:theorem} below.

Rewriting the symplectic form (\ref{presymfinal1}) in terms of the new variables (\ref{newset}) gives
\begin{eqnarray}
  \label{presymfinal2}
\Omega_{\scri^-} &=&  - \frac{1}{e^2} \int dv \int d^2 \Omega  \left[
  \partial_v {\tilde \Psi}^{\rm e}_{1\,\ms} D^2 {\tilde \Psi}^{\rm
    e}_{2\,\ms}
+  \partial_v \Psi^{\rm m}_{1\,\ms} D^2 \Psi^{\rm
    m}_{2\,\ms}
- ( 1 \leftrightarrow 2) \right] \nonumber \\
 && - \frac{1}{e^2} \int d^2 \Omega  \left\{
  D^2 \Delta \Psi^{\rm e}_{1\,\ms} \left[ {\bar \Psi}^{\rm
    e}_{2\,\ms} + 2 \int dv g^\prime {\tilde \Psi}^{\rm e}_{2 \ms} \right]
         - ( 1 \leftrightarrow 2) \right\} \nonumber \\
&&  + \int dv \int d^2 \Omega  \left[
  \partial_v \delta_1
  \chi_{\ms}^* \delta_2 \chi_{\ms}
+    \partial_v \delta_1
  \chi_{\ms} \delta_2 \chi_{\ms}^*
- (1 \leftrightarrow 2)  \right].
\end{eqnarray}
We see that the soft and hard phase space variables (\ref{newset}) are not symplectically orthogonal,
that is, there are nonvanishing Poisson brackets between the hard and soft variables.  This can be remedied
by defining the new soft variable
\be
   {\hat \Psi}^{\rm e}_{\ms}(\bftheta) = {\bar \Psi}^{\rm e}_{\ms}(\bftheta) + 2 \int dv g'(v) {\tilde \Psi}^{\rm e}_{\ms}(v,\bftheta).
\label{hatPsi}
   \ee
   Then the soft variables $\Delta \Psi^{\rm  e}_{\ms}(\bftheta)$, ${\hat \Psi}^{\rm e}_{\ms}(\bftheta)$ and the hard variables 
$ {\tilde \Psi}^{\rm e}_{\ms}(v,\bftheta), \Psi^{\rm
    m}_{\ms}(v,\bftheta), \chi_{\ms}(v,\bftheta)$ are symplectically orthogonal, from Eq.\ (\ref{presymfinal2}). 
   The corresponding nonzero Poisson brackets can be obtained
from Eqs.\ (\ref{presymfinal2}) and (\ref{hatPsi}) 
by
expanding in spherical harmonics and using the boundary conditions
(\ref{assumedfalloff}), (\ref{bc33}), (\ref{kkk1}) and (\ref{kkk3}), 
and are\footnote{Note that many of these
  Poisson bracket relations are not continuous in the limits $v, v' \to
  \pm \infty$.  That is, for example, the limit $v \to \pm \infty$ of
  a Poisson bracket of a field will not coincide with the Poisson
  bracket of the limit of the field.  However, there is no inconsistency.}\ \footnote{These Poisson brackets agree with those of Ref.\ \cite{Strominger:lectures} but not those of Ref.\ \cite{Ashtekar:1987tt}.  We believe that the right hand side of Eq.\ (C.5) of Ref.\ \cite{Ashtekar:1987tt} should be proportional to the derivative of a delta function instead of a step function.}  
\begin{subequations}
  \label{pbs}
\begin{eqnarray}
\label{pbs1}
  \left\{ {\tilde \Psi}^{\rm e}_{\ms}(v,\bftheta) \ , \ D^2 {\tilde \Psi}^{\rm e}_{\ms}(v',\bftheta') \right\}  &=&
   \frac{e^2}{2} \left[ \Theta(v-v') - \frac{1}{2} \right]
   \left[\delta^{(2)}(\bftheta,\bftheta') - \frac{1}{4 \pi} \right],  \\
   \label{pbs2}
    \left\{ \Psi^{\rm m}_{\ms}(v,\bftheta) \ , \ D^2 \Psi^{\rm m}_{\ms}(v',\bftheta') \right\}  &=&
 \frac{e^2}{2} \left[ \Theta(v-v') - \frac{1}{2} \right] \left[
   \delta^{(2)}(\bftheta,\bftheta') - \frac{1}{4 \pi} \right],  \\
 \label{pbs3}
 \left\{
      \Delta \Psi^{\rm e}_{\ms}(\bftheta)     
     \ , \
D^2 {\hat \Psi}^{\rm e}_{\ms}(\bftheta')
     \right\}
 &=&  e^2 \left[ \delta^{(2)}(\bftheta,\bftheta') - \frac{1}{4 \pi}
  \right], \\
\label{pbs4}
  \left\{ \chi_{\ms}(v,\bftheta) \ , \chi_{\ms}^{*}(v',\bftheta') \right\}  &=&
   -\frac{1}{2} \left[ \Theta(v-v') - \frac{1}{2} \right]
  \delta^{(2)}(\bftheta,\bftheta').
\end{eqnarray}
\end{subequations}
Here $\Theta(x)=1$ for $x>0$ and $\Theta(x)=0$ for $x<0$, and
$\delta^{(2)}(\bftheta,\bftheta') = \delta(\theta-\theta') \delta(\varphi - \varphi') /\sqrt{h}$ is the covariant delta function on the unit sphere.  The
formulae (\ref{pbs1}) -- (\ref{pbs3}) contain factors of $D^2$ inside the Poisson
brackets, but they determine the corresponding formulae without
factors of $D^2$ since  
all of the functions of $\bftheta$ (except $\chi_{\ms}$) have no $l=0$ components.

Note that it follows from Eqs.\ (\ref{transform}) and (\ref{pbs}) that
the field $\Psi^{\rm  
  e}_{\ms}(v,\bftheta)$ satisfies the same Poisson bracket relation
(\ref{pbs1}) as the field ${\tilde \Psi}^{\rm e}_{\ms}(v,\bftheta)$.
However, one cannot replace Eqs.\ (\ref{pbs1}) and (\ref{pbs3}) with
the single equation (\ref{pbs1}) with ${\tilde \Psi}^{\rm e}_{\ms}(v,\bftheta)$
replaced by $\Psi^{\rm    e}_{\ms}(v,\bftheta)$, because limits $u \to \pm \infty$ of the
Poisson brackets do not coincide with Poisson brackets of limits \cite{Strominger:lectures}.

We remark that the soft degrees of freedom 
$(\Delta \Psi^{\rm e}_{\ms}, {\hat \Psi}^{\rm e}_{\ms})$
can alternatively be described in the language
of edge modes \cite{Donnelly:2016auv,Speranza:2017gxd,2017NuPhB.924..312G,Harlow:2016vwg},
as was conjectured in Ref.\ \cite{Strominger:lectures}.
This equivalence is outlined in Appendix \ref{app:edge}.

Turn now to the situation at future null infinity $\scri^+$.  An
analogous analysis yields versions of the formulae (\ref{presymfinal2}) --
(\ref{pbs}) with $v$ replaced by $u$ everywhere, and with the subscripts $+$
replaced by subscripts $-$.

Finally, we note that the main differences between the construction of the phase
space given here and previous treatments
\cite{Strominger:lectures,Ashtekar:1987tt,Ashtekar:1981bq} are
as follows:
\begin{itemize}

\item Ashtekar \cite{Ashtekar:1987tt,Ashtekar:1981bq} 
  constructs the phase space by imposing $\Psi^{\rm
  e}_{\psms} = \Psi^{\rm e}_{\msps} = 0$ instead of the matching
  condition (\ref{matching2}) suggested by Strominger.  This eliminates one of the physical
  degrees of freedom (the edge mode) and also the symmetry that underlies the
  conservation laws (\ref{Echarge}).
  Including the extra degree of
  freedom will modify the character of the quantum theory constructed in
  \cite{Ashtekar:1987tt}. Similarly, the analysis of infrared
  scattering of the recent paper \cite{Prabhu:2022mcj} is based on an
  algebra obtained by excluding this degree of freedom.  See Appendix
  \ref{app:edge} for further discussion.

\item In fixing the gauge we restrict to degeneracy directions of the presymplectic form,
and demonstrate that the gauge conditions (\ref{gaugec1}) used are
degeneracy directions\footnote{An exception is the gauge condition
  (\ref{matching2}) that we have adopted which is not a degeneracy
  direction of the presymplectic forms (\ref{Omegapm}).  However
  $\Omega_{\scri^+}$ and $\Omega_{\scri^-}$ do not coincide until
  after this condition is imposed, so one can argue that
  (\ref{Omegapm}) is not the correct presymplectic form until after
  the condition is imposed.}.  

\item In separating out hard and soft variables, we use a coordinate
system on phase space in which all the variables are independent,
unlike the set of variables in Refs.\   
\cite{Strominger:lectures,Bousso}.  This change does not affect the computation of Poisson brackets,
but will be important in the decoupling discussions in
Secs.\ \ref{sec:arguments} and \ref{sec:higherorders} 
below.

\end{itemize}

\section{Classical scattering map: foundations, definition and parameterization}
\label{sec:scattering map}

Having completed our analysis of the symmetries, charges and
asymptotics of the theory, and the definition of the phase space, we now turn to an exploration of the dynamics of the theory in the deep infrared.
Our goal is to determine the extent to which the conserved charges (\ref{Echarge})
constrain in a nontrivial way the dynamics of the theory. 
In this section we will define the scattering map and derive some of
its properties as well as some useful parameterizations.
We also set up the framework for computing the scattering map in
perturbation theory, and compute the free field scattering.
In later sections we will explicitly compute higher order
contributions, and we will show
that the scattering map cannot be factored into maps that act
individually on hard and soft sectors.

\subsection{Asymptotic field expansions in a more general class of gauges}
\label{sec:gen-gauge}

It will be convenient to use \Lorentz gauge
for our explicit computations, since the equations of motion reduce to simple wave equations
in this gauge.  
However, the form (\ref{Aexpand}) of the asymptotic expansions that we have
assumed are insufficiently general for this purpose, as \Lorentz gauges
generically requires logarithmic terms when sources are
present\cite{Himwich:2019dug,Satishchandran:2019pyc}.  Indeed,   
starting from the expansions (\ref{Aexpand})
we obtain
\begin{eqnarray}
\nabla_a A^a &=& -\frac{2}{r} {\cal A}_{\ps u} + \frac{1}{r^2} \left[ -
  \partial_u {\cal A}_{\ps r} - {\hat {\cal A}}_{\ps u} + D^A
  {\cal A}_{\ps A} \right] + O \left( \frac{1}{r^3} \right) \nonumber
\\
&=&  -\frac{2}{r} {\cal A}_{\ps u} + \frac{1}{r^2} \left[
  e^2 \int_u^\infty d u' {\cal J}_{\ps u}(u') + D^A {\cal A}_{\psps A}
- \int_u^\infty d u' D^2 {\cal A}_{\ps u}  \right]
+ O \left( \frac{1}{r^3} \right),
\label{nolorentz}
\end{eqnarray}
where we have used the asymptotic Maxwell's equations (\ref{maxwellexpand})
and the boundary
conditions (\ref{falloffiplus}) to rewrite the coefficient of the $1/r^2$ term.
Thus our assumed expansions are incompatible with \Lorentz gauges, since
when the first term in Eq.\ (\ref{nolorentz}) vanishes the second term is generically nonzero and cannot be made to vanish
using the gauge transformations (\ref{gggs}).

We therefore generalize the form of the expansion (\ref{Aexpand}) to
encompass \Lorentz gauges, following
Refs.\ \cite{Himwich:2019dug,Satishchandran:2019pyc}.  In the limit to $\scri^{+}$, we assume
\begin{subequations}
  \label{Aexpandnew}
  \begin{eqnarray}
    \label{AexpandnewI}
    A_A &=& {\cal A}_{\ps A}
    - \frac{\ln  r}{r} D_A {\tilde {\cal A}}_{\ps r}
    + \frac{1}{r}  {\hat {\cal A}}_{\ps A} 
    - \frac{\ln  r}{2 r^2} D_A  {\tilde {\tilde  {\cal
            A}}}_{\ps r}
        + \frac{1}{r^2}  {\hat {\hat {\cal A}}}_{\ps A} 
    + O
    \left( \frac{\ln r}{r^3} \right), \\
        \label{AexpandnewII}
         A_u &=& {\cal A}_{\ps u}
    - \frac{\ln  r}{r} \partial_u {\tilde {\cal A}}_{\ps r}
         + \frac{1}{r}
         {\hat {\cal A}}_{\ps u}
             - \frac{\ln  r}{2 r^2} \partial_u  {\tilde
                 {\tilde  {\cal A}}}_{\ps r}
    + \frac{1}{r^2} {\hat {\hat  {\cal A}}}_{\ps u}
    + O\left( \frac{\ln r}{r^3} \right), \\
        \label{AexpandnewIII}
    A_r &=& \frac{Q}{r} +
     \frac{\ln  r}{r^2}  {\tilde {\cal A}}_{\ps r}
    + \frac{1}{r^2} {\cal
        A}_{\ps r} +
     \frac{\ln  r}{r^3}   {\tilde {\tilde {\cal A}}}_{\ps r}
    + \frac{1}{r^3}  {\hat {\cal A}}_{\ps r} + O
    \left( \frac{\ln r}{r^4} \right), \\
        \label{AexpandnewIV}
     \Phi &=& e^{i Q \ln r} \left[ \frac{1}{r} \chi_{\ps} -
i \frac{\ln r}{r^2} {\tilde {\cal A}}_{\ps r}\, \chi_{\ps} + 
       \frac{1}{r^2}  {\hat \chi}_{\ps} + O\left( \frac{1}{r^3} \right) \right].
      \end{eqnarray}
\end{subequations}
Here the notational conventions are as follows.
Caligraphic font quantities are coefficients in the double expansion
in $1/r$ and $\ln r/r$, functions of $u$ and $\theta^A$.  The 
quantities ${\cal A}_{\ps A}$, ${\cal
    A}_{\ps u}$ and ${\cal A}_{\ps r}$ without any
tildes or carets are the leading order fields discussed in
Sec.\ \ref{sec:found} above.  Quantities with one or more carets like $
  {\hat {\cal  A}}_{\ps r}$ are coefficients of subleading terms in
the $1/r$ expansion, while quantities with one or more tildes like 
${\tilde {\cal  A}}_{\ps r}$ are coefficients of log terms,
new in this section.  Finally Eqs.\ (\ref{AexpandnewIII}) and (\ref{AexpandnewIV}) depend
on a quantity $Q$ which is a constant (the total ingoing or outgoing charge multiplied by $4 \pi$).

The specific relations between the coefficients of the log terms in
Eqs.\ (\ref{Aexpandnew}) are chosen to ensure that the expansions
(\ref{Fexpand}) of the Maxwell tensor
and (\ref{currentexpand}) of the current are
still valid in
this context.  The formulae (\ref{related}) for the expansion coefficients
are replaced by
\begin{subequations}
  \label{relatednew}
  \begin{eqnarray}
    \label{Furfnew}
    {\cal F}_{\ps ur} &=& \partial_u {\cal A}_{\ps r} + \partial_u
    {\tilde {\cal A}}_{\ps r} + {\hat {\cal
        A}}_{\ps u}, \ \ \ \     {\hat {\cal F}}_{\ps ur} = \partial_u
    {\hat {\cal A}}_{\ps r} + 2 {\hat {\hat {\cal
          A}}}_{\ps u} + \frac{1}{2} \partial_u {\tilde {\tilde {\cal A}}}_{\ps r},     \\
    {\cal F}_{\ps uA} &=& \partial_u {\cal A}_{\ps A} - D_A {\cal
        A}_{\ps u}, \ \ \         {\hat {\cal F}}_{\ps uA} =
    \partial_u {\hat {\cal A}}_{\ps A} - D_A {\hat {\cal
        A}}_{\ps u}, \ \ \        {\hat {\hat {\cal F}}}_{\ps uA} =
    \partial_u {\hat {\hat {\cal A}}}_{\ps A} - D_A {\hat {\hat {\cal
        A}}}_{\ps u}, \\
        \label{FrAfnew}
    {\cal F}_{\ps rA} &=& -D_A {\cal A}_{\ps r} - {\hat {\cal A}}_{\ps
      A} - D_A {\tilde {\cal A}}_{\ps r}, \ \ \ \
        {\hat {\cal F}}_{\ps rA} = -D_A {\hat {\cal A}}_{\ps r} - 2
        {\hat {\hat {\cal A}}}_{\ps
      A} - \frac{1}{2} D_A {\tilde {\tilde {\cal A}}}_{\ps r},     \\
\label{FABformulanew}
    {\cal F}_{\ps AB} &=& D_A {\cal A}_{\ps B} - D_B {\cal A}_{\ps A},
    \ \ \ \
        {\hat {\cal F}}_{\ps AB} = D_A {\hat {\cal A}}_{\ps B} - D_B
        {\hat {\cal A}}_{\ps A}.
      \end{eqnarray}
  \end{subequations}

The gauge transformation expansion (\ref{gaugeexpand}) is replaced by the more
general version
\be
\label{gaugeexpandnew}
\varepsilon = \delta Q \, \ln r + \varepsilon_{\ps} +
\frac{\ln r}{r}     {\tilde \varepsilon}_{\ps}
+\frac{1}{r} {\hat \varepsilon}_{\ps}
+ \frac{\ln r}{r^2}     {\tilde {\tilde \varepsilon}}_{\ps}
+ \frac{1}{r^2} {\hat
  {\hat \varepsilon}}_{\ps} + O\left( \frac{1}{r^3} \right),
\ee
under which we have
\begin{subequations}
\label{gggsnew}
  \begin{eqnarray}
\label{gg1new}
Q &\to& Q + \delta Q, \\ 
{\cal A}_{\ps u} &\to& {\cal A}_{\ps u} + \partial_u \varepsilon_{\ps},
    \ \ \ \ {\hat {\cal A}}_{\ps u} \to {\hat {\cal A}}_{\ps u} +
    \partial_u {\hat \varepsilon}_{\ps}, \ \ \ \ {\hat {\hat {\cal A}}}_{\ps u}
    \to {\hat {\hat {\cal A}}}_{\ps u} +
    \partial_u {\hat {\hat \varepsilon}}_{\ps}, \\ 
        {\cal A}_{\ps A} &\to& {\cal A}_{\ps A} + D_A \varepsilon_{\ps},
    \ \ \ \ {\hat {\cal A}}_{\ps A} \to {\hat {\cal A}}_{\ps A} +
    D_A {\hat \varepsilon}_{\ps},
        \ \ \ \ {\hat {\hat {\cal A}}}_{\ps A} \to {\hat {\hat {\cal A}}}_{\ps A} +
    D_A {\hat {\hat \varepsilon}}_{\ps},   \\
    \label{gg3new}
{\cal A}_{\ps r} &\to& {\cal A}_{\ps r} + {\tilde \varepsilon}_{\ps} - {\hat \varepsilon}_{\ps},
\ \ \ \ {\hat {\cal A}}_{\ps r} \to {\hat {\cal A}}_{\ps r} + {\tilde
  {\tilde \varepsilon}}_{\ps} - 2
        {\hat {\hat \varepsilon}}_{\ps}, \\
{\tilde {\cal A}}_{\ps r} &\to& {\tilde {\cal A}}_{\ps r} - {\tilde
  \varepsilon}_{\ps},
\ \ \ \ {\tilde {\tilde {\cal A}}}_{\ps r} \to {\tilde {\tilde {\cal
      A}}}_{\ps r} -2  {\tilde {\tilde
  \varepsilon}}_{\ps},
\\
               \chi_{\ps} & \to& e^{i \varepsilon_{\ps}} \chi_{\ps},
               \ \ \ \
                      {\hat \chi}_{\ps}  \to e^{i \varepsilon_{\ps}}( {\hat  \chi}_{\ps}
        + i {\hat \varepsilon}_{\ps} \chi_{\ps} ).
  \end{eqnarray}
  \end{subequations}

The \Lorentz gauge condition can be written using the expansion
(\ref{Aexpandnew}) as 
\begin{eqnarray}
  0 &=& \nabla_a A^a = \frac{1}{r} \left[ - 2 {\cal A}_{\ps u} \right] +
  \frac{1}{r^2} \left[Q
 -   {\cal A}_{\ps r,u} - {\hat {\cal A}}_{\ps u} + D^A
 {\cal A}_{\ps A} + {\tilde {\cal A}}_{\ps r,u} \right] \nonumber \\
  && + \frac{\ln r}{r^3} \left[ - {\tilde {\tilde {\cal A}}}_{\ps r,u}
    - D^2 {\tilde {\cal A}}_{\ps r} \right]
  + \frac{1}{r^3} \left[ {\tilde A}_{\ps r} -   {\hat {\cal A}}_{\ps r,u} - {\hat {\cal A}}_{\ps u} + D^A
 {\hat {\cal A}}_{\ps A} + \frac{1}{2} {\tilde {\tilde {\cal A}}}_{\ps
   r,u} \right] \nonumber \\
  && + O \left( \frac{\ln r}{r^4} \right),
\end{eqnarray}
where $D^2 = h^{AB} D_A D_B$.
Setting the coefficients of this expansion to zero gives four
conditions for \Lorentz gauge, and starting from a general gauge of
the form (\ref{Aexpandnew}) one can check that it is possible to use the
transformation freedom (\ref{gggsnew}) to satisfy these conditions.

\subsection{Transformation to preferred asymptotic gauge}
\label{sec:pag}

We now discuss the transformation from \Lorentz gauge to the
preferred asymptotic gauge discussed in Sec.\ \ref{U-1} above, which
is defined by the 
expansion (\ref{Aexpand}) and the conditions (\ref{gaugec1}) and
(\ref{matching2}).  We start from an expansion of the form
(\ref{Aexpandnew}), specialized to \Lorentz gauge (which implies ${\cal A}_{\ps u}=0$):
\begin{subequations}
  \label{AexpandLorentz}
  \begin{eqnarray}
    \label{AexpandLorentzI}
    {\underline A}_A &=& {\underline {\cal A}}_{\ps A}
    - \frac{\ln  r}{r} D_A {\underline {\tilde {\cal A}}}_{\ps r}
    + \frac{1}{r}  {\underline {\hat {\cal A}}}_{\ps A} 
    - \frac{\ln  r}{2 r^2} D_A  {\underline {\tilde {\tilde  {\cal
            A}}}}_{\ps r}
        + \frac{1}{r^2}  {\underline {\hat {\hat {\cal A}}}}_{\ps A} 
    + O
    \left( \frac{\ln r}{r^3} \right), \\
        \label{AexpandLorentzII}
         {\underline A}_u &=& 
     \frac{\ln  r}{r} \partial_u {\underline {\tilde {\cal A}}}_{\ps r}
         + \frac{1}{r}
         {\underline {\hat {\cal A}}}_{\ps u}
             - \frac{\ln  r}{2 r^2} \partial_u  {\underline {\tilde
                 {\tilde  {\cal A}}}}_{\ps r}
    + \frac{1}{r^2} {\underline {\hat {\hat  {\cal A}}}}_{\ps u}
    + O\left( \frac{\ln r}{r^3} \right), \\
        \label{AexpandLorentzIII}
    {\underline A}_r &=& \frac{Q}{r} +
     \frac{\ln  r}{r^2}  {\underline {\tilde {\cal A}}}_{\ps r}
    + \frac{1}{r^2} {\underline {\cal
        A}}_{\ps r} +
     \frac{\ln  r}{r^3}   {\underline {\tilde {\tilde {\cal A}}}}_{\ps r}
    + \frac{1}{r^3}  {\underline {\hat {\cal A}}}_{\ps r} + O
    \left( \frac{\ln r}{r^4} \right), \\
        \label{AexpandLorentzIV}
     {\underline \Phi} &=& e^{i Q \ln r} \left[ \frac{1}{r}
       {\underline \chi}_{\ps} -
i \frac{\ln r}{r^2}  {\underline {\tilde {\cal A}}}_{\ps r} {\underline \chi}_{\ps} + 
       \frac{1}{r^2}  {\underline {\hat \chi}}_{\ps} + O\left( \frac{\ln r }{r^3} \right) \right].
      \end{eqnarray}
\end{subequations}
Here and throughout 
underlined quantities
refer to quantities in \Lorentz gauge.
We now make a gauge transformation of the form (\ref{gaugeexpandnew}) with the
expansion coefficients chosen to be
\be    \label{gc1}
    \delta Q = - Q, \ \ \ \ 
    \varepsilon_{\ps} = 0, \ \ \ \ 
        {\tilde \varepsilon}_{\ps} =  {\underline {\tilde A}}_{\ps
          r}, \ \ \ \ 
            {\hat \varepsilon}_{\ps} = {\underline {\cal A}}_{\ps r}
            + {\underline {\tilde {\cal A}}}_{\ps r}, \ \ \ \ {\tilde
              {\tilde \varepsilon}}_{\ps} = \frac{1}{2} {\underline {\tilde
              {\tilde {\cal A}}}}_{\ps r},
\ee
which enforces the required conditions (\ref{Aexpand}) and the first
two equations of (\ref{gaugec1}) by Eqs.\ (\ref{gggsnew}). 
We make a similar gauge transformation near $\scri^-$ to enforce
(\ref{Aexpandminus}) and the last two equations of (\ref{gaugec1}).

We have not yet enforced the condition (\ref{matching2}) of the preferred
asymptotic gauge.  To do so we use an odd transformation of the form
(\ref{sigmaodd}).  The gauge transformation function is determined by
the condition (\ref{matching2}) of preferred asymptotic gauge,
together with the condition (\ref{ccdd2}) of asymptotic \Lorentz
gauge,
which is valid for interacting solutions as well as free solutions as
discussed in Appendix \ref{app:intLor}.
The resulting transformation between \Lorentz gauge fields and
preferred asymptotic gauge fields is
\begin{subequations}
	\label{newgauge-both}
  \begin{eqnarray}
    {\Psi}^{\rm e}_{\ps} &=& {\underline {\Psi}}^{\rm e}_{\ps} +
    \varepsilon_{\ps}, \ \ \ \ 
    {\Psi}^{\rm e}_{\ms} = {\underline {\Psi}}^{\rm e}_{\ms} -
    {\cal P}_* \varepsilon_{\ps},\\
     {\Psi}^{\rm m}_{\ps} &=& {\underline {\Psi}}^{\rm m}_{\ps},
     \ \ \ \ \ \ \ \ \ \ 
         {\Psi}^{\rm m}_{\ms} = {\underline {\Psi}}^{\rm m}_{\ms},\\
             \chi_{\ps} &=& e^{i \varepsilon_{\ps}} {\underline \chi}_{\ps},
    \ \ \ \ \ \ \chi_{\ms} = e^{- i {\cal P}_* \varepsilon_{\ps}} {\underline \chi}_{\ms}.
  \end{eqnarray}
  \end{subequations}
with
\be
\varepsilon_{\ps} = \frac{1}{2} \left(
{\cal P}_* {\underline {\Psi}}^{\rm e}_{\msps}
-{\underline {\Psi}}^{\rm e}_{\psms}
\right).
\ee
The inverse transformation is given by the same formulae
(\ref{newgauge-both}) but with $\varepsilon_{\ps}$
now expressed in terms of the preferred asymptotic gauge fields:
\be
\label{ve44}
\varepsilon_{\ps} = \frac{1}{2} \left(
{\Psi}^{\rm e}_{\psps}-{\cal P}_* {\Psi}^{\rm e}_{\msms}
           \right).
\ee

\subsection{Perturbative framework}

The scattering map (\ref{scatter0}) can be written schematically as
\begin{subequations}
  \label{scatter000}
  \begin{eqnarray}
    \chi_{\ps}(u,\bftheta) &=&     \chi_{\ps}[u,\bftheta; \chi_{\ms},
      {\cal A}_{\ms A} ],\\
        {\cal A}_{\ps A}(u,\bftheta) &=&     \chi_{\ps}[u,\bftheta; \chi_{\ms},
      {\cal A}_{\ms A} ],
  \end{eqnarray}
  \end{subequations}
where the functional dependence on the initial data is indicated by
the square brackets.  We will compute this map perturbatively by
considering general \Lorentz gauge solutions, and by transforming from
\Lorentz gauge to preferred asymptotic gauge using
Eqs.\ (\ref{newgauge-both}).

We make the following ansatz for the scalar
field and vector potential  
\begin{subequations}
  \label{pert-ansatz}
  \begin{eqnarray}
\Phi &= \tepsilon \Phi^{(1)} + \tepsilon^{2} \Phi^{(2)} + \tepsilon^{3} \Phi^{(3)}
+ O(\tepsilon^4),\\
A^{a} &= \tepsilon A^{(1)\,a}  + \tepsilon^{2} A^{(2)\,a} + \tepsilon^{3} A^{(3)\,a}
+ O(\tepsilon^4),
	\end{eqnarray}
\end{subequations}
where $\tepsilon$ is the perturbative expansion parameter\footnote{This expansion is equivalent
  to expanding in powers of the charge $e$ at fixed $\Phi$ and fixed
  $A_a/e$.}. There is a corresponding expansion for the initial data
${\cal A}_{\ms A}$, $\chi_{\ms}$ on $\scri^-$ in the preferred
asymptotic gauge given by Eqs.\ (\ref{Aexpandminus}), (\ref{gaugec1}) and (\ref{matching2}):
\begin{subequations}
  \label{pert-ansatz1}
	\begin{eqnarray}
\chi_{\ms} &= \tepsilon \chi_{\ms}^{(1)} + \tepsilon^{2} \chi_{\ms}^{(2)} + \tepsilon^{3} \chi_{\ms}^{(3)}
+ O(\tepsilon^4),\\
{\cal A}_{\ms A}
&= \tepsilon {\cal A}_{\ms A}^{(1)}  + \tepsilon^{2} {\cal A}_{\ms A}^{(2)} + \tepsilon^{3} {\cal A}_{\ms A}^{(3)}
+ O(\tepsilon^4).
	\end{eqnarray}
\end{subequations}
We will take the second order initial data $\chi_{\ms}^{(2)}$ and ${\cal A}_{\ms A}^{(2)}$ and higher order initial data to vanish.  This is done for convenience and incurs no loss in generality, since terms linear, quadratic, cubic etc. in the first order fields
$\chi_{\ms}^{(1)}$ and ${\cal A}_{\ms A}^{(1)}$
give complete information about the scattering map.
The corresponding expansion of the final data at $\scri^+$ in
preferred asymptotic gauge is
\begin{subequations}
  \label{pert-ansatz2}
  \begin{eqnarray}
      \label{pert-ansatz2a}
\chi_{\ps} &= \tepsilon \chi_{\ps}^{(1)} + \tepsilon^{2} \chi_{\ps}^{(2)} + \tepsilon^{3} \chi_{\ps}^{(3)}
+ O(\tepsilon^4),\\
{\cal A}_{\ps A}
&= \tepsilon {\cal A}_{\ps A}^{(1)}  + \tepsilon^{2} {\cal A}_{\ps A}^{(2)}  + \tepsilon^{3} {\cal A}_{\ps A}^{(3)}
+ O(\tepsilon^4).
	\end{eqnarray}
\end{subequations}

We will denote the Lorenz-gauge fields with
underlines, and write them as $\underline{\Phi}$, $\underline{A}_a$. The general equations of motion
(\ref{eom00}) reduce in this gauge to
\begin{subequations}
  \label{eqns-lg}
	\begin{eqnarray}
	\label{scalar-eqn-lg}
	\Box \underline{\Phi} &=& 2 i  \underline{A}_{a} \nabla^{a} \underline{\Phi} + \underline{A}^{a} \underline{A}_{a} \underline{\Phi}\,,\\
	\Box \underline{A}^{a} &=& - i e^{2} (\underline{\Phi} \nabla^{a} \underline{\Phi}^* - \underline{\Phi}^* \nabla^{a} \underline{\Phi}) + 2 e^{2} \underline{A}^{a} \underline{\Phi}^* \underline{\Phi} \,.
	\label{vector-eqn-lg}
	\end{eqnarray}
\end{subequations}
Now using the expansions (\ref{pert-ansatz})
yields the leading order equations of motion 
\begin{subequations}
  \label{leading-order-eom}
  \begin{eqnarray}
\Box\underline{\Phi}^{(1)} &= 0 \,,\\
\Box \underline{A}^{(1)\,a} &= 0\,,
	\end{eqnarray}
\end{subequations}
the subleading order equations 
\begin{subequations}
	\label{eq:subleading-equations}
	\begin{eqnarray}
  \label{eqn-subleading-scalar}
\Box \underline{\Phi}^{(2)} &=& 2  i  \underline{A}^{(1)\,a} \nabla_{a} \underline{\Phi}^{(1)}\,, \\
\Box \underline{A}^{(2)\,a} &=& -ie^{2}  \underline{\Phi}^{(1)} \nabla^{a} \underline{\Phi}^{(1)*} + i e^{2} \underline{\Phi}^{(1)*} \nabla^{a} \underline{\Phi}^{(1)},
\label{eqn-subleading-vec} 
	\end{eqnarray}
\end{subequations}
and the subsubleading equations
\begin{subequations}
	\label{eq:subsubleading-equations}
	\begin{eqnarray}
  \label{eqn-subsubleading-scalar}
  \Box \underline{\Phi}^{(3)} &=&
  2  i  \underline{A}^{(1)\,a} \nabla_{a} \underline{\Phi}^{(2)}+
  2  i  \underline{A}^{(2)\,a} \nabla_{a} \underline{\Phi}^{(1)}\,
+ \underline{A}^{(1)\,a} \underline{A}^{(1)}_{a} \underline{\Phi}^{(1)}\,,\\
\Box \underline{A}^{(3)\,a} &=& -ie^{2}  \underline{\Phi}^{(1)}
\nabla^{a} \underline{\Phi}^{(2)*} + i e^{2} \underline{\Phi}^{(1)*}
\nabla^{a} \underline{\Phi}^{(2)}
-ie^{2}  \underline{\Phi}^{(2)} \nabla^{a} \underline{\Phi}^{(1)*} + i
e^{2} \underline{\Phi}^{(2)*} \nabla^{a} \underline{\Phi}^{(1)}
\nonumber \\
&&
+ 2 e^{2} \underline{A}^{(1)\,a} \underline{\Phi}^{(1)*} \underline{\Phi}^{(1)} \,.
\label{eqn-subsubleading-vec} 
	\end{eqnarray}
\end{subequations}

\subsection{First order solutions and scattering map}
\label{sec:fos}

Appendix \ref{app:freesolns} reviews the general solutions of the leading order
\Lorentz gauge equations of motion (\ref{leading-order-eom}) which
have nontrivial soft charges.  These solutions
satisfy 
our assumed expansions (\ref{Aexpand}) and (\ref{Aexpandminus}) near
$\scri^+$ and $\scri^-$, and our asymptotic
gauge conditions (\ref{gaugec1}).  They do not satisfy the matching condition
(\ref{matching2}), a reflection of the fact that
the \Lorentz and preferred
asymptotic gauges do not coincide in general.

From these solutions one can evaluate the free field scattering map.
One might expect this map to be
trivial for free solutions, and to reduce essentially to the identity
map (up to antipodal identification).
However the presence of the nontrivial soft charges makes the
situation slightly more complicated, and in particular the identity
map would not be consistent with the matching condition (\ref{matching2}) at
spatial infinity.
The scattering map when written
in terms of the potentials $\Psi^{\rm e}$ and $\Psi^{\rm m}$, and
denoting \Lorentz gauge fields with underlines, is
[cf.\ Eq.\ (\ref{eq:free-map11})]
\begin{subequations}
	\label{eq:free-map00}
        \begin{eqnarray}
    \label{eq:free-map00a}          
          {\underline \Psi}^{\rm e}_{\ps}(u,\bftheta) &=&  {\cal P}_*
\left[{\underline \Psi}^{\rm e}_{\ms}(u,\bftheta)
  - {\underline \Psi}^{\rm e}_{\msps}(\bftheta)
  + {\underline \Psi}^{\rm e}_{\msms}(\bftheta)
  \right] + O(\alpha^2),\\
          {\underline \Psi}^{\rm m}_{\ps}(u,\bftheta) &=&  -{\cal P}_*
{\underline \Psi}^{\rm m}_{\ms}(u,\bftheta) + O(\alpha^2)
  ,\\
    {\underline \chi}_{\ps}(u,\bftheta) &=& - {\cal P}_* \, {\underline \chi}_{\ms}(u,\bftheta) + O(\alpha^2).
\end{eqnarray}
\end{subequations}

We can rewrite this scattering map in terms of the preferred
asymptotic gauge fields using the gauge transformation
(\ref{newgauge-both}) applied to both the initial and final fields.
The result for the scalar field is
\be
\chi_{\ps} = - \exp \left[  2 i {\cal P}_* (\Psi^{\rm e}_{\msps} - \Psi^{\rm e}_{\msms}) \right] {\cal P}_* \chi_{\ms} + O(\alpha^2).
\ee
However the phase factor here is a nonlinear effect that should be discarded
in a perturbative expansion.
Discarding this phase factor the
full free field scattering map is, from Eqs.\ (\ref{newgauge-both})
and (\ref{eq:free-map00}),  
\begin{subequations}
  \label{eq:var-redef-3}
  \begin{eqnarray}
    \label{eq:var-redef-3a}
    \Psi^{\rm e}_{\ps} &=& {\cal P}_* \left[ \Psi^{\rm e}_{\ms} - \Psi^{\rm e}_{\msms} + \Psi^{\rm e}_{\msps} \right] + O(\alpha^2), \\
    \Psi^{\rm m}_{\ps} &=& - {\cal P}_* \, \Psi^{\rm m}_{\ms} + O(\alpha^2), \\
    \label{eq:chiaa}
    \chi_{\ps} &=& - {\cal P}_* \chi_{\ms} + O(\alpha^2).
  \end{eqnarray}
\end{subequations}
Note that some of the signs in Eq.\ (\ref{eq:var-redef-3a}) differ from those in
Eq.\ (\ref{eq:free-map00a}).   The scattering map (\ref{eq:var-redef-3})
manifestly satisfies the
matching condition (\ref{matching2}).  It also
preserves the symplectic form (\ref{presymfinal1}), as it should,
which provides a nontrivial
consistency check of some of the coefficients of the $u$-independent terms.

We can rewrite the free field scattering map (\ref{eq:var-redef-3}) in terms of the
hard variables $({\tilde \Psi}^{\rm e}, \Psi^{\rm m}, \chi)$
and
soft variables $(\Delta \Psi^{\rm e}, {\bar \Psi}^{\rm e})$
defined in
Sec.\ \ref{sec:Dirac} above. The result is
\begin{subequations}
      	\label{eq:freee}
      	\begin{eqnarray}
        \label{eq:freee-b}
      	\pb \tilde{\Psi}^{\rm e}_{\ps} &=& \tilde{\Psi}^{\rm e}_{\ms} + O(\alpha^2), \\
\label{eq:freee-c}
      	\pb \Psi^{\rm m}_{\ps} &=& - \Psi^{\rm m}_{\ms} + O(\alpha^2), \\
          \label{eq:freee-a}
      	\pb \chi_{\ps} &=& - \chi_{\ms} + O(\alpha^2), \\
        \label{eq:freee-d}
      	\pb \Delta \Psi^{\rm e}_{\ps} &=&  \Delta \Psi^{\rm
          e}_{\ms} + O(\alpha^2), \label{eq:d44}\\
        \label{eq:freee-e}
      	\pb {\bar \Psi}^{\rm e}_{\ps} &=& 
        {\bar \Psi}^{\rm e}_{\ms} + \Delta
          \Psi^{\rm e}_{\ms} + O(\alpha^2).
\end{eqnarray}
\end{subequations}
Note that the hard and soft sectors are decoupled to this
order\footnote{This remains true if we express the mapping in terms of 
the symplectically orthogonal variables
$(
{\tilde \Psi}^{\rm e},
\Psi^{\rm m},
\chi,
\Delta \Psi^{\rm e},
{\hat \Psi}^{\rm e}
)$
obtained by
replacing ${\bar \Psi}^{\rm e}$ with ${\hat \Psi}^{\rm e}$ using Eq.\ (\ref{hatPsi}),
instead of the variables
(\ref{eq:freee}).}, with
the soft sector evolving via Eqs.\ (\ref{eq:freee-d}) --
(\ref{eq:freee-e}) and the hard sector via Eqs.\ (\ref{eq:freee-a}) -- (\ref{eq:freee-c}).

\subsection{General parameterization of the scattering map}
\label{sec:generalform}

The scattering map ${\cal S} : \Gamma_- \to \Gamma_+$ must satisfy a number of constraints.
In this section we derive the most general form of the map that obeys
all the constraints, as a foundation for the non-decoupling analysis of the later sections.

The various constraints are:

\begin{itemize}

\item The scattering map must satisfy the conservation law
  (\ref{Echarge}) for the soft charges $Q_\varepsilon$.
This
  conservation law can be written as \cite{Strominger:lectures}
  \be
   \label{eq:cons-law}
  Q_{\ps}(\bftheta) = \pb Q_{\ms}(\bftheta),
\ee
where the charges can be decomposed into hard and soft pieces as
\be \label{eq:cons-law00}
Q_{\ps}(\bftheta) =   Q^{\rm hard}_{\ps}(\bftheta) + \frac{1}{e^2} D^2 \Delta \Psi^{\rm
    e}_{\ps}(\bftheta), \ \ \ \  
  Q_{\ms}(\bftheta) =  Q^{\rm hard}_{\ms}(\bftheta) + \frac{1}{e^2} D^2 \Delta \Psi^{\rm
    e}_{\ms}(\bftheta),
  \ee
from Eqs.\ (\ref{maxwellexpand1}), (\ref{c1}), (\ref{fc1}), (\ref{FaAf}) 
  (\ref{gaugec1}), (\ref{AAdecompose}) and (\ref{DeltaPsidef}).
Here
  \be
  Q^{\rm hard}_{\ms} = \int dv {\cal J}_{\ms v}, \ \ \ \ Q^{\rm hard}_{\ps} = \int du
  {\cal J}_{\ps u}
\label{currents}
  \ee
  are the total ingoing and outgoing hard charges per unit
angle.  Since we have specialized to the sector where the 
magnetic charges (\ref{magcharges}) vanish, there is no corresponding constraint
from the magnetic charges.

\item It must be compatible with the transformations of initial and
  final data associated with the residual even transformations
  $\varepsilon_{\rm e}$ discussed in
  Sec.\ \ref{sec:gauge-specialization} above, which act on the
  physical phase space since they are not degeneracy directions of the
  presymplectic form. Specifically, under the transformation of
  initial data
  \be
  \label{tr1}
  \chi_{\ms} \to e^{i \varepsilon} \chi_{\ms}, \ \ {\cal A}_{\ms A}
  \to {\cal A}_{\ms A}  + D_A \varepsilon,
  \ee
  where $\varepsilon = \varepsilon(\bftheta)$, the final data must
  transform as
  \be
  \label{tr2}
  \chi_{\ps} \to e^{i {\cal P}_* \varepsilon} \chi_{\ps}, \ \ {\cal A}_{\ps A}
  \to {\cal A}_{\ps A}  + D_A {\cal P}_* \varepsilon.
  \ee

\item It must obey the gauge specialization condition
  (\ref{matching2}) that was imposed in order to correlate the gauge
  freedom on $\scri^-$ and $\scri^+$ in such a way as to allow the
  scattering map to be a symplectomorphism, as discussed in
  Sec.\ \ref{sec:gauge-specialization}. 
  
\item It must transform appropriately under Poincar\'e symmetries.
  While this is an important constraint we will not make it explicit
  in this section.  

  \item It must be a symplectomorphism on phase space.  We will impose
    this requirement in Sec.\ \ref{sec:gf} below where we parameterize the
    scattering map in terms of a generating functional.  

\end{itemize}
Taken together, these requirements strongly constrain the scattering
map.

For the analysis in this section it will be convenient to use as the
basic variables particular combinations of the preferred asymptotic
gauge fields, namely
$\Psi^{\rm m}$, $\chi$, ${\bar \Psi}^{\rm e}$ and [cf.\ Eqs.\ (\ref{transform}) above]
\be
   {\breve \Psi}^{\rm e} \equiv \Psi^{\rm e} - {\bar \Psi}^{\rm e} = {\tilde
     \Psi}^{\rm e} + g \Delta \Psi^{\rm e}.
   \label{brevePsidef}
   \ee
The general scattering map (\ref{scatter000}) can be written in terms of these variables as
\begin{subequations}
\label{eq:general0}
\begin{eqnarray}
\label{eq:generalcheckpsie0}
\pb  {\breve \Psi}^{\rm e}_{\ps} &=& \,{\breve \Psi}^{\rm e}_{\ms} 
+ {\cal H}^{\rm e}
\left[u,
  {\breve \Psi}^{\rm e}_{\ms},
  {\bar \Psi}^{\rm e}_{\ms},
  \Psi^{\rm m}_{\ms},
  \chi_{\ms} \right], \\
\label{eq:generalbarpsie0}
\pb  {\bar \Psi}^{\rm e}_{\ps} &=& \,{\bar
  \Psi}^{\rm e}_{\ms} +
{\Delta \Psi}^{\rm e}_{\ms} +
{\cal I}
\left[  {\breve \Psi}^{\rm e}_{\ms},
  {\bar \Psi}^{\rm e}_{\ms},
  \Psi^{\rm m}_{\ms},
  \chi_{\ms} \right], \\
\label{eq:generalpsim0}
\pb \Psi^{\rm m}_{\ps} &=& \,-\Psi^{\rm m}_{\ms} 
+ {\cal H}^{\rm m}\left[u,
  {\breve \Psi}^{\rm e}_{\ms},
  {\bar \Psi}^{\rm e}_{\ms},
  \Psi^{\rm m}_{\ms},
  \chi_{\ms} \right], \\ 
\label{eq:generalchi0}
\pb \chi_{\ps} &=& - \chi_{\ms} 
+  {\cal K}
\left[u,
  {\breve \Psi}^{\rm e}_{\ms},
  {\bar \Psi}^{\rm e}_{\ms},
  \Psi^{\rm m}_{\ms},
  \chi_{\ms} \right], 
\end{eqnarray}
\end{subequations}	
in terms of some functionals ${\cal H}^{\rm e}$, ${\cal H}^{\rm m}$,
${\cal I}$ and ${\cal K}$, where the functional dependence on the
initial data is indicated by the square brackets.
Here for convenience we have separated out the terms that arise in
the free evolution (\ref{eq:freee}), so that the functionals
parameterize the nonlinear 
interactions. The functionals do depend on the angles $\bftheta$ but
we have suppressed this dependence for simplicity.

We start by imposing the transformation property (\ref{tr1}) and (\ref{tr2}).
The functionals ${\cal H}^{\rm e}$, ${\cal H}^{\rm m}$ and ${\cal I}$
need to be invariant under the transformation, while ${\cal K}$ needs
to transform by a phase. Choosing $\varepsilon = - {\bar \Psi}^{\rm
  e}_{\ms}$, the invariance implies for example that
\be
{\cal H}^{\rm e}
\left[u,
  {\breve \Psi}^{\rm e}_{\ms},
  {\bar \Psi}^{\rm e}_{\ms},
  \Psi^{\rm m}_{\ms},
  \chi_{\ms}\right] =
{\cal H}^{\rm e}
\left[u,
  {\breve \Psi}^{\rm e}_{\ms},
  0,
  \Psi^{\rm m}_{\ms},
  e^{- i {\bar \Psi}^{\rm e}_{\ms}} \chi_{\ms}\right].
 \ee
 Hence by redefining
the functionals the scattering map can be written
in the general form
\begin{subequations}
\label{eq:general1}
\begin{eqnarray}
\label{eq:generalcheckpsie1}
\pb  {\breve \Psi}^{\rm e}_{\ps} &=& \,{\breve \Psi}^{\rm e}_{\ms} 
+ {\cal H}^{\rm e}
\left[u,
  {\breve \Psi}^{\rm e}_{\ms},
  \Psi^{\rm m}_{\ms},
  e^{- i{\bar \Psi}^{\rm e}_{\ms}} \chi_{\ms}\right], \\
\label{eq:generalbarpsie1}
\pb  {\bar \Psi}^{\rm e}_{\ps} &=& \,{\bar
  \Psi}^{\rm e}_{\ms} +
{\Delta \Psi}^{\rm e}_{\ms} +
{\cal I}
\left[  {\breve \Psi}^{\rm e}_{\ms},
  \Psi^{\rm m}_{\ms},
  e^{- i{\bar \Psi}^{\rm e}_{\ms}} \chi_{\ms}\right],\\
\label{eq:generalpsim1}
\pb \Psi^{\rm m}_{\ps} &=& \,-\Psi^{\rm m}_{\ms} 
+ {\cal H}^{\rm m}
\left[u,
  {\breve \Psi}^{\rm e}_{\ms},
  \Psi^{\rm m}_{\ms},
  e^{- i{\bar \Psi}^{\rm e}_{\ms}} \chi_{\ms}\right],\\
\label{eq:generalchi1}
\pb \chi_{\ps} &=& - \chi_{\ms} 
+ \exp \left[ i {\bar \Psi}^{\rm e}_{\ms} \right] 
{\cal K}
\left[u,
  {\breve \Psi}^{\rm e}_{\ms},
  \Psi^{\rm m}_{\ms},
  e^{- i{\bar \Psi}^{\rm e}_{\ms}} \chi_{\ms}\right].
\end{eqnarray}
\end{subequations}	

Next we impose the matching condition (\ref{matching2}).  From the
definition
(\ref{brevePsidef}) of $\breve{\Psi}^{\rm e}$ we have that
$\breve{\Psi}^{\rm e}_{\ps}(u=\infty) = - \breve{\Psi}^{\rm
  e}_{\ps}(u=-\infty)$, and ${\cal H}^{\rm e}$ must also have this
property by Eq.\ (\ref{eq:generalcheckpsie1}).  Defining the functional
\be
\label{Hinftydef}
{\cal H}^{\rm e}_\infty = \lim_{u \to \infty} {\cal H}^{\rm e}(u) =
-\lim_{u \to -\infty} {\cal H}^{\rm e}(u),
\ee
we find from Eqs.\ (\ref{matching2})
and (\ref{eq:general1}) that ${\cal I} = {\cal H}^{\rm e}_\infty$.
Therefore the scattering map can be written as
\begin{subequations}
\label{eq:general2}
\begin{eqnarray}
\label{eq:generalcheckpsie2}
\pb  {\breve \Psi}^{\rm e}_{\ps} &=& \,{\breve \Psi}^{\rm e}_{\ms} 
+ {\cal H}^{\rm e}
\left[u,
  {\breve \Psi}^{\rm e}_{\ms},
  \Psi^{\rm m}_{\ms},
  e^{- i{\bar \Psi}^{\rm e}_{\ms}} \chi_{\ms}\right], \\
\label{eq:generalbarpsie2}
\pb  {\bar \Psi}^{\rm e}_{\ps} &=& \,{\bar
  \Psi}^{\rm e}_{\ms} +
{\Delta \Psi}^{\rm e}_{\ms} +
{\cal H}^{\rm e}_\infty
\left[  {\breve \Psi}^{\rm e}_{\ms},
  \Psi^{\rm m}_{\ms},
  e^{- i{\bar \Psi}^{\rm e}_{\ms}} \chi_{\ms}\right],\\
\label{eq:generalpsim2}
\pb \Psi^{\rm m}_{\ps} &=& \,-\Psi^{\rm m}_{\ms} 
+ {\cal H}^{\rm m}
\left[u,
  {\breve \Psi}^{\rm e}_{\ms},
  \Psi^{\rm m}_{\ms},
  e^{- i{\bar \Psi}^{\rm e}_{\ms}} \chi_{\ms}\right],\\
\label{eq:generalchi2}
\pb \chi_{\ps} &=& - \chi_{\ms} 
+ \exp \left[ i {\bar \Psi}^{\rm e}_{\ms} \right] 
{\cal K}
\left[u,
  {\breve \Psi}^{\rm e}_{\ms},
  \Psi^{\rm m}_{\ms},
  e^{- i{\bar \Psi}^{\rm e}_{\ms}} \chi_{\ms}\right].
\end{eqnarray}
\end{subequations}

Finally we impose the conservation laws (\ref{eq:cons-law}).
Taking the limits $u \to \pm \infty$ of Eq.\ (\ref{eq:generalcheckpsie2}) and
using the definitions
(\ref{DeltaPsidef}), (\ref{brevePsidef})
and (\ref{Hinftydef}) yields
\be
\label{eq:generalcheckpsie2soft}
\pb  {\Delta \Psi}^{\rm e}_{\ps} = \,{\Delta \Psi}^{\rm e}_{\ms} 
+ 2{\cal H}^{\rm e}_\infty
\left[
  {\breve \Psi}^{\rm e}_{\ms},
  \Psi^{\rm m}_{\ms},
  e^{- i{\bar \Psi}^{\rm e}_{\ms}} \chi_{\ms}\right].
\ee
It follows that ${\cal H}^{\rm e}_\infty$ is essentially the change in the
electromagnetic memory $\Delta \Psi^{\rm e}$
\cite{EMmemory-1,EMmemory-2,EMmemory-3}
between $\scri^-$ and $\scri^+$.  Now combining this with the
conservation law  (\ref{eq:cons-law}) gives
\be
   {\cal H}^{\rm e}_\infty = - \frac{1}{2} e^2 \pb D^{-2} \Delta
   Q^{\rm hard}(\bftheta),
   \label{cons23}
\ee
where the total change in the charge per unit angle is $\Delta Q^{\rm hard} = 
Q_{\ps}^{\rm hard}(\bftheta) - \pb Q^{\rm hard}_{\ms}(\bftheta)$, which can be computed from the
functional ${\cal K}$ from Eqs.\ (\ref{calJu}), (\ref{eq:cons-law}), (\ref{currents})
and (\ref{eq:generalchi2}).
This tells us that the functionals are not all independent, as ${\cal
  H}^{\rm e}_\infty$ can be computed from ${\cal K}$.

To summarize, we have derived a general parameterization of the
scattering map that is consistent with all of the constraints listed
above, given by Eq.\ (\ref{eq:general2}), assuming that the various functionals 
transform appropriately under Poincar\'e transformations.

We can rewrite the scattering map (\ref{eq:general2}) in terms of the
symplectically orthogonal variables
$(
{\tilde \Psi}^{\rm e},
\Psi^{\rm m},
\chi,
\Delta \Psi^{\rm e},
{\hat \Psi}^{\rm e}
)$  introduced in Sec.\ \ref{sec:Dirac} using the definitions
(\ref{transform}), (\ref{hatPsi}) and (\ref{brevePsidef}).  The resulting map is given
by Eqs.\ (\ref{eq:generalpsim2}), (\ref{eq:generalchi2}), and (\ref{eq:generalcheckpsie2soft}) together with
\begin{subequations}
\label{eq:general2n}
\begin{eqnarray}
\label{eq:generaltildepsie2n}
\pb  {\tilde \Psi}^{\rm e}_{\ps} &=& \,{\tilde \Psi}^{\rm e}_{\ms} 
+ {\cal H}^{\rm e}(u) - 2 g(u) {\cal H}^{\rm e}_\infty, \\
\label{eq:generalhatpsie2n}
\pb  {\hat \Psi}^{\rm e}_{\ps} &=& \,{\hat
  \Psi}^{\rm e}_{\ms} +
{\Delta \Psi}^{\rm e}_{\ms} +
{\cal H}^{\rm e}_\infty + 2 \int du g'(u) {\cal H}^{\rm e}(u),
\end{eqnarray}
\end{subequations}	
where for convenience we have suppressed the arguments of the functionals
${\cal H}^{\rm e}$ and ${\cal H}^{\rm e}_\infty$.

We can also rewrite the scattering map (\ref{eq:general2}) in terms of 
asymptotic \Lorentz gauge fields using the gauge transformation (\ref{newgauge-both}).
The arguments of the functionals in Eq.\ (\ref{eq:general2}) are
invariant under the transformation, and so we can simply insert
underlines on all of these arguments to indicate asymptotic \Lorentz gauge fields.  The gauge transformation
function (\ref{ve44}) evaluates to $\varepsilon_{\ps} = {\cal P}_* \Delta
\Psi^{\rm e}_{\ms} + {\cal P}_* {\cal H}^{\rm e}_\infty$, and 
the final result is 
\begin{subequations}
\label{eq:general3}
\begin{eqnarray}
\label{eq:generalcheckpsie3}
\pb  {\underline {\breve \Psi}}^{\rm e}_{\ps} &=& \,{\underline {\breve \Psi}}^{\rm e}_{\ms} 
+ {\cal H}^{\rm e}
\left[u,
  {\underline {\breve \Psi}}^{\rm e}_{\ms},
  {\underline \Psi}^{\rm m}_{\ms},
  e^{- i{\underline {\bar \Psi}}^{\rm e}_{\ms}} {\underline
    \chi}_{\ms}\right]
, \\
\label{eq:generalbarpsie3}
\pb  {\underline {\bar \Psi}}^{\rm e}_{\ps} &=& \,{\underline {\bar
    \Psi}}^{\rm e}_{\ms}
- \Delta {\underline \Psi}^{\rm e}_{\ms}
- {\cal H}^{\rm e}_\infty
\left[ {\underline {\breve \Psi}}^{\rm e}_{\ms},
  {\underline \Psi}^{\rm m}_{\ms},
  e^{- i{\underline {\bar \Psi}}^{\rm e}_{\ms}} {\underline
    \chi}_{\ms}\right],\\
\label{eq:generalpsim3}
\pb {\underline \Psi}^{\rm m}_{\ps} &=& \,-{\underline \Psi}^{\rm m}_{\ms} 
+ {\cal H}^{\rm m}
\left[u,
  {\underline {\breve \Psi}}^{\rm e}_{\ms},
  {\underline \Psi}^{\rm m}_{\ms},
  e^{- i{\underline {\bar \Psi}}^{\rm e}_{\ms}} {\underline
    \chi}_{\ms}\right],\\
\label{eq:generalchi3}
\pb {\underline \chi}_{\ps} &=&
\exp \left\{ -2 i {\cal H}^{\rm e}_\infty
\left[ {\underline {\breve \Psi}}^{\rm e}_{\ms},
  {\underline \Psi}^{\rm m}_{\ms},
  e^{- i{\underline {\bar \Psi}}^{\rm e}_{\ms}} {\underline
    \chi}_{\ms}\right] - 2 i \Delta {\underline \Psi}^{\rm e}_{\ms} \right\}
\left\{- {\underline \chi}_{\ms} 
+ \exp \left[ i {\underline {\bar \Psi}}^{\rm e}_{\ms} \right] 
{\cal K}
\left[u,
  {\underline {\breve \Psi}}^{\rm e}_{\ms},
  {\underline \Psi}^{\rm m}_{\ms},
  e^{- i{\underline {\bar \Psi}}^{\rm e}_{\ms}} {\underline
    \chi}_{\ms}\right] \right\}. \nonumber \\
\end{eqnarray}
\end{subequations}	
This has the same form as the original scattering map 
(\ref{eq:general2}) except for sign flips in two of the terms in
Eq.\ (\ref{eq:generalbarpsie3}) and the overall phase factor in the scalar field (\ref{eq:generalchi3}).

\subsection{Scattering map in terms of a generating functional}
\label{sec:gf}

In this section we discuss a more efficient parameterization of the
scattering map, in terms of a generating functional on phase space.
This representation will be used extensively in the discussions in
Secs.\ \ref{sec:triviald}, \ref{sec:theorem} and \ref{sec:secondversioncoupled} below.

We start by decomposing the scattering map
  ${\cal S} : \Gamma_{\ms} \to \Gamma_{\ps}$ as described in the introduction:
  \be
     {\cal S} = {\cal S}_2 \circ {\cal S}_0,
     \label{eq:scdecompose}
     \ee
     where   ${\cal S}_0 : \Gamma_{\ms} \to \Gamma_{\ps}$ is the
     linear order scattering map (\ref{eq:freee}) and ${\cal S}_2 :
     \Gamma_{\ps} \to \Gamma_{\ps}$ encodes the nontrivial part of the
     scattering\footnote{This is similar to using the interaction
     representation in quantum mechanics.}.
Here $\Gamma_{\ms}$ and $\Gamma_{\ps}$ are the spaces of initial and final
data on $\scri^-$ and $\scri^+$ defined in
Sec.\ \ref{sec:gauge-specialization}.
Since ${\cal S}$ and ${\cal S}_0$ are both symplectomorphisms, so is
${\cal S}_2$.
     We now define the generating functional $G : \Gamma_{\ps} \to {\bf R}$
     by demanding that for any functional $f : \Gamma_{\ps} \to {\bf R}$ we have
     \be
     f \circ {\cal S}_2 = \exp \bigg[ \left\{ \ldots , G \right\} \bigg] f \equiv f + \left\{ f, G \right\} + \frac{1}{2} \left\{ \left\{ f, G \right\} , G \right\} + \ldots
     \label{generatedef}
     \ee
Here the notation $\left\{ \ldots , G \right\} $ means the
differential operator that acts on functionals $f$ and returns $\{ f, G
\}$.
The definition (\ref{generatedef}) uses the fact that a general
symplectomorphism can be obtained by exponentiating, as for any Lie group. 
The functional $G$ is similar to a Hamilton-Jacobi functional in that
it encodes the dynamics (except that it is a functional of original
coordinates and original momenta instead of being a functional of original
coordinates and new momenta).

The results of Secs.\ \ref{sec:secondorder} and \ref{sec:higherorders} below
imply that $G = O(\alpha^3)$, and it follows that through $O(\alpha^5)$
it is sufficient to use a truncated version of the formula (\ref{generatedef}) that retains only the first two terms.
In particular, if $y^a$ are arbitrary coordinates on phase space, the scattering map ${\cal S}_2$ can be represented by
\be
\label{eq:gf}
     y^a \circ {\cal S}_2 = y^a + \left\{ y^a, G \right\} + O(\alpha^6).
       \ee

In Appendix \ref{app:symplectomorphism} we
relate the generating functional $G$ to the various functionals
${\cal H}^{\rm e}$, ${\cal H}^{\rm m}$ and ${\cal K}$ defined in Sec.\ \ref{sec:generalform} above.
We also show in Appendix \ref{app:symplectomorphism} that $G$ has the specific form
\be
\label{eq:mathscrGdef0}
G\left[
  {\tilde \Psi}^{\rm e}_{\ps},
    \Psi^{\rm m}_{\ps},
  \chi_{\ps},
  \Delta \Psi^{\rm e}_{\ps},
  {\hat \Psi}^{\rm e}_{\ps}
    \right] = {\mathscr G}\left[{\breve \Psi}^{\rm
      e}_{\ps},  \Psi^{\rm
      m}_{\ps}, e^{-i {\bar \Psi}^{\rm e}_{\ps}} e^{i {\Delta \Psi}^{\rm e}_{\ps}/2}
  \chi_{\ps}
  \right].
\ee
Here on the left hand side the arguments
\be
\label{goodcoords}
\left(
  {\tilde \Psi}^{\rm e}_{\ps},
    \Psi^{\rm m}_{\ps},
  \chi_{\ps},
  \Delta \Psi^{\rm e}_{\ps},
  {\hat \Psi}^{\rm e}_{\ps}
     \right)
\ee
are the independent phase space
coordinates defined in Sec.\ \ref{sec:Dirac},
on the right hand side ${\mathscr G}$ is some functional, and
its arguments are given in terms of the fields on the left hand side by
\begin{subequations}
\begin{eqnarray}
   {\breve \Psi}^{\rm e}_{\ps}(u,\bftheta) &=&  {\tilde
     \Psi}^{\rm e}_{\ps}(u,\bftheta) + g(u) \Delta \Psi^{\rm
     e}_{\ps}(\bftheta), \\
   {\bar \Psi}^{\rm e}_{\ps}(\bftheta) &=&    {\hat \Psi}^{\rm e}_{\ps}(\bftheta) - 2 \int du g'(u) {\tilde \Psi}^{\rm e}_{\ps}(u,\bftheta),
   \end{eqnarray}
\end{subequations}
[cf.\ Eqs.\ (\ref{hatPsi}) and (\ref{brevePsidef})].
The functional ${\mathscr G}$ must obey the constraint (\ref{eq:fa}), whose
interpretation is that its functional derivative with respect to its
first argument ${\breve \Psi}^{\rm e}_{\ps}$ involves just a bulk term
and no boundary term.  It must also obey the constraint (\ref{Gconstraint1}).
These constraints together with the form (\ref{eq:mathscrGdef0}) of the generating
functional guarantee that the corresponding symplectomorphism (\ref{generatedef})
satisfies the required conditions discussed in
Sec.\ \ref{sec:generalform}:
it preserves the conserved quantities (\ref{eq:generalcheckpsie2soft}) and (\ref{matching2})
and transforms appropriately under the symmetry (\ref{tr2}).

\section{Coupling of soft and hard degrees of freedom: general
  arguments and their validity}
\label{sec:arguments}

We now turn to an assessment of some of the arguments that have been
made in the literature about the nature of the dynamics of the
soft degrees of freedom, using as a foundation properties
of the scattering map derived in the previous section.
In the following three subsections, we will discuss three different 
lines of reasoning for why the soft dynamics is either trivial or completely
decoupled from the dynamics of a hard sector. We will argue that these conclusions
are unfounded.

\subsection{Argument for triviality of soft dynamics}
\label{sec:triviald}

One argument that might be made about the dynamics of the soft degrees
of freedom is that they are a trivial extension of the dynamics of the
hard degrees of freedom.  Specifically, the full dynamics can  be
derived from the generating functional [cf.\ Eq.\ (\ref{eq:mathscrGdef0}) above]
\be
\label{eq:mathscrGdef1}
  {\mathscr G}\left[{\breve \Psi}^{\rm
      e}_{\ps},  \Psi^{\rm
      m}_{\ps}, e^{-i {\bar \Psi}^{\rm e}_{\ps}} e^{i {\Delta \Psi}^{\rm e}_{\ps}/2}
    \chi_{\ps} \right],
  \ee
  where we recall that the definition (\ref{brevePsidef}) of the field ${\breve \Psi}^{\rm
    e}_{\ps}$ requires the boundary condition
  \be
  \label{eq:requiredbc}
  \lim_{u \to \pm \infty} \, {\breve \Psi}^{\rm
    e}_{\ps}(u,\bftheta) = \pm \Delta \Psi^{\rm e}_{\ps}(\bftheta)/2.
\ee
Suppose now that one considers scattering with purely hard initial
data,
for which ${\bar \Psi}^{\rm e}_{\ps} = \Delta \Psi^{\rm e}_{\ps}
=0$\footnote{This condition is equivalent to 
${\bar \Psi}^{\rm e}_{\ms} = \Delta \Psi^{\rm e}_{\ms} =0$, since we
  are working within a framework where the $+$ and $-$ variables
  are related by the linear order scattering map (\ref{eq:freee}),
  cf.\ Eqs.\ (\ref{eq:scdecompose}) and (\ref{generatedef})
  above.}.  Such scattering is defined by a functional 
${\mathscr G}[{\breve \Psi}^{\rm
      e}_{\ps},  \Psi^{\rm
      m}_{\ps}, \chi_{\ps}]$,
and fully general scattering is defined by the same functional
(\ref{eq:mathscrGdef1}) except that one of the arguments is modified
by a phase factor.  Therefore fully general scattering is a trivial
extension of purely hard scattering.

A potential loophole in this argument is the fact that the functionals in the
two different cases are defined on different function spaces.  The
purely hard scattering functional is defined on a function space where
${\breve \Psi}^{\rm e}_{\ps}$ satisfies the boundary conditions
(\ref{eq:requiredbc}) with vanishing right hand side, whereas the fully
general functional is defined on the larger function space where the
right hand side is nonzero.  However, it turns out that within the
context of perturbative classical scattering, the general functional
${\mathscr G}$ is determined by continuity by its restriction to the
smaller function space.  This follows from the constraint (\ref{eq:fa})
that the generating functional must satisfy.  See Sec.\ \ref{sec:secondorder} below
for further discussion of this point within the context of second order perturbation theory.
Thus the more general functional is determined uniquely by the
restricted functional.

There is, however, a flaw in the argument.  Scattering with initial
data with ${\bar \Psi}^{\rm e}_{\ps} = \Delta \Psi^{\rm e}_{\ps} =0$
is not ``purely hard'' scattering, since these variables will
generically be nonzero after the scattering due to the electromagnetic
memory effect (see Sec.\ \ref{sec:fourth-order} and Appendix
\ref{app:memory} below for more details).  Thus, even within the
restricted functional, the hard and soft degrees of freedom are intertwined.

\subsection{Argument for decoupling based on theorem in symplectic geometry}
\label{sec:theorem}

A more fruitful approach, following Bousso and Porrati
\cite{Bousso} (henceforth BP), is to focus on the quantities that are
preserved under the scattering map (\ref{calS2:def}), instead of on $\Delta
\Psi^{\rm e}_{\ps}$ and ${\bar \Psi}^{\rm e}_{\ps}$.  These quantities
are the charge $Q_{\ps}(\bftheta)$ given by Eq.\ (\ref{eq:cons-law}), and the
quantity $\Psi^{\rm e}_{\psms}(\bftheta)$ given by
Eqs.\ (\ref{transform})
and (\ref{eq:phasespace-variables})
which is
preserved by Eq.\ (\ref{matching2}).  From Eqs.\ (\ref{pbs}) these
quantities have the Poisson brackets
\begin{subequations}
\label{pbs22}
  \begin{eqnarray}
  \left\{ Q_{\ps}(\bftheta), Q_{\ps}(\bftheta') \right\} &=& 0, \\
   \left\{ \Psi^{\rm e}_{\psms}(\bftheta), \Psi^{\rm e}_{\psms}(\bftheta') \right\} &=& 0, \\
  \left\{ Q_{\ps}(\bftheta), \Psi^{\rm e}_{\psms}(\bftheta') \right\} &=& \delta^{(2)}(\bftheta,\bftheta') - \frac{1}{4 \pi}.
\end{eqnarray}
\end{subequations}

One can now attempt to make an argument for the decoupling of hard and soft degrees of freedom based on the following theorem in symplectic geometry, which is proved in Appendix \ref{app:theorem}:

\medskip
\noindent
{\bf Theorem:} {\it Let $(M, \Omega_{ab})$ be a symplectic manifold of finite dimension $n$,
with associated Poisson bracket defined by $\left\{ f, g \right\} =
\Omega^{ab} \nabla_a f \nabla_b g$ for any functions $f$, $g$ on $M$,
where $\Omega^{ab} \Omega_{bc} = \delta^a_{\ c}$.  Let ${\cal S}_2: M
\to M$ be a symplectomorphism.  Suppose that there exists a set of functions
$s^A : M \to {\bf R}$ for $1 \le A \le t$ which satisfies the two
properties:
\begin{enumerate}
\item The functions are preserved under the symplectomorphism,
\be
s^A \circ {\cal S}_2 = s^A.
\label{preserve}
\ee
\item The Poisson brackets
  \be
  \omega^{AB} = \left\{ s^A, s^B \right\}
  \label{spb}
  \ee
  are
  constants on $M$ and form an invertible $t \times t$ matrix.
\end{enumerate}
Then locally the symplectomorphism factorizes, that is, locally one
can find coordinates $(s^1, \ldots, s^t, h^1, \ldots h^{n-t})$ for
which the symplectomorphism takes the form
\be
s^A \to s^A, \ \ \ \ h^\Gamma \to {\bar h^\Gamma}(h^\Sigma),
\label{simpleans}
\ee
for $1 \le A \le t$ and $1 \le \Gamma,\Sigma \le n-t$.}

\medskip
Interpreting $s^A$ as soft and $h^\Gamma$ as hard
degrees of freedom, the theorem would seem to give sufficient conditions for the
two sets of degrees of freedom to decouple from one another.
Indeed, if we take $s^A$ to be the conserved quantities $Q_{\ps}(\bftheta)$ and
$\Psi^{\rm e}_{\psms}(\bftheta)$ discussed above,
and use the definition (\ref{calS2:def}) of the scattering map ${\cal S}_2$,
then both conditions
of the theorem are satisfied, from Eqs.\ (\ref{pbs22}).   It should
follow that there exists a complementary set of hard variables that are
completely decoupled from the soft dynamics, as in
Eq.\ (\ref{simpleans}).
This is essentially the decoupling argument of BP\footnote{BP present their
  argument in somewhat different terms: instead of invoking the
  theorem they attempt to directly derive the change of phase space coordinates to
  the decoupled set satisfying Eq.\ (\ref{simpleans}).  They also
  focus on the gravitational case rather than the electromagnetic
  case.  However the
  theorem captures the essential idea of their argument.}.

This argument fails for a subtle reason: the theorem is valid in finite
dimensions but not in the infinite dimensional context in which we
want to apply it.
We can demonstrate this failure as follows.  If the
theorem were applicable, it would establish the existence of a set of
coordinates on phase space which satisfy the following three conditions:
(i) The coordinates are independent; (ii) They are
symplectically orthogonal (canonical); and (iii) They include the
conserved charges $Q_{\ps}(\bftheta)$ and $\Psi^{\rm
  e}_{\psms}(\bftheta)$.  Such coordinates do not in fact exist. To
see this, we can start from the coordinates (\ref{goodcoords}) which satisfy (i) and
(ii) but not (iii).  Assuming that coordinates exist which satisfy (i),
(ii) and (iii), it follows that there exists
a symplectomorphism
that maps $\Delta \Psi^{\rm e}_{\ps}$ onto $Q_{\ps}$ and ${\hat
  \Psi}^{\rm e}_{\ps}$ onto $\Psi^{\rm e}_{\psms}$.
We attempt to solve for this symplectomorphism
by using a type II generating functional following BP.
This yields the following coordinates that satisfy (ii) and (iii):
\begin{subequations}
  \label{bcanonical}
\begin{eqnarray}
\label{eq:tildePsinew}
  {\tilde \Psi}^{\rm e}_{\rm new}(u,\bftheta) &=& {\tilde \Psi}^{\rm
    e}_{\ps}(u,\bftheta) + g(u) \Delta \Psi^{\rm e}_{\ps}(\bftheta), \\
  \label{eq:Psimnew}
{\Psi}^{\rm m}_{\rm new}(u,\bftheta) &=& {\Psi}^{\rm m}_{\ps}(u,\bftheta),
\\
\label{eq:chinew}
\chi_{\rm new}(u,\bftheta) &=& \exp \left\{- i \left[ {\hat \Psi}^{\rm e}_{\ps}(\bftheta) - \frac{1}{2} \Delta \Psi^{\rm e}_{\ps}(\bftheta) - 2 \int d{\bar u} g'({\bar u}) {\tilde \Psi}^{\rm e}_{\ps}({\bar u},\bftheta) \right] \right\}\,
\chi_{\ps}(u,\bftheta), \\
\label{eq:DeltaPsinew}
{\Delta \Psi}^{\rm e}_{\rm new}(\bftheta) &=& {\Delta \Psi}^{\rm e}_{\ps}(\bftheta) + e^2 D^{-2} Q^{\rm hard}(\bftheta)[\chi_{\ps}], \\
\label{eq:hatPsinew}
{\hat \Psi}^{\rm e}_{\rm new}(\bftheta) &=& {\hat \Psi}^{\rm
  e}_{\ps}(\bftheta) - \frac{1}{2} \Delta \Psi^{\rm e}_{\ps}(\bftheta) - 2 \int
du g'(u) {\tilde \Psi}^{\rm e}_{\ps}(u,\bftheta), 
\end{eqnarray}
\end{subequations}
where $Q^{\rm hard}$ is defined by Eqs.\ (\ref{currents}) and (\ref{eq:current-leading}).
One can check that these new coordinates have the same Poisson brackets
as the original $+$ coordinates, and so are symplectically orthogonal.
However these coordinates do not satisfy (i), i.e.\ they are not
independent, since we have
\be
\Delta \Psi^{\rm e}_{\rm new}(\bftheta)  = 2 \lim_{u \to \infty} {\tilde \Psi}^{\rm e}_{\rm
  new}(u,\bftheta) + e^2 D^{-2} Q^{\rm hard}(\bftheta)[\chi_{\rm
    new}],
\label{eq:notindep}
\ee
from Eqs.\ (\ref{gbvals}) and (\ref{bc33}).
What has happened is that the original ${\tilde \Psi}$ coordinate
satisfied the boundary conditions ${\tilde \Psi}(u) \to 0$ as $u \to
\pm \infty$, whereas the new ${\tilde \Psi}$ coordinate does not and
so effectively encodes more information\footnote{  
A similar phenomenon arises if one tries to diagonalize the
Hamiltonian functional
\be
H = \int du \int d^2 \Omega (\partial_u \Psi)^2.
\label{eq:ham}
\ee
In finite dimensions one can always find canonical coordinates that
diagonalize a positive definite quadratic Hamiltonian (Williamson's
theorem \cite{williamson1936algebraic,arnold1978mathematical}).
However Williamson's theorem does not apply in our infinite
dimensional context: applying standard
methods to find the canonical phase space coordinates that diagonalize
the Hamiltonian (\ref{eq:ham}) gives rise to coordinates that are not
independent, just as in Eq.\ (\ref{eq:notindep}). The origin of the
difficulty in both cases is that the inverse (\ref{pbs}) of the symplectic form
(\ref{presymfinal2}) does not lie in $\Gamma \otimes \Gamma$ as it
would if the phase space $\Gamma$ were finite dimensional, instead it
lies in a larger space. See Ashtekar \cite{Ashtekar:1981bq} for a rigorous treatment
of $\Gamma$ as a Fr\'echet space.}.
Thus the requirements of
  an autonomous soft sector, independence of
phase space coordinates, and symplectic orthogonality of hard and soft
sectors are incompatible.

A useful perspective on the dynamics can be obtained by defining
an enlarged phase space ${\tilde \Gamma}$ in which the would-be
symplectomorphism (\ref{bcanonical}) is actually a symplectomorphism.
This space is parameterized by the coordinates on the left hand
sides of Eqs.\ (\ref{bcanonical}), where we demand that
\be
\lim_{u \to \infty} {\tilde \Psi}^{\rm e}_{\rm new}(u,\bftheta) = -
\lim_{u \to -\infty} {\tilde \Psi}^{\rm e}_{\rm new}(u,\bftheta),
\ee
but we do not demand that this limit satisfy the constraint
(\ref{eq:notindep}).   The enlarged phase space factors as ${\tilde \Gamma} =
{\tilde \Gamma}_{\rm hard} \times 
{\tilde \Gamma}_{\rm soft}$, where ${\tilde \Gamma}_{\rm hard}$ is parameterized by the
coordinates (\ref{eq:tildePsinew}) -- (\ref{eq:chinew}), and ${\tilde \Gamma}_{\rm soft}$ by the
coordinates (\ref{eq:DeltaPsinew}) and (\ref{eq:hatPsinew}).
We can define a scattering map ${\tilde {\cal S}}_2 : {\tilde \Gamma} \to
{\tilde \Gamma}$, which factorizes, acts on ${\tilde \Gamma}_{\rm soft}$ as the identity,
and acts on ${\tilde \Gamma}_{\rm hard}$ as determined by the generating
functional (\ref{eq:mathscrGdef0}) [note that the arguments of the generating
  functional ${\mathscr G}$ coincide with the coordinates 
  (\ref{eq:tildePsinew}) -- (\ref{eq:chinew})].
The physical phase
space $\Gamma$ is then given by restricting to the subspace of
${\tilde \Gamma}$ given by
the constraint (\ref{eq:notindep}), and the physical scattering map
${\cal S}_2$ is given by 
restricting the action of ${\tilde {\cal S}}_2$ to this
subspace.
Thus, although the scattering map does not factorize as argued by BP,
it descends from a mapping on a larger phase space which does.

In the enlarged phase space, even though the scattering map ${\tilde
  {\cal S}}_2$ factorizes, the map on the hard factor ${\tilde \Gamma}_{\rm
  hard}$ is constrained by the existence of the soft conservation
laws.  In particular, the right hand side of Eq.\ (\ref{eq:notindep})
is preserved, because of the the constraint (\ref{Gconstraint1}) on
the generating functional ${\mathscr G}$.  This is necessary for the
mapping ${\tilde {\cal S}}_2$ to preserve the physical phase space
$\Gamma$.  Thus, the argument of BP for the triviality of the soft
conservation laws --- that they act only on a soft sector and do not
affect the dynamics in a hard sector --- does not apply.

\subsection{Argument for decoupling based on factorization theorem for S matrix elements}
\label{sec:softtheorem}

Another kind of argument for the triviality of scattering of soft
degrees of freedom was given by Mirbabayi and Porrati (henceforth MP)
\cite{Mirbabayi:2016axw}, based on the soft theorems for S-matrix
elements \cite{Weinberg-soft}.  MP assumed that
\begin{enumerate}
\item The system including soft degrees of freedom can be described by
  a Hilbert space ${\cal H}$.
  \item There is a factorization of the Hilbert space into hard and
    soft factors
    \be
    \label{Hfactor}
          {\cal H} = {\cal H}_{\rm hard} \otimes {\cal H}_{\rm soft},
       \ee
       a basis $\left| a \right>$ of ${\cal
         H}_{\rm hard}$, and a set of unitary operators 
       $\Omega(a)$ on ${\cal H}_{\rm soft}$ for which the $S$ matrix
       can be expressed as
       \be
       \label{Smatrixsoft}
       S = S_{ab} \left| a \right> \left< b \right| \Omega(a)
         \Omega(b)^\dagger.
         \ee
\end{enumerate}
MP defined the factorization (\ref{Hfactor}) in terms of an infrared cutoff on mode energy,
and took the basis $\left| a \right>$ of hard states to be the usual
particle number eigenstates used in Feynman diagrams.  They then
argued for the decomposition (\ref{Smatrixsoft}) from soft theorems
\cite{Weinberg-soft}, with the operators $\Omega(a)$ providing the
universal soft factors for $S$ matrix elements when some of the
ingoing and outgoing particles become soft.

The assumptions (\ref{Hfactor}) and (\ref{Smatrixsoft}) are equivalent
to a decoupling of two sectors.  We can write
\be
S = U ( S_{\rm hard} \otimes 1 ) U^\dagger
\ee
with $U$ being the unitary operator $U = \sum_a \Omega(a) \left| a
\right> \left< a \right|$ and $S_{\rm hard} = S_{ab} \left| a \right>
\left< b \right|$.  It follows that there exists another factorization
of the Hilbert space ${\cal H} = {\bar {\cal H}}_{\rm hard} \otimes
{\bar {\cal H}}_{\rm soft}$ for which the scattering matrix $S$
factorizes
and acts as the identity on ${\bar {\cal H}}_{\rm soft}$.  Taking the
classical limit it then follows that the scattering map ${\cal S}_2$
is of the factorized form (\ref{simpleans}) discussed earlier,
for some choice of hard and soft variables.

This conclusion is in conflict with explicit computations within the
classical theory detailed in Secs.\ \ref{sec:secondorder} and
\ref{sec:higherorders} below (where we use an exact decomposition into
hard and soft degrees of freedom rather than a decomposition based on
letting a cutoff go to zero).
We conclude that the assumptions 1 and 2 above are not valid in such a
context:
the correct description of the soft theorems within the quantum theory
must be more involved than Eqs.\ (\ref{Hfactor}) and (\ref{Smatrixsoft}).
It is possible that no Hilbert space description of the
dynamics exists when soft degrees of freedom are included, as argued by
Prabhu, Satishchandran, and Wald \cite{Prabhu:2022mcj} (see
Sec.\ \ref{conclusion} below for further discussion).
Another possibility is that the Hilbert space cannot be expressed as a
tensor product of hard and soft factors, the appropriate
structure might be something like the fusion product defined by
Donnelly and Freidel\cite{Donnelly:2016auv}.

\section{Explicit computation of scattering map: second order}
\label{sec:secondorder}

\subsection{Preamble}

In the previous section we refuted some general arguments for the
existence of decoupled hard and soft sectors.
However to show that the sectors are actually always coupled requires
an explicit computation of the dynamics.  Such an explicit computation in the context
of perturbation theory is the goal of the rest of this paper.
We start in this section by considering the second order scattering map,
and proceed to third and fourth orders in the following section.

\subsection{Second order dynamics}

The scattering at second order is trivial in the sense that all the
functionals defined in Eq.\ (\ref{eq:general2}), that parameterize the nonlinear
interactions in preferred asymptotic gauge, vanish to this order:
\be
{\cal H}^{\rm e} = O(\alpha^3), \ \ \ 
{\cal H}^{\rm e}_\infty = O(\alpha^3), \ \ \
{\cal H}^{\rm m} = O(\alpha^3), \ \ \
{\cal K} = O(\alpha^3).
\label{eq:secondordertrivial}
\ee
As a consequence, 
the explicit form of the scattering map coincides 
with the linear order scattering map (\ref{eq:freee}),
\begin{subequations}
      	\label{eq:freee2}
      	\begin{eqnarray}
        \label{eq:freee2-b}
      	\pb \tilde{\Psi}^{\rm e}_{\ps} &=& \tilde{\Psi}^{\rm e}_{\ms}
        + O(\alpha^3) , \\
\label{eq:freee2-c}
      	\pb \Psi^{\rm m}_{\ps} &=& - \Psi^{\rm m}_{\ms} + O(\alpha^3), \\
          \label{eq:freee2-a}
      	\pb \chi_{\ps} &=& - \chi_{\ms} + O(\alpha^3), \\
\label{eq:freee2-d}
      	\pb \Delta \Psi^{\rm e}_{\ps} &=&  \Delta \Psi^{\rm
          e}_{\ms} + O(\alpha^3), \\
        \label{eq:freee2-e}
      	\pb {\bar \Psi}^{\rm e}_{\ps} &=& 
        {\bar \Psi}^{\rm e}_{\ms} + \Delta
          \Psi^{\rm e}_{\ms} + O(\alpha^3), 
        \end{eqnarray}
\end{subequations}
from Eqs.\ (\ref{transform}) and
(\ref{brevePsidef}).  In particular, the
hard sector $(\tilde{\Psi}^{\rm e}, \Psi^{\rm m}, \chi)$
and
the soft sector $(\Delta \Psi^{\rm e}, {\bar \Psi}^{\rm e})$
are decoupled from one another
and evolve independently.  Thus we will need go to higher order in
perturbation theory to find the leading order couplings between the
soft and hard sectors.  We will return to this story in
Sec.\ \ref{sec:higherorders} below.  In the remainder of this section
we will discuss some properties of the second order scattering and its derivation.

Although the scattering is trivial in preferred asymptotic gauge, it
is not trivial in asymptotic \Lorentz gauge, where there is a nonlinear
soft-hard interaction at quadratic order.  This can be seen by substituting
the result (\ref{eq:secondordertrivial}) into the general \Lorentz gauge
scattering map (\ref{eq:general3}), which yields
from Eqs.\ (\ref{transform}) and 
(\ref{brevePsidef}) that
\begin{subequations}
      	\label{eq:freee2l}
      	\begin{eqnarray}
        \label{eq:freee2-bl}
      	\pb \tilde{{\underline \Psi}}^{\rm e}_{\ps} &=&
        \tilde{{\underline \Psi}}^{\rm e}_{\ms}
        + O(\alpha^3) , \\
        \label{eq:freee2-cl}
      	\pb {\underline \Psi}^{\rm m}_{\ps} &=& - {\underline \Psi}^{\rm m}_{\ms} + O(\alpha^3), \\
          \label{eq:freee2-al}
      	\pb {\underline \chi}_{\ps} &=& - {\underline \chi}_{\ms} + 2
        i \Delta {\underline \Psi}^{\rm e}_{\ms} {\underline
          \chi}_{\ms} + O(\alpha^3), \\
          \label{eq:freee2-dl}
      	\pb \Delta {\underline \Psi}^{\rm e}_{\ps} &=&  \Delta
            {\underline \Psi}^{\rm
          e}_{\ms} + O(\alpha^3),\\
            \label{eq:freee2-el}
      	\pb {\bar {\underline \Psi}}^{\rm e}_{\ps} &=& 
        {\bar {\underline \Psi}}^{\rm e}_{\ms} - \Delta
          {\underline \Psi}^{\rm e}_{\ms} + O(\alpha^3).
        \end{eqnarray}
\end{subequations}
The nonlinear term in Eq.\ (\ref{eq:freee2-al}) arises from the phase factor in
Eq.\ (\ref{eq:generalchi3}), and can be written in terms of the expansion
coefficients (\ref{pert-ansatz1}) and (\ref{pert-ansatz2}) as
\be
      	\pb {\underline \chi}^{(2)}_{\ps}(u,\bftheta) =  2
        i \Delta {\underline \Psi}^{{\rm e}(1)}_{\ms}(\bftheta)\, {\underline
          \chi}^{(1)}_{\ms}(u,\bftheta).
        \label{nonlinearexplicit}
\ee
The nonlinear scattering map (\ref{eq:freee2l}), which is equivalent to the result
(\ref{eq:secondordertrivial}), is derived in Appendix \ref{app:secondorder}.

The nonlinearity in the second order \Lorentz gauge scattering map (\ref{eq:freee2l})
arises only because of the presence of soft degrees of freedom.  It
vanishes if
we specialize to incoming data that is purely hard 
($\Delta {\underline \Psi}^{\rm e}_{\ms}= {\bar {\underline
    \Psi}}^{\rm e}_{\ms} = 0$).
This is a well-known result, familiar in the quantum context
from the Feynman rules for scalar QED \cite{2014qfts.book.....S}.
The argument is that quadratic scattering corresponds to a three
particle interaction and to the operator $A^a \Phi^* \nabla_a
\Phi + $ c.c. in the Lagrangian.  If we combine the conservation of
four-momentum and the fact that the final momentum must be null to be
on-shell, this forces all the momenta to be collinear.  Finally the
\Lorentz gauge condition forces the final result to vanish.
In Appendix \ref{app:secondorder} we translate this argument into the
language of 
perturbative classical scattering used in this paper, and explain how
it breaks down in the presence of soft degrees of freedom.

One consequence of the nonlinear soft-hard interaction is that the
scattering map (\ref{eq:freee2l}) is not 
continuous, in the following sense.  
Consider a sequence ${}^{(n)} {\underline \Psi}^{\rm e}_{\ms}$
of incoming configurations, each of which has no soft part
[${}^{(n)}\Delta {\underline \Psi}^{\rm e}_{\ms}= {}^{(n)}{\bar {\underline
    \Psi}}^{\rm e}_{\ms} = 0$], which converges pointwise to
${\underline \Psi}^{\rm e}_{\ms}$:
\be
\lim_{n \to \infty} {}^{(n)} {\underline \Psi}^{\rm e}_{\ms}
(v,\bftheta) =
{\underline \Psi}^{\rm e}_{\ms}
(v,\bftheta).
\label{limm}
\ee
For each element in the sequence, the result
(\ref{nonlinearexplicit}) vanishes, and so the $n \to \infty$
limit of the scattered fields also has this property.
However the result (\ref{nonlinearexplicit}) does not vanish for the
scattering of the $n \to \infty$ pointwise limit (\ref{limm}) of 
the initial data. Thus,
the scattering of soft degrees of freedom cannot be
obtained by simply taking a naive limit of hard scattering.
Equivalently, the scattering map is not determined by continuity by
its restriction to initial 
data that consists of smooth wavepacket states.
This property is specific to asymptotic \Lorentz gauge, and does not occur in preferred asymptotic gauge where the scattering map is continuous,
as can be seen from Eqs.\ (\ref{eq:fa}) and (\ref{eq:mathscrGdef}) of Appendix \ref{app:symplectomorphism}.

\section{Explicit computation of scattering map: third and fourth orders}
\label{sec:higherorders}

\subsection{Overview}

In the previous section we showed that the classical scattering map
factorizes into decoupled hard and soft sectors 
at quadratic order.
We now proceed to third and fourth orders in perturbation theory.
We will show in this section that no factorization into decoupled
sectors is possible at these orders.

It will be sufficient for our arguments to consider only two 
specific pieces of the fourth order scattering.  The first piece is the change in
electromagnetic memory between $\scri^-$ and $\scri^+$.  In  
Sec.\ \ref{sec:fourth-order} and Appendix \ref{app:memory} we compute
this quantity explicitly and show that it is nonzero, as would be
expected on general grounds \cite{Strominger:lectures}.  The second
piece is the contribution to $\chi_{\ps}$ that is linear in $
{\bar \Psi}^{\rm e}_{\ms}$ and cubic in $\chi_{\ms}$, which we show in
Sec.\ \ref{sec:fourth-order} is nonzero.

Next, in order to show that the hard and soft sectors are always dynamically
coupled, we need to address an ambiguity in the definitions of these
sectors.  Roughly speaking the soft sector consists of zero energy
modes, while the hard sector consists of finite energy modes, but these
requirements allow for considerable leeway in the definitions.
Our strategy will be to show that for {\it any} definitions of hard
and soft sectors, the two sectors are dynamically coupled.
However, we do require that the same
definitions of hard and soft be used at $\scri^-$ and at $\scri^+$
(otherwise it is always trivially possible to find definitions which
decouple the dynamics).
As discussed in the introduction, we adopt two different conventions
for identifying the phase spaces $\Gamma_-$ at $\scri^-$ and
$\Gamma_+$ at $\scri^+$, that allow us to enforce this requirement.

Section \ref{sec:firstversioncoupled} considers the first convention,
based on identifying $\scri^-$ and $\scri^+$ using null geodesics.
First, in
Sec.\ \ref{sec:defic}, we consider cubic and 
quartic terms in the scattering map which couple together the soft and
hard sectors.
Some of these terms are invariant under
linear field redefinitions and perturbative field redefinitions that alter the definitions
of the two sectors. 
We will call such terms {\it invariant interactions}, as they cannot be
removed by field redefinitions, and we derive criteria for interactions to be invariant.
Then, in Sec.\ \ref{sec:memoryinvariant}, we show 
that the two pieces of the fourth order scattering discussed in Sec.\ \ref{sec:fourth-order} are such
invariant interactions.  Since these quantities are generically nonzero,
the hard and soft sectors are always coupled.  These invariant interactions also disallow
scattering maps of the forms (\ref{eq:factorizes1}) and (\ref{eq:factorizes2}).

In Sec.\ \ref{sec:secondversioncoupled} we turn to the second
convention, which is based on the free field scattering map.
For this case there exist definitions of
the soft variables which are conserved, discussed in Sec.\ \ref{sec:theorem} above,
and it is natural to restrict
to definitions of soft variables which preserve this property.  With
this assumption we
show that it is not possible to find definitions
of soft and hard variables for which the scattering map has the
uncoupled form (\ref{eq:factorizes}), again using
the results of Sec.\ \ref{sec:fourth-order}.

\subsection{Two nonzero pieces of fourth order scattering}
\label{sec:fourth-order}

In this section we focus attention on two particular pieces of the fourth
order scattering which we will show are nonzero.

The first piece is the total change
$\Delta \Psi^{\rm e}_{\ps} - \pb \Delta \Psi^{\rm e}_{\ms}$
in electromagnetic memory between
$\scri^-$ and $\scri^+$.  From the conservation law (\ref{eq:cons-law})
this quantity is proportional to
\be
\label{deltaqhard}
\Delta Q^{\rm hard}(\bftheta) = Q^{\rm hard}_{\ps}(\bftheta ) - \pb Q^{\rm
  hard}_{\ms}(\bftheta),
\ee
the change in the total hard charge per unit angle in the scalar field.
It also corresponds to the the functional ${\cal H}^{\rm e}_\infty$
in the general parameterization (\ref{eq:general2}) of the scattering
map, from Eq.\ (\ref{cons23}). 
It is clear that this quantity is
nonzero in general \cite{Strominger:lectures}.  For example, for
electromagnetism coupled to massive charged 
particles, if one has two incoming charged particles, there will in
general be a nontrivial scattering and the outgoing particles will
have different directions of propagation from the incoming ones.
In this paper our source is a
massless scalar field and so we deal instead with wave packet ingoing
and outgoing states, but it is clear that generically there will be
nontrivial scattering.  In Appendix \ref{app:memory} we verify this
explicitly by computing the leading order contribution to
$\Delta Q^{\rm hard}(\bftheta)$, which is quartic in the initial
scalar field $\chi^{(1)}_{\ms}$ and is given by the expressions
(\ref{cm-tchannel1}) and (\ref{cm-schannel1}).

The second piece of the fourth order scattering that we consider is the contribution to
$\chi_{\ps}(u,\bftheta)$ that is linear in ${\bar \Psi}^{\rm e}_{\ms}$
and cubic in $\chi_{\ms}(v,\bftheta)$.  We can show that this
contribution is nonzero as follows.  Combining the leading order formula (\ref{eq:gf})
for the scattering map ${\cal S}_2$ in terms of the generating
functional $G$ together with the Poisson bracket (\ref{pbs4}) gives
that the change in $\chi_{\ps}(u,\bftheta)^*$ is
\be
- \frac{1}{2} \int du' K(u-u') \frac{ \delta G} {\delta
  \chi_{\ps}(u',\bftheta)},
\ee
where $K(u) = \Theta(u)-1/2$.  Combining this with the specific form 
(\ref{eq:mathscrGdef0}) of the generating functional, the definition
(\ref{calS2:def}) of the scattering map ${\cal S}_2$ and the free field
scattering map (\ref{eq:freee}) we obtain
\be
\pb \chi_{\ps} = - \chi_{\ms} - \frac{1}{2} \int du' K(u-u') e^{i
  {\bar \Psi}^{\rm e}_{\ms}} e^{i \Delta \Psi^{\rm e}_{\ms}/2}
\frac{\delta {\mathscr G}}{\delta \chi_{\ps}}\left[ {\tilde \Psi}^{\rm
    e}_{\ms} + g \Delta \Psi^{\rm e}_{\ms}, - \Psi^{\rm m}_{\ms}, -
  e^{- i {\bar \Psi}^{\rm e}_{\ms}} e^{- i \Delta \Psi^{\rm
      e}_{\ms}/2} \chi_{\ms} \right]^*.
\label{ansg}
\ee
If we now specialize to a vanishing incoming electromagnetic field we
obtain
\be
\pb \chi_{\ps} = - \chi_{\ms} - \frac{1}{2} \int du' K(u-u') 
\frac{\delta {\mathscr G}}{\delta \chi_{\ps}}\left[ 0, 0, -\chi_{\ms}
  \right]^*.
\label{ansg1}
\ee
The second term here is computed explicitly in the last paragraph of Appendix
\ref{app:memory} and is cubic in $\chi_{\ms}$ and nonzero.
Consider now the contribution to the general result (\ref{ansg}) that
is linear in ${\bar \Psi}^{\rm e}_{\ms}$ evaluated at ${\Delta \Psi}^{\rm
  e}_{\ms} = {\tilde \Psi}^{\rm e}_{\ms}=0$. Because of the form of the
expression (\ref{ansg}), this contribution can be computed
from the known expression (\ref{ansg1}) by inserting the complex
exponential factors.
Therefore we conclude that the contribution which is linear
in ${\bar \Psi}^{\rm e}_{\ms}$ and cubic in $\chi_{\ms}$ is nonzero.
We write this contribution as
\be
\chi_{\ps} = {\hat F}[{\bar \Psi}^{\rm e}_{\ms}, \chi_{\ms},
  \chi_{\ms}, \chi_{\ms} ]
\label{hatFdef}
\ee
where the function ${\hat F}$ is linear in its first, second and third
arguments and antilinear in its fourth (see Appendix \ref{app:memory}).

\subsection{Hard and soft sectors are coupled for first version of
  scattering map}
\label{sec:firstversioncoupled}

We now specialize to the first version (\ref{calS1:def}) of the scattering maps
discussed in the introduction.  For this version the phase spaces $\Gamma_-$ and
$\Gamma_+$ are identified using the mapping $\varphi_0$ from $\scri^-$ to
$\scri_+$ defined with null geodesics, allowing us to define the scattering map ${\cal S}_1$ from
$\Gamma_+$ to $\Gamma_+$.  We will allow general definitions of 
hard and soft sectors of $\Gamma_+$, and will not require these
sectors to be symplectically orthogonal.

\subsubsection{Definition of invariant interactions}
\label{sec:defic}

The general form (\ref{eq:general2}) of the scattering map can be written
schematically as
\be
\label{eq:scattering-map-abstract}
\bar{y}^{a}  = L^{a}_{\ b} y^{b} + M^{a}_{\ bcd} y^{b} y^{c} y^{d} +
N^{a}_{\ bcde} y^{b} y^{c} y^{d} y^e + O(y^5),
\ee
where $y^a$ are abstract phase space coordinates
and the map is
parameterized in terms of some phase space tensors 
$L^{a}_{\ b}$, $M^{a}_{\ bcd}$ and 
$N^{a}_{\ bcde}$.  Here we are using a
notation where the indices $a, b, \ldots $ run over
the various fields
and also encode the dependence of these fields on the coordinates $u,
\theta^A$, so that contractions over these indices encompass integrals
over these variables.  The barred coordinates ${\bar y}^a$ refer to the final data on
$\scri^+$,
\be
{\bar y}^a = \left( {\tilde \Psi}_{\ps}^{\rm e}, \Psi_{\ps}^{\rm m}, \chi_{\ps}, \Delta
\Psi_{\ps}^{e}, {\bar \Psi}_{\ps}^{\rm e} \right),
\ee
while the unbarred coordinates $y^a$ refer to the
data on $\scri^+$ obtained from the identification map
(\ref{firstid}) acting on the initial data on $\scri^-$.  This mapping
amounts to evaluating the initial data at $v=u$ and applying the
pullback $\pb$, so the functions on $\scri^+$ are
\be
y^a = \pb \left( {\tilde \Psi}_{\ms}^{\rm e}, \Psi_{\ms}^{\rm m}, \chi_{\ms}, \Delta
\Psi_{\ms}^{e}, {\bar \Psi}_{\ms}^{\rm e} \right).
\label{eq:listof}
\ee
Finally the schematic
scattering map (\ref{eq:scattering-map-abstract}) encodes the fact that 
there are no quadratic terms when using preferred asymptotic gauge,
cf.\ Eqs.\ (\ref{eq:freee2}) above, so the leading
nonlinearities arise at cubic order.

As in the previous section the phase space coordinates can be
decomposed into hard and soft components,
\be
y^a = ( h^A, s^\Gamma ),
\label{hardsoft}
\ee
where the hard variables $h^A$ refer to the first three fields in
(\ref{eq:listof}), which depend on $u$ and $\theta^A$, while the soft
variables $s^\Gamma$ refer to the last two fields in
(\ref{eq:listof}), which depend only on $\theta^A$.  As discussed in
Sec.\ \ref{sec:fos}, the two sectors are uncoupled in the linear order
scattering map (\ref{eq:freee}), so the tensor $L^a_{\ b}$ is block
diagonal with vanishing off-diagonal blocks:  
\be
L^\Gamma_{\ A} = 0, \ \ \ L^A_{\ \,\Gamma} = 0.
\label{ld}
\ee
The diagonal block $L^A_{\ B}$ in the hard sector is given by
Eqs.\ (\ref{eq:freee-c}) -- (\ref{eq:freee-a}), with the pullback
operators $\pb$ removed because of Eq.\ (\ref{eq:listof}), while the diagonal
block $L^{\Gamma}_{\ \Sigma}$ in the soft sector is similarly given 
by Eqs.\ (\ref{eq:freee-d}) and (\ref{eq:freee-e}) with $\pb$ removed.

As we showed in Sec.\ \ref{sec:fourth-order} above, the hard and soft sectors are
coupled via the higher order terms 
in Eq.\ (\ref{eq:scattering-map-abstract}) which mix the two sectors together.
Our goal here is to determine when such interactions can be removed by
field redefinitions that alter the definitions of the hard
and soft sectors.  We first consider perturbative field redefinitions of the form
\be
y^a = z^a + \Upsilon^a_{\ bcd} z^b z^c z^d + \Xi^a_{\ bcde} z^b z^c
z^d z^e+ O(z^5),
\label{pstransf}
\ee
which defines new phase space coordinates $z^a$, together with an
identical transformation for the barred variables. 
Here we have assumed that the transformation is the identity to linear
order, as linear transformations are considered separately below.
We also exclude any quadratic terms in the transformation
(\ref{pstransf}), since such terms would not generally maintain the
property of the scattering map 
(\ref{eq:scattering-map-abstract}) of having no quadratic
terms\footnote{\label{longf}
Terms of the form $\rho^a_{\ bc} z^b z^c$ in
Eq.\ (\ref{pstransf}) with $\rho^a_{\ bc}$ chosen to satisfy
\be
L^a_{\ e} \rho^e_{\ bc} = \rho^a_{\ ef} L^e_{\ b} L^f_{\ c}
\label{cc33}
\ee
will maintain the property of no quadratic terms in the scattering
map. However the contributions of such terms to
Eqs.\ (\ref{eq:transflawBC}) vanish when $\Upsilon^a_{\ bcd} =
\Xi^a_{\ bcde} =0$, except for the contribution
\be
3 M^a_{\ f(bc} \rho^f_{\ de)} -2 \rho^a_{\ fg} M^f_{\ (bcd}L^g_{\ e)}
\label{anom}
\ee
to
Eq.\ (\ref{eq:transflawBC3}).  These terms do not contribute to the interaction
(\ref{candidate}).
This follows from the fact that 
the condition (\ref{cc33}) together with Eq.\ (\ref{eq:freee})
allow the following components of $\rho$ to be freely specified: 
$\rho^1_{\ 11}, \rho^1_{\ 14}, \rho^1_{\ 12},\rho^1_{\ 33}, \rho^1_{\ 44}, \rho^1_{\ 42},\rho^1_{\ 22},
\rho^3_{\ 13}, \rho^3_{\ 34}, \rho^3_{\ 32},
\rho^4_{\ 14}, \rho^4_{\ 44}, \rho^4_{\ 42},
\rho^5_{\ 11}, \rho^5_{\ 14}, \rho^5_{\ 12}, \rho^5_{\ 33}, \rho^5_{\ 44}, \rho^5_{\ 42}, \rho^5_{\ 22},
\rho^5_{\ 11},\rho^5_{\ 14}$,
$\rho^2_{\ 12}, \rho^2_{\ 33}, \rho^2_{\ 44}, \rho^2_{\ 42},\rho^2_{\ 22}$,
with the remaining components constrained to vanish except for
$\rho^5_{\ 15} = \rho^4_{\ 14}$, $\rho^5_{\ 45} = \rho^4_{\ 44}/2$, $\rho^5_{\ 52} = \rho^4_{\ 42}$
and those determined by $\rho^a_{\ bc} = \rho^a_{\ cb}$.
Here we are using the convention (\ref{eq:listof}) for ordering the fields.
Also the components
$M^4_{\ 133} = M^{{\Delta \Psi}^{\rm e}}_{{\tilde \Psi}^{\rm e} \chi
  \chi}$
and $M^4_{\ 533} = M^{{\Delta \Psi}^{\rm e}}_{{\bar \Psi}^{\rm e} \chi \chi}$
in Eq.\ (\ref{anom}) vanish from
Eq.\ (\ref{final444}) and an equation analogous to Eq.\ (\ref{f33}) at
one lower order.
Since general transformations can be obtained by  
composing those with $\rho^a_{\ bc}=0$ and those with  
$\Upsilon^a_{\ bcd} = \Xi^a_{\ bcde} =0$, we can without loss of
generality neglect those with nonzero $\rho^a_{\ bc}$.}.

Using the transformation (\ref{pstransf}) and its inverse, we can
write the scattering map (\ref{eq:scattering-map-abstract}) in terms
of the new phase space variables $z^a$.  The result is
\be \label{eq:scattering-map-abstract1}
\bar{z}^{a}  = {\hat L}^{a}_{\ b} z^{b} + {\hat M}^{a}_{\ bcd} z^{b} z^{c} z^{d} +
{\hat N}^{a}_{\ bcde} z^{b} z^{c} z^{d} z^e + O(z^5), 
\ee
where the transformed tensors are
\begin{subequations}
    \label{eq:transflawBC}
    \begin{eqnarray}
{\hat L}^a_{\ b} &=& L^a_{\ b},\\      
{\hat M}^a_{\ bcd} &=&  M^a_{\ bcd}  + L^a_{\ e} \Upsilon^e_{\ bcd} -
\Upsilon^a_{\ efg} L^e_{\ b} L^f_{\ c} L^g_{\ d},\\
\label{eq:transflawBC3}
{\hat N}^a_{\ bcde} &=&  N^a_{\ bcde}  + L^a_{\ f} \Xi^f_{\ bcde} -
\Xi^a_{\ fghi} L^f_{\ b} L^g_{\ c} L^h_{\ d} L^i_{\ e}.
  \end{eqnarray}
\end{subequations}
    
We denote by ${\cal V}$ the linear space of tensors $(M^a_{\ bcd},
N^a_{\ bcde})$.
Consider now linear maps $\ell : {\cal V} \to {\bf R}$,
elements of the dual space ${\cal V}^*$.  For example
$\ell(M,N)$ could be a particular component of the tensor $N$.
We define ${\cal W}_{\rm mixed}$ to be the subspace of ${\cal V}^*$
that is spanned by components of $M$ and $N$ with both hard and soft
indices, excluding the purely soft components 
$(M^\Sigma_{\ \,\Gamma\Delta\Upsilon},
N^\Sigma_{\ \,\Gamma\Delta\Upsilon\Lambda})$ and purely hard
components
$(M^A_{\ \,BCD},
N^A_{\ \,BCDE})$.
We will call maps $\ell$ in ${\cal W}_{\rm mixed}$ {\it interactions},
since they couple the hard and soft sectors together in the dynamics.
An example of such an interaction is the change in electromagnetic
memory discussed in Sec.\ \ref{sec:fourth-order} above.
We
define a linear map   $ {\frak A} : {\cal V} \to {\cal V}$
that takes
\be
\label{frakAdef}
{\frak A}   : 
( \Upsilon^a_{\ bcd}, \Xi^a_{\ bcde} ) \to 
\left( L^a_{\ e} \Upsilon^e_{\ bcd} - \Upsilon^a_{\ efg} L^e_{\ b}
L^f_{\ c} L^g_{\ d}\ ,\  L^a_{\ f} \Xi^f_{\ bcde} -
\Xi^a_{\ fghi} L^f_{\ b} L^g_{\ c} L^h_{\ d} L^i_{\ e}\right).
\ee
We define the subspace ${\cal W}_{\rm invariant}$ of ${\cal V}^*$ to be the set of maps
$\ell$ for which $\ell \circ \frak{A} =0$, which is the kernel of the
transpose of ${\frak A}$.
The space ${\cal W}_{\rm invariant}$ depends on the linear order
scattering map $L^a_{\ b}$, for example if $L^a_{\ b} = \delta^a_{\ b}$
then it is the entire space ${\cal V}^*$.  
The key property of this definition is that
nonzero interactions in ${\cal W}_{\rm invariant} \cap {\cal W}_{\rm mixed}$
cannot be set to zero
using
the field redefinitions
(\ref{pstransf}), from Eqs.\ (\ref{eq:transflawBC}) and (\ref{frakAdef}).

Turn now to linear field redefinitions of the form
\be
y^a = \Omega^a_{\ b} z^b.
\label{redeflin}
\ee
The transformed scattering map is again of the form
(\ref{eq:scattering-map-abstract1})
with the transformed tensors being
\begin{subequations}
    \label{eq:transflawBC1}
    \begin{eqnarray}
      \label{hatA11}
{\hat L}^a_{\ b} &=& \left( \Omega^{-1} \right)^a_{\ c} L^c_{\ d}
\Omega^d_{\ b},\\      
      {\hat M}^a_{\ bcd} &=&  \left( \Omega^{-1} \right)^a_{\ e} M^e_{\ fgh}
      \Omega^f_{\ b} \Omega^g_{\ c} \Omega^h_{\ d},\\
      \label{hatNtransform}
{\hat N}^a_{\ bcde} &=&  \left( \Omega^{-1} \right)^a_{\ f} N^f_{\ ghij}
\Omega^g_{\ b} \Omega^h_{\ c} \Omega^i_{\ d} \Omega^j_{\ e}.
  \end{eqnarray}
\end{subequations}
We restrict attention to transformations $\Omega^a_{\ b}$ which
preserve the decoupling (\ref{ld}) of the two sectors at linear order:
\be
{\hat L}^\Gamma_{\ A} = 0, \ \ \ {\hat L}^A_{\ \,\Gamma} = 0.
\label{ld1}
\ee
Without loss of generality for analyzing decoupling we can assume that
\be
\Omega^\Gamma_{\ \Sigma} = \delta^\Gamma_{\ \Sigma},
\ \ \ \Omega^A_{\ B} = \delta^A_{\ B}.
\label{condtt4}
\ee
The conditions (\ref{ld1}) are then equivalent
to, from Eqs.\ (\ref{ld}) and (\ref{hatA11}),
\be
\Omega^A_{\ \Sigma} L^\Sigma_{\ \Gamma} = L^A_{\ B}
\Omega^B_{\ \Gamma}, \ \ \ \ \
\Omega^\Gamma_{\ B} L^B_{\ A} = L^\Gamma_{\ \Sigma}
\Omega^{\Sigma}_{\ A}.
\label{ld2}
\ee

For each such map $\Omega^a_{\ b}$, we define the map ${\frak
  A}_\Omega : {\cal V} \to {\cal V}$ by
\be
\label{frakAdef1}
{\frak A}_\Omega   : 
( M^a_{\ bcd}, N^a_{\ bcde} ) \to 
\left( 
      M^a_{\ bcd} - \Omega^{-1\,a}_{\ \ e} M^e_{\ fgh}
      \Omega^f_{\ b} \Omega^g_{\ c} \Omega^h_{\ d} \ , \ 
N^a_{\ bcde} -  \Omega^{-1\,a}_{\ \ f} N^f_{\ ghij}
\Omega^g_{\ b} \Omega^h_{\ c} \Omega^i_{\ d} \Omega^j_{\ e} \right).
\ee
We define the subspace ${\cal W}^\prime_{\rm invariant}$ of ${\cal V}^*$ to be the set of maps
$\ell$ for which
$\ell \circ \frak{A}_\Omega =0$
for all $\Omega^a_{\ b}$ satisfying the conditions (\ref{condtt4}) and (\ref{ld2}).
Equivalently, ${\cal W}^\prime_{\rm invariant}$ is the intersection of
the kernels of the transposes of the maps $ {\frak A}_\Omega$. 
From Eqs.\ (\ref{eq:transflawBC1}) and (\ref{frakAdef1}) we see
that 
nonzero interactions $\ell$ in ${\cal W}^\prime_{\rm invariant} \cap {\cal W}_{\rm mixed}$
cannot be set to zero using
the field redefinitions (\ref{redeflin}).

We will refer to the interactions in
${\cal W}_{\rm invariant} \cap {\cal W}_{\rm invariant}^\prime \cap
{\cal W}_{\rm mixed}$
as {\it invariant interactions}.
These interactions are not altered by either type of transformation,
linear or nonlinear.
Which specific cubic and quartic interactions are invariant depends on
the details of the free field scattering map $L^a_{\ b}$.

\subsubsection{Two invariant nonzero interactions at quartic order}
\label{sec:memoryinvariant}

We next show that the two quartic interactions discussed in
Sec.\ \ref{sec:fourth-order} above are invariant interactions.

Consider first the change (\ref{deltaqhard}) in electromagnetic memory, specialized 
to the case when there is an
incoming scalar field but no electromagnetic field.
We show in Appendix \ref{app:memory} that this quantity is of
order $O(\alpha^4)$, so it contributes to the tensor $N^a_{\ bcde}$.
Since the incoming scalar
field $\chi_{\ms}$ is assumed to have no soft part [cf.\ Eq\. (\ref{assumedfalloff}) above]
the corresponding component of $N$ is
\be
\ell(M,N) = N^\Lambda_{\ \,ABCD}
\label{candidate}
\ee
where the $\Lambda$ index corresponds to the field $\Delta \Psi^{\rm
  e}$, 
and the indices
$A$, $B$, $C$ and $D$ 
correspond to the field 
$\chi$, from Eqs.\  (\ref{eq:generalbarpsie2}),
(\ref{eq:generalcheckpsie2soft}) and (\ref{hardsoft}).

We now consider the action of the map (\ref{frakAdef})
on the interaction 
(\ref{candidate}).  From Eqs.\ (\ref{ld}), (\ref{frakAdef}) and (\ref{candidate}) we have
that $\ell \circ {\frak A} =0$ if
\be
L^\Lambda_{\ \Sigma} \Xi^\Sigma_{\ ABCD} = \Xi^\Lambda_{\ EFGH} L^E_{\ A}
L^F_{\ B} L^G_{\ C} L^H_{\ D},
\label{condtqq}
\ee
where the indices $\Lambda$, $A$, $B$, $C$ and $D$ have the values
discussed after Eq.\ (\ref{candidate}), and $\Xi^a_{\ bcde}$ is the
transformation tensor defined in Eq.\ (\ref{pstransf}).
By using the linear order scattering map
(\ref{eq:freee}) together with Eq.\ (\ref{eq:listof}),
we see that the effect of the mappings $L^E_{\ A}$ on the right hand
side is to replace $\chi_{\ms}$ with $- \chi_{\ms}$.
Similarly the mapping $L^\Lambda_{\ \Sigma}$ on the left
hand side acts as the identity map.
Therefore
the condition (\ref{condtqq}) can be written as  
\be
  F[\chi_{\ms}] =  
F[-  \chi_{\ms}],
\label{condtxx}
\ee
where we have defined the quartic functional $F[\chi_{\ms}]$ to be
$ \Xi^\Lambda_{\ ABCD} y^A y^B y^C y^D$ with the same values of $\Lambda$ and $A,B,C,D$.
On the right hand side,
the minus sign in the
argument will cancel out, since the term is a quartic
function of this argument.  Therefore the condition (\ref{condtxx}) is satisfied, and so
the interaction (\ref{candidate}) is invariant under the perturbative redefinitions
(\ref{pstransf}) and is an element of ${\cal W}_{\rm invariant} \cap
{\cal W}_{\rm mixed}$.

We now turn to invariance under the linear phase space transformations (\ref{redeflin}).
These transformations are strongly constrained by the conditions
(\ref{condtt4}) and (\ref{ld2}) and by the explicit form (\ref{eq:freee}) of the free field scattering
map.  The most general linear map consistent with these conditions is\footnote{These transformations encompass transformations caused by changes
in the choice of function $g(v)$ in the definition (\ref{third}),
and also the change of variables (\ref{hatPsi}) used to diagonalize the
Poisson brackets.}
\begin{subequations}
  \label{lt22}
  \begin{eqnarray}
    {\tilde \Psi}^{\rm e} &\to& {\tilde \Psi}^{\rm e} + \beta[\Delta
      \Psi^{\rm e}], \\
    \Psi^{\rm m} &\to& \Psi^{\rm m}, \\
    \label{eq:inv9-c}
    \chi &\to& \chi, \\
    \label{eq:inv9-a}
    \Delta \Psi^{\rm e} &\to& \Delta \Psi^{\rm e}, \\
    {\bar \Psi}^{\rm e} &\to& {\bar \Psi}^{\rm e} + \sigma[{\tilde
        \Psi}^{\rm e}], 
  \end{eqnarray}
  \end{subequations}
where $\beta$ and $\sigma$ are linear functionals of their arguments.
The electromagnetic memory
produced by an incoming scalar field with no incoming electromagnetic field
is
invariant under these transformations.  This follows from the fact
that it is the change in the quantity $\Delta \Psi^{\rm e}$, which is
invariant by Eq.\ (\ref{eq:inv9-a}), and it is a functional only of
$\chi$, which is invariant by (\ref{eq:inv9-c}).

Since the change in memory $\Delta Q^{\rm hard}(\bftheta)$ is both generically
nonzero from Sec.\ \ref{sec:fourth-order} and an invariant interaction, and since it constitutes a
transformation of incoming hard degrees of freedom to outgoing soft
degrees of freedom, we conclude that there is an unavoidable coupling
between the hard and soft sectors for the scattering map ${\cal S}_1$.
Furthermore this interaction rules out the form (\ref{eq:factorizes2}) of the
scattering map.

A similar argument applies to the interaction (\ref{hatFdef}) and rules out
the form (\ref{eq:factorizes1}) of the scattering map.  By paralleling
the derivation of Eq.\ (\ref{condtxx}), we find that this interaction
will be invariant under the perturbative redefinition (\ref{pstransf})
if
\be
   {\tilde F}[{\bar\Psi}^{\rm
       e}_{\ms},\chi_{\ms}] =  
 -  {\tilde F}[{\bar \Psi}^{\rm
     e}_{\ms},-\chi_{\ms} ].
 \label{condtxx1}
\ee
Here we have defined the quartic functional ${\tilde F}[{\bar
  \Psi}^{\rm e}_{\ms}, \chi_{\ms}]$ to be
$ \Xi^A_{\Lambda BCD} y^\Lambda y^B y^C y^D$,
where the index $A$ corresponds to the field $\chi_{\ps}$, $\Lambda$
to the field
${\bar \Psi}^{\rm  e}_{\ms}$ and $B$, $C$ and $D$ to the field $\chi_{\ms}$.
The condition (\ref{condtxx1}) is satisfied since ${\tilde F}$ is
cubic in $\chi_{\ms}$.  Similarly one can check that the interaction (\ref{hatFdef}) 
is invariant under the linear field redefinition
(\ref{hatNtransform}) using the explicit formula (\ref{lt22}).
Finally one can check that the conditions of footnote \ref{longf} are
satisfied using the conditions on $\rho^a_{\ bc}$ listed there and the properties of third order scattering
\be
M^{\chi}_{\ {\bar \Psi}^{\rm e} {\bar \Psi}^{\rm e} \chi} =
M^{\chi}_{\ {\tilde \Psi}^{\rm e} {\bar \Psi}^{\rm e} \chi} =
M^{\Delta \Psi^{\rm e}}_{\ {\bar \Psi}^{\rm e} \chi \chi} =
M^{\tilde \Psi^{\rm e}}_{\ {\bar \Psi}^{\rm e} \chi \chi} = 0,
\ee
which can be derived using the methods of Appendices
\ref{app:secondorder} and \ref{app:memory}.

\subsection{Hard and soft sectors are coupled for second version of
  scattering map}
\label{sec:secondversioncoupled}

We next turn to the second version (\ref{calS2:def}) of the scattering maps
discussed in the introduction.  For this case there exist definitions of
the soft variables which are conserved, and it is natural to restrict
to definitions of soft variables which preserve this property.  We
will show that in this context it is not possible to find definitions
of hard and soft variables for which the scattering map has the
uncoupled form (\ref{eq:factorizes}).

We take our initial definition of hard and soft sectors
to be given by the phase space coordinates
\be
y^a = ( h^A, s^\Gamma ) = 
\left( {\tilde \Psi}^{\rm e}_{\ps}, \Psi^{\rm m}_{\ps}, {\breve \chi}_{\ps},
Q_{\ps}, {\Psi}^{\rm e}_{\psms} \right).
\label{hardsoft0}
\ee
Here the hard variables $h^A$ consist of the first three fields 
which depend on $u$ and $\bftheta$, while the soft
variables $s^\Gamma$ consist of the last two fields, 
which depend only on $\bftheta$ and which are conserved in the
scattering
as discussed in Sec.\ \ref{sec:theorem}.
Note that these phase space coordinates are not symplectically
orthogonal but they are independent, cf.\ the discussion in
Sec.\ \ref{sec:theorem} above. 
Also we have defined
\be
   {\breve \chi}_{\ps} = e^{-i \Psi^{\rm e}_{\psms}} \chi_{\ps},
   \label{hatchidef}
   \ee
motivated by the form (\ref{eq:mathscrGdef0}) of the generating functional;
this definition removes some (but not all) of the soft-hard coupling.

We denote by $\Sigma_s$ the subspace of the phase space $\Gamma_{\ps}$
given by fixing values of the soft variables $s$.
Taking the pullback of the symplectic form (\ref{presymfinal2}) to $\Sigma_s$
yields a symplectic form $\Omega_{AB}$ whose inverse yields a Dirac
bracket on $\Sigma_s$ which we denote by $\left\{ \ ,\ 
\right\}_{\rm D}$.  This Dirac bracket is easiest to compute starting
from the Poisson brackets of the variables (\ref{bcanonical}).  The
result is that the Dirac brackets of the hard variables 
$( {\tilde \Psi}^{\rm e}_{\ps}, \Psi^{\rm m}_{\ps}, {\breve \chi}_{\ps})$
coincide with the Poisson brackets (\ref{pbs}) (with $\chi_{\ps}$
replaced by ${\breve \chi}_{\ps}$) except for the
bracket
\be
\label{diracbracket}
  \left\{ D^2 {\tilde \Psi}^{\rm e}_{\ps}(u,\bftheta) \ , \ 
  {\breve \chi}_{\ps}(u',\bftheta') \right\}_{\rm D}  =
   i e^2  g(u) 
  \delta^{(2)}(\bftheta,\bftheta') {\breve \chi}_{\ps}(u',\bftheta').
\ee

Since the soft variables $s$
are conserved, the restriction of the scattering map ${\cal S}_2$ to $\Sigma_s$ gives a
map ${\cal S}_s : \Sigma_s \to \Sigma_s$, which we can
write as
\be
h^A  \to {\bar h}^A(h,s).
\label{restrictedscattering}
\ee
When the right hand side is independent of $s$ then the two sectors
are uncoupled.  If we make an $s$-dependent change of the coordinates
$h^A$ on
$\Sigma_s$ (effectively changing the definition of the hard sector),
the effect is to make the replacement
\be
\label{Sigmastransform}
{\cal S}_s \to {\bar {\cal S}}_s = \varphi_s \circ {\cal S}_s \circ
\varphi_s^{-1},
\ee
for some diffeomorphism $\varphi_s : \Sigma_s \to \Sigma_s$.
We do not require that $\varphi_s$ be a symplectomorphism.
We would like to show that there is no choice of diffeomorphism that
removes the $s$ dependence from the right hand side of
Eq.\ (\ref{restrictedscattering}).
We will show that attempting to derive such a diffeomorphism runs into
the same kind of obstacle as in Sec.\ \ref{sec:theorem} above: one obtains
a unique result but it is not a diffeomorphism, since the resulting
coordinates are not independent.

To show that no such diffeomorphism exists it is sufficient to work to linear order in $s$ and to
linear order in  the deviation of ${\cal S}_s$ from the identity map
(appropriate to the small $h$ limit),
since if it is impossible to remove the coupling in this regime
then it is impossible in general.  Within this limit we can write the
mapping (\ref{restrictedscattering}) as
\be
h^A \to h^A + \xi^A(h) + v^A(h,s),
\label{restrictedscattering1}
\ee
where $\xi^A$ is independent  of $s$ and $v^A$ is linear in $s$.  The
effect of the transformation (\ref{Sigmastransform}) to linear order is to add to the right hand side
of Eq.\ (\ref{restrictedscattering1}) a term $({\cal L}_{\vec \eta}
{\vec \xi})^A$  for some vector field $\eta^A$ on $\Sigma_s$.  Thus
the condition for the dependence on $s$ to be removable is that
there exists a $\eta^A$ which depends linearly on $s$  for which
\be
{\vec v} = - {\cal L}_{\vec \eta} {\vec \xi}.
\label{condtnocouple}
\ee

Next, the generating function (\ref{eq:mathscrGdef0}) for the scattering map can be written
in terms of the coordinates (\ref{hardsoft0}) used in this section as
\be
\label{eq:mathscrGdef00}
 {\mathscr G}\left[{\tilde \Psi}^{\rm 
     e}_{\ps} + g e^2 D^{-2} Q_{\ps} - g e^2 D^{-2} Q^{\rm hard}[{\breve
       \chi}_{\ps}]
   ,  \Psi^{\rm
      m}_{\ps}, 
  {\breve \chi}_{\ps}
  \right],
 \ee
from Eqs.\ (\ref{transform}), (\ref{eq:cons-law}), (\ref{brevePsidef})
and (\ref{hatchidef}).  Note that this expression depends on only one
of the two soft variables, $Q_{\ps}$.
Evaluating this generating functional at $Q_{\ps}=0$ and inserting
into the linearized transformation\footnote{Using the Dirac brackets discussed around
Eq.\ (\ref{diracbracket}).} (\ref{eq:gf}) 
yields the vector
field $\xi^A$.
Evaluating the piece of ${\mathscr G}$
linearized in
$Q_{\ps}$, which is 
\be
e^2 \int du \int d^2 \Omega \, g(u) \frac{ \delta {\mathscr G}}{\delta
  {\breve \Psi}^{\rm e}_{\ps}(u,\bftheta) }
D^{-2} Q_{\ps}(\bftheta),
\label{key}
\ee
gives instead the vector field $v^A$.  Note that the quantity
(\ref{key}) is generally nonzero, by
Eqs.\ (\ref{Hinftydef}), (\ref{cons23}), (\ref{eq:fnalformulae3})
and the result of Sec.\ \ref{sec:fourth-order}.
With these inputs we find that the solution to Eq.\ (\ref{condtnocouple})
for $\eta^A$ is
\be
{\vec \eta} = -e^2 \int  du \int d^2 \Omega \, g(u) [D^{-2}
Q_{\ps}(\bftheta) ]\frac{\delta }{\delta {\tilde \Psi}^{\rm
    e}_{\ps}(u,\bftheta)}.
\ee
However the corresponding coordinate transformation on $\Sigma_s$ is
similar to Eq.\ (\ref{eq:tildePsinew}), 
and leads to a set of coordinates which
are not independent.  Thus, just as in Sec.\ \ref{sec:theorem}, it is not
possible to decouple the two sectors.

\section{Discussion and conclusions}
\label{conclusion}

In this paper, we computed the perturbative classical scattering map from $\scrm$
to $\scrp$ of electromagnetism coupled to a massless, charged
scalar field in four-dimensional Minkowski spacetime.
We consider various definitions hard and soft sectors of the
theory, and adopt two different conventions for how to ensure that the
same definitions of hard and soft are used at $\scri^-$ and at
$\scri^+$.  For both conventions, we showed that
the soft and hard sectors of the theory evolve independently to
quadratic order, but that at higher orders the two
sectors are coupled.
In the first convention, this coupling between the hard and soft sectors cannot be eliminated
using linear or perturbative redefinitions of the two sectors, which includes any possible  
``dressing'' of the hard degrees of freedom.  In the second
convention, the coupling cannot be eliminated by any field
redefinitions which preserve the fact that the soft variables are
conserved by the scattering process.

Our conclusions disagree with those of Ref.\ \cite{Bousso}, which
argued that there exist definitions of the two sectors for which there is
an exact decoupling.  We showed that the explanation for the disagreement is a
property of symplectic geometry which holds in finite dimensions but
fails in infinite dimensions.

We expect this nontrivial coupling of soft and hard sectors to
generalize
to nonabelian gauge theories and to general relativity, and to the
quantum
as well as classical regimes\footnote{As discussed in Sec.\ \ref{sec:softtheorem}, the statement
that the quantum theory does not admit 
a decomposition into two decoupled sectors is not in conflict with the
factorization theorems that constrain the soft behavior of $S$ matrix
elements \cite{Feige:2014wja,Nande:2017dba}.}.
As a result,
the infinitely many conserved soft hair charges should yield non-trivial
constraints on gravitational scattering and on black
hole formation and evaporation, as argued in
Refs.\ \cite{HPS-1,HPS-2}.

To probe this issue further,
it would be interesting to study the quantization
of the gauge theory studied in this paper, while carefully accounting
for the soft degrees of freedom.  A recent detailed study of infrared
issues in scattering by Prabhu, Satishchandran and Wald (PSW) \cite{Prabhu:2022mcj}
shows that for massless fields it is not possible to find in and out 
Hilbert spaces that accommodate nontrivial changes in memory.
However, as mentioned in Sec.\ \ref{sec:Dirac} above, PSW
start from an algebra of observables which is different from the
algebra obtained from our phase space $\Gamma$, due to our inclusion
of the edge mode (\ref{hatPsi}) that is canonically
conjugate to memory which PSW exclude as being pure gauge.
It would be interesting to study the consequences for the analysis of
PSW of adopting the different algebra used here.

\section*{Acknowledgments}

We thank Raphael Bousso, Mihalis Dafermos, Andrzej Herdegen,
Shahin Sheikh Jabbari, Mehrdad Mirbabayi, Kartik Prabhu,
Gautam Satishchandran, Matthew
Schwartz and Bob Wald for helpful discussions and/or correspondence.
We thank an anonymous referee for some useful comments and suggestions.
This work was supported in part by NSF grants PHY-1707800 and PHY-2110463
to Cornell University.  IS was supported in part by the John and David Boochever
prize fellowship in fundamental theoretical physics.
EF thanks the Department of Applied Mathematics and 
Theoretical Physics at Cambridge University, where part of this paper
was written, for its hospitality.

\appendix

\section{Symplectically orthogonal hard and soft sectors exclude certain forms of scattering map}
\label{app:so}

In this appendix we show that a scattering map ${\cal S} : \Gamma \to \Gamma$ on a phase space
$\Gamma$ cannot have the forms (\ref{eq:factorizes1}) or
(\ref{eq:factorizes2}) if the soft and hard variables are
symplectically orthogonal.  We assume a finite dimensional phase
space although we expect the result to generalize.

By assumption the phase space can be expressed as a product of hard and soft factors
$\Gamma = \Gamma_{\rm soft} \times \Gamma_{\rm hard}$, and we denote 
by $y^a = (s^A, h^\Gamma)$ corresponding coordinates.  Defining
$\Omega^{ab} = \{y^a, y^b\}$, we have by assumption that
\be
0 = \left\{ s^A, h^\Gamma \right\} = \Omega^{ab} \partial_a s^A
\partial_b h^\Gamma.
\label{so1}
\ee
Applying the pullback operator ${\cal S}_*$ to this equation
gives 
\be
({\cal S}_* \Omega)^{ab} \partial_a ({\cal S}_* s^A)
\partial_b ({\cal S}_* h^\Gamma) =0.
\ee
We now use the fact that the scattering map is a symplectomorphism so
that ${\cal S}_* \Omega^{ab} = \Omega^{ab}$, and use the definitions
${\bar s}^A = {\cal S}_* s^A = s^A \circ {\cal S}$, ${\bar h}^\Gamma =
{\cal S}_* h^\Gamma = h^\Gamma \circ {\cal S}$.  This gives
$\Omega^{ab} \partial_a {\bar s}^A \partial_b {\bar h}^\Gamma=0$, and
expanding this out and using Eq.\ (\ref{so1}) again gives
\be
\label{central}
\Omega^{BC}
\frac{ \partial {\bar s}^A}{\partial s^B}
\frac{ \partial {\bar h}^\Gamma}{\partial s^C}
 = - \Omega^{\Sigma\Lambda} 
\frac{ \partial {\bar s}^A}{\partial h^\Sigma}
\frac{ \partial {\bar h}^\Gamma}{\partial h^\Lambda}.
\ee

We now assume a scattering map of the form (\ref{eq:factorizes1}),
which implies that
\be
\frac{\partial {\bar h}^\Gamma }{ \partial s^C}=0,
\label{uuud}
\ee
so the left hand side of Eq.\ (\ref{central}) vanishes.
On the right hand side the first factor is
invertible by Eq.\ (\ref{so1}) since $\Omega^{ab}$ is invertible, and
similarly the third factor is invertible by 
Eq.\ (\ref{uuud}). This implies that 
\be
\frac{\partial {\bar s}^A }{ \partial h^\Sigma}=0,
\label{uuud1}
\ee
which contradicts the assumed form (\ref{eq:factorizes1}).
A similar argument excludes the form (\ref{eq:factorizes2}).

\section{Free \Lorentz gauge solutions with nontrivial soft charges}
\label{app:freesolns}

In this appendix we review the solutions of the free, noninteracting
theory which satisfy everywhere the \Lorentz gauge condition.
We show that these solutions satisfy the
gauge-invariant asymptotic properties at $i^+$ and $i^-$
given in Eqs.\ (\ref{falloffiminus}) and (\ref{falloffiplus}),
but fail to satisfy the matching condition (\ref{matching2}) [cf.\ Eq.\ (\ref{signflip}) below].
We then generalize the solutions to a larger class which satisfy
\Lorentz gauge only asymptotically, and deduce the general form of the
\Lorentz gauge scattering map for free solutions.

\subsection{Global \Lorentz gauge}

In inertial coordinates $(t,x^i)$ the homogeneous global \Lorentz gauge solutions are
\begin{subequations}
\label{freesoln}
  \begin{eqnarray}
    {\underline A}_0 &=& 0, \\
    {\underline A}_i &=& \sum_{l \ge 0} \left\{ \partial_{ijL} \left[ \frac{D_{jL}(t-r)}{r} - \frac{D_{jL}(t+r)}{r} \right]
- \partial_{L} \left[ \frac{D_{iL}''(t-r)}{r} - \frac{D_{iL}''(t+r)}{r} \right]
    \right\} \nonumber \\
    && + \sum_{l \ge 0}  \epsilon_{ipq} \partial_{pL} \left[ \frac{C_{qL}(t-r)}{r} - \frac{C_{qL}(t+r)}{r} \right] .
  \end{eqnarray}
  \end{subequations}
The notation here follows Ref.\ \cite{RevModPhys.52.299} and is as follows.
The symbol $L$ is the multi-index $L = (i_1,i_2, \ldots i_l)$, and
$D_{iL}$, $C_{iL}$ are Cartesian tensors which are symmetric and
trace-free on all of their indices.  The symbol $\partial_L$ means
$\partial_{i_1} \ldots \partial_{i_l}$. Here and throughout
underlines mean that the corresponding quantities are in \Lorentz gauge.
Note that, for a given solution, the tensors $D_{iL}$ and $C_{iL}$
are not unique, since the solution (\ref{freesoln}) is invariant under transformations
of the form
\be
\label{invariance1}
D_{iL}(x) \to D_{iL}(x) + \delta D_{iL}(x), \ \ \ \ C_{iL}(x) \to C_{iL}(x) + \delta C_{iL}(x),
\ee
for $l \ge 0$, where $\delta D_{iL}$ and $\delta C_{iL}$ are a polynomials in $x$ of
degree $2 l +2$.

We will restrict attention to solutions for which the asymptotic
behavior of the symmetric tracefree tensors $D_{iL}$ and $C_{iL}$ as $x \to  \infty$ is
given by
\be
\label{assume1}
    D_{iL}(x) = {\tilde D}_{+iL}(x) + P_{+iL}(x), \ \ \ \ C_{iL}(x) =
    {\tilde C}_{+iL}(x) + Q_{+iL}(x),
    \ee
where ${\tilde D}_{+iL}, {\tilde C}_{+iL}$ go to $0$ as $x \to \infty$,
and $P_{+iL}$ and $Q_{+iL}$ are
polynomials in $x$ of degree $l+2$ and $l+1$
respectively.  Similarly as $x \to -\infty$ we 
require that
\be
\label{assume2}
    D_{iL}(x) = {\tilde D}_{-iL}(x) + P_{-iL}(x), \ \ \ \ C_{iL}(x) =
    {\tilde C}_{-iL}(x) + Q_{-iL}(x),
    \ee
where ${\tilde D}_{-iL}, {\tilde C}_{-iL} \to 0$ as $x \to -\infty$,
and $P_{-iL}$ and $Q_{-iL}$ are again 
polynomials in $x$ of degree $l+2$ and $l+1$ respectively.  Because of the invariance property
(\ref{invariance1}),
the solution (\ref{freesoln}) depends only on the differences
$\Delta P_{iL} = P_{+iL} - P_{-iL}$ and
$\Delta Q_{iL} = Q_{+iL} - Q_{-iL}$ between these polynomials and not on $P_{\pm iL}$ or $Q_{\pm iL}$  individually.
The assumptions (\ref{assume1}) and (\ref{assume2}) are compatible
with the large $r$ field expansions (\ref{Aexpand}) and
(\ref{Aexpandminus}) that we have assumed\footnote{This would no longer be
true if the polynomials $P_{\pm iL}$ and $Q_{\pm iL}$ were of higher
degree than $l+2$ and $l+1$.}, and also yield solutions with 
finite energy.

The coefficients of the large $r$ expansions of the fields near $\scri^+$ are given by
${\underline {\cal A}}_{+u}= {\underline {\cal A}}_{+r} = 0$, and
\be
\label{ans17}
{\underline {\cal A}}_{+A}(u,\bftheta) = \sum_{l} \, (-1)^{l+1} n^L e^i_A \left[ {\tilde D}_{+iL}^{(l+2)}(u) 
+ \epsilon_{ipq} n^p  {\tilde C}_{+qL}^{(l+1)}(u) \right],
\ee
where $n^i = x^i/r$, $e^i_A = D_A n^i$, $n^L = n^{i_1} \ldots n^{i_l}$, and the superscripts $(l+2)$ and $(l+1)$ indicate the number of derivatives taken.
This expression satisfies the condition
\be
\label{ccdd}
   {\underline {\cal A}}_{+A} \to 0
   \ee
 as $u \to
\infty $ at $i^+$,
   which yields the condition (\ref{fc4}), from Eq.\ (\ref{FABformula}).
In the other limit $u \to -\infty$ at $i^0$ we have, from Eqs.\ (\ref{assume1}) and (\ref{assume2}),
   \be
   \label{limit11}
   {\underline {\cal A}}_{+A}(u,\bftheta) \to {\underline {\cal A}}_{\psms A}(\bftheta) 
= \sum_{l} \, (-1)^{l} n^L e^i_A \left[ \Delta P_{iL}^{(l+2)} 
+ \epsilon_{ipq} n^p  \Delta Q_{qL}^{(l+1)} \right].
\ee
Here the right hand side is independent of $u$ since $\Delta P_{iL}$
is a polynomial of order $l+2$ and $\Delta Q_{qL}$ is a polynomial of
order $l+1$.

Similar results apply for the limiting behavior of the solutions near
$\scri^-$.  The expansion coefficients are 
${\underline {\cal A}}_{-v}= {\underline {\cal A}}_{-r} = 0$, and
\be
\label{ans18}
{\underline {\cal A}}_{-A}(v,\bftheta) = \sum_{l} \, n^L e^i_A \left[ {\tilde D}_{-iL}^{(l+2)}(v) 
- \epsilon_{ipq} n^p  {\tilde C}_{-qL}^{(l+1)}(v) \right].
\ee
This satisfies the condition as $v \to -\infty $ at $i^-$
\be
\label{ccdd1}
   {\underline {\cal A}}_{-A} \to 0,
   \ee
yielding the condition (\ref{c4}),  while as $v \to \infty$ at $i^0$ we have, from Eqs.\ (\ref{assume1}) and (\ref{assume2})
   \be
   \label{limit22}
   {\underline {\cal A}}_{-A}(v,\bftheta) \to {\underline {\cal A}}_{\msps A}(\bftheta) 
= \sum_{l} \, n^L e^i_A \left[ \Delta P_{iL}^{(l+2)} 
- \epsilon_{ipq} n^p  \Delta Q_{qL}^{(l+1)} \right].
\ee

One can check that these \Lorentz gauge solutions (\ref{freesoln})
satisfy the matching conditions (\ref{matching}) and (\ref{Bmatch})
[or equivalently (\ref{matching22})] discussed in the body
of the paper.
Note, however, that they do not
satisfy the matching condition (\ref{matching2}), instead this relation
is satisfied with a sign flip, since from Eqs.\ (\ref{limit11}) and
(\ref{limit22}) we have
\be
{\underline {\cal A}}_{\psms\,A} = - {\cal P}_* {\underline {\cal
    A}}_{\msps\,A}.
\label{signflip}
\ee
This is discussed further in Sec.\ \ref{sec:pag} above.

By expanding the solution (\ref{freesoln}) to subleading order in
$1/r$ near $\scri^+$ and $\scri^-$ and reading off the subleading
coefficients ${\hat {\cal A}}_{+u}$, ${\hat {\cal A}}_{+A}$,
${\hat {\cal A}}_{-v}$ and ${\hat {\cal A}}_{-A}$,
one can verify the limiting
behavior at $i^-$ and $i^+$ given by Eqs.\ (\ref{c1}), (\ref{c2}),
(\ref{fc1}) and (\ref{fc2}).  Similar analyses for free massless scalar field
solutions establishes (\ref{c3}) and (\ref{fc3}).

\subsection{Asymptotic \Lorentz gauge}

The solutions (\ref{freesoln}) are the most general free solutions that obey
the \Lorentz gauge condition everywhere in spacetime.   However one
can obtain a more general class of solutions in {\it asymptotic \Lorentz
gauge}, in which one imposes the \Lorentz gauge condition only at $r
\ge R$ for some $R$.  Specifically for $r \ge R$ one can transform the
solutions 
according to
\be
{\underline A}_a \to {\underline A}_a + \nabla_a \varepsilon, \ \ \ \ {\underline \Phi} \to e^{i \varepsilon}
{\underline \Phi}, \ \ \ \ \varepsilon = \sum_{l \ge 1} D_L \partial_L \left( \frac{ u^l + v^l}{r} \right),
\label{gaugegg}
\ee
where $D_L$ for $l \ge 1$ are some constant traceless symmetric
tensors.  For $r < R$ one can use any smooth extension of $\varepsilon$.
This transformation preserves the asymptotic gauge conditions (\ref{gaugec1}),
and the initial and final data transform as
${\underline {\cal A}}_{\ps A} \to {\underline {\cal A}}_{\ps A} + D_A \varepsilon_{\ps}$,
${\underline {\cal A}}_{\ms A} \to {\underline {\cal A}}_{\ms A} + D_A \varepsilon_{\ms}$,
where $\varepsilon_{\ps}(\bftheta)$ is a freely specifiable function with only
$l \ge 1$ components\footnote{This function is given explicitly
in terms of the coefficients $D_L$ in Eq.\ (\ref{gaugegg}) by
$\varepsilon_{\ps} = \sum_{l \ge 1} c_l D_L n^L$ with $c_l = 2
\sum_{i=0,\ i \ {\rm even}}^l 2 (i-1) (i -3) \ldots (i - 2l + 1)$.} and $\varepsilon_{\ms} = {\cal P}_* \varepsilon_{\ps}$.
The transformation is of the even form (\ref{sigmaeo}),
and thus is not gauge but instead is a mapping from solutions to
physically distinct solutions, as discussed in Sec.\ \ref{sec:gauge-specialization}.

The general asymptotic \Lorentz gauge solutions will no longer satisfy
the conditions (\ref{ccdd}) and (\ref{ccdd1}) at past and future timelike infinity.
However there is a combination of these conditions which is unaffected
by the transformation (\ref{gaugegg}), and which is still valid for
the general solutions, namely
\be
{\underline \Psi}^{\rm e}_{\psps} = {\cal P}_* {\underline \Psi}^{\rm
  e}_{\msms}.
\label{ccdd2}
\ee

\subsection{Scattering map for free solutions}

The scattering map that relates the initial data on $\scri^-$ to
the final data on $\scri^+$ for the free global \Lorentz gauge solutions can be read off from
Eqs.\ (\ref{ans17}) and (\ref{ans18}) and their analogs for scalar fields, and is
\begin{subequations}
	\label{eq:free-map}
  \begin{eqnarray}
    {\underline \chi}_{\ps}(u,\bftheta) &=& - {\cal P}_* \, {\underline \chi}_{\ms}(u,\bftheta), \\
{\underline {\cal A}}_{\ps A}(u,\bftheta) &=&  {\cal P}_*
\left[{\underline {\cal A}}_{\ms A}(u,\bftheta)
  - {\underline {\cal A}}_{\msps A}(\bftheta) \right].
\end{eqnarray}
\end{subequations}
This scattering map can be generalized to the class of asymptotic
\Lorentz gauge solutions generated by the transformation
(\ref{gaugegg})
by expressing the gauge transformation function $\varepsilon_{\ps}$
in terms of the new initial data.  Writing the result in terms of the
potentials (\ref{AAdecompose}), making use of the vanishing
magnetic charges condition (\ref{kkk3}) and using ${\cal P}_*
\varepsilon_{AB} = - \varepsilon_{AB}$ gives
\begin{subequations}
	\label{eq:free-map11}
        \begin{eqnarray}
          {\underline \Psi}^{\rm e}_{\ps}(u,\bftheta) &=&  {\cal P}_*
\left[{\underline \Psi}^{\rm e}_{\ms}(u,\bftheta)
  - {\underline \Psi}^{\rm e}_{\msps}(\bftheta)
  + {\underline \Psi}^{\rm e}_{\msms}(\bftheta)
  \right],\\
          {\underline \Psi}^{\rm m}_{\ps}(u,\bftheta) &=&  -{\cal P}_*
{\underline \Psi}^{\rm m}_{\ms}(u,\bftheta)
  ,\\
    {\underline \chi}_{\ps}(u,\bftheta) &=& - {\cal P}_* \, {\underline \chi}_{\ms}(u,\bftheta).
\end{eqnarray}
\end{subequations}

\section{Properties of interacting \Lorentz gauge solutions}
\label{app:intLor}

In this appendix we deduce some properties of nonlinear \Lorentz gauge
solutions.  Consider first solutions in global \Lorentz gauge.  We
claim that the conditions (\ref{ccdd}) and (\ref{ccdd1}) at future and past
timelike infinity are still satisfied by these solutions.  To see
this, consider the version of the theory on the Einstein static
universe obtained by making a conformal transformation.  In this version
spacetime is compact and all the fields are bounded.  Consider now a
small neighborhood ${\cal V}$ of future timelike infinity $i^+$.
At each order in perturbation theory, one can obtain the fields inside
${\cal V}$ by specifying the initial conditions on the Cauchy surface
obtained by taking the intersection of
$\partial {\cal V}$ with the image of Minkowski spacetime on the
Einstein static universe cylinder.  The fields inside ${\cal V}$ (and
in particular the limit to $i^+$ of the fields on $\scri^+$) are given
as a sum of a homogeneous solution determined by the initial data,
and an inhomogeneous solution determined by the sources inside ${\cal
  V}$ with zero initial data.  However, we know from Appendix
\ref{app:freesolns}
that the homogeneous solutions must satisfy the vanishing
condition (\ref{ccdd}) at $i^+$, since they are free \Lorentz gauge
solutions.  The inhomogeneous solution can give a nonvanishing
contribution to the limit, however this can be made arbitrarily small
by taking the size of the neighborhood to zero, since the sources are bounded.
We conclude that the conditions (\ref{ccdd}) and (\ref{ccdd1}) are satisfied.

Just as for the free solutions, more general nonlinear solutions in asymptotic
\Lorentz gauge obtained from the transformation (\ref{gaugegg})
will no longer satisfy the conditions (\ref{ccdd}) and (\ref{ccdd1}),
but will satisfy the condition (\ref{ccdd2}).

\section{Presymplectic form is independent of choice of Cauchy surface}
\label{app:extend}

In the body of the paper,
we showed that the presymplectic form (\ref{presym}) is independent
of the choice of Cauchy surface $\Sigma$, for the cases $\Sigma = \scri^-$ and $\Sigma = \scri^+$, and 
for field perturbations of the special form (\ref{puregauge}) given the condition (\ref{matching2}).
In this appendix we review and modify slightly the results of Campiglia and Eyheralde (CE) \cite{Campiglia:2017mua},
that give sufficient conditions for extending the independence of
Cauchy surface to arbitrary Cauchy surfaces and to arbitrary field perturbations.
A closely related treatment has been given by Henneaux and Troessaert \cite{Henneaux:2018gfi}.

Consider first spacelike Cauchy surfaces $\Sigma$ that limit to
spatial infinity $i^0$.
We denote by ${\cal H}$ the unit hyperboloid of the space of directions of approach to the point of spatial infinity.
This hyperboloid acts like a spacetime boundary, and
the presymplectic form $\Omega_\Sigma$ can potentially depend on the
Cauchy surface $\Sigma$ through the location of the two-surface ${\cal
S}$ where $\Sigma$ intersects ${\cal H}$.
Under the appropriate assumptions of the asymptotic fall offs of the fields near ${\cal H}$ [Eq.\ (4.2) of CE)],
one can compute the pullback $\theta$ of the presymplectic potential
of the theory to ${\cal H}$.  This pullback can be decomposed as
\be
\theta =  - \delta \ell  + d \beta + {\cal E},
\label{decompos}
\ee
the sum of a total variation boundary term $-\delta \ell$, an exact corner term $d \beta$, and a residual term ${\cal E}$ called the flux term.
The general theory for obtaining a presymplectic form that is independent of Cauchy surface \cite{Campiglia:2017mua,Harlow:2019yfa}
consists of (i) Modifying the definition (\ref{presym}) of the presymplectic form to exclude the corner term,
\be
\Omega_{\Sigma} = \delta \int_\Sigma \theta - \delta \int_{\cal S} \beta;
\label{eq:psf}
\ee
and (ii) Specializing the decomposition (\ref{decompos}) and the definition of the field configuration space to make the flux term vanish.

CE find a nonzero flux term in general, the second term in their Eq.\ (4.5).  This term can be eliminated
by specializing the definition of the field configuration space.
CE impose an asymptotic \Lorentz gauge condition, their Eq.\ (4.8).
However that \Lorentz gauge condition is incompatible with our condition (\ref{matching2}), and instead yields
Eq.\ (\ref{matching2}) with a sign flip [cf.\ Eq.\ (\ref{signflip}) above].  Therefore we instead impose a modified version
of their Eq.\ (4.8) with the first term omitted.  That modified condition can always be enforced
by making an appropriate ``gauge transformation'' of the form of their Eq.\ (2.38), although 
as explained after Eq.\ (\ref{matching2}) above the transformation should not be thought of
as a gauge specialization, and instead should be thought of as a specialization of the definition
of the field configuration space.  Our modified version of their
condition (4.8) eliminates the flux term and also implies
Eq.\ (\ref{matching2})\footnote{This follows from the fact that, in
  their notation, the
  even parity piece of $A^{0}_\alpha$ is of the form $D_\alpha
  \lambda$ for some $\lambda$ with $D_\alpha D^\alpha \lambda =0$, and
  their Eq.\ (3.2).}.

CE also find a nonzero corner term proportional to the pullback to
${\cal H}$ of $\epsilon_{abcd} A^{[c} \delta A^{d]}$, their Eq.\ (4.6)\footnote{Note
that they use different terminology and call it a boundary term, and they also work with $\omega =
  \delta \theta$ rather than $\theta$.}.  With the corner term
subtraction    
included, the presymplectic form (\ref{eq:psf}) is independent of the choice of
spacelike Cauchy surface $\Sigma$ that limits to $i^0$.  The corner
term on ${\cal H}$ vanishes when one takes the limit
to the boundaries of ${\cal H}$, using their Eqs.\ (4.8) [the modified version],
(4.6), (3.9), (B.2), footnote 3, and assuming the specialization given in their footnote 9.
It also vanishes on null infinity, with the gauge specializations (\ref{gaugec1}).
It follows that the corner term
subtraction
does not affect
the presymplectic forms
$\Omega_{\scri^-}$ and $\Omega_{\scri^+}$, and that
$\Omega_{\scri^-}$ and $\Omega_{\scri^+}$
coincide with the common
value of $\Omega_\Sigma$ for all the spacelike Cauchy surfaces $\Sigma$.

\section{Field configuration space in the magnetic sector}
\label{app:fieldconfig-magsector}

In this appendix we discuss a difference between the space of solutions for the
electric potential $\Psi_{\ms}^{\rm e}$ and the magnetic potential
$\Psi^{\rm m}_{\ms}$.  For the class of \Lorentz gauge free solutions discussed in
Appendix \ref{app:freesolns} these potentials vanish at $i^-$, from
Eq.\ (\ref{ccdd1}):
\be
{\underline \Psi}_{\msms}^{\rm e} = 0, \ \ \ \ {\underline \Psi}_{\msms}^{\rm m}=0,
\label{vv}
\ee
and this generalizes to the solutions of the interacting theory.
However, by using the transformation freedom (\ref{sigmaeven}) we can
make $\Psi_{\msms}^{\rm e}$ be nonzero, from Eq.\ (\ref{ggg1}).
Thus in the full space of
solutions with our preferred asymptotic gauge conditions
(\ref{gaugec1}) and (\ref{matching2}) [which do 
  not include \Lorentz gauge] we have that both
$\Psi_{\msms}^{\rm e}$ and $\Psi_{\msps}^{\rm e}$ are nonzero in general.

For the magnetic variables the story is somewhat different.
Starting from the class of solutions which satisfy (\ref{vv}),
one can attempt to obtain a larger class of solutions by analogy with
the procedure for the electric variables, by making a transformation
of the initial data on $\scri^-$ of the form
\be
\label{notgauge}
\Psi_{\ms}^{\rm e} \to \Psi_{\ms}^{\rm e}, \ \ \ \ \ \Psi_{\ms}^{\rm
  m} \to \Psi_{\ms}^{\rm m} + {\tilde \varepsilon}_{\ms},
\ee
where ${\tilde \varepsilon}_{\ms}$ is a function on $\scri^-$ with no $l=0$
component that is independent of $v$.
Although the transformation (\ref{notgauge}) is not a gauge transformation, it is a kind of magnetic
analog of the gauge transformation given by Eqs.\ (\ref{ggg0})
\cite{Strominger:2015bla}.  By solving the equations of motion one can
determine the effects of this transformation of the initial data on
the entire solution, and in particular the data on $\scri^+$
transforms as [cf.\ Eq.\ (\ref{sigmaeven}) above]
\be
\label{mgt}
\Psi_{\ps}^{\rm e} \to \Psi_{\ps}^{\rm e}, \ \ \ \ \ \Psi_{\ps}^{\rm
  m} \to \Psi_{\ps}^{\rm m} + {\tilde \varepsilon}_{\ps}.
\ee
where ${\tilde \varepsilon}_{\ps} = {\cal P}_* {\tilde \varepsilon}_{\ms}$.
The magnetic charges
\be
   {\tilde Q}_{\tilde \varepsilon} = \frac{1}{e^2} \int d^2 \Omega
   {\tilde \varepsilon}_{\ps} {\cal F}_{\psms AB} 
   = \frac{1}{e^2} \int d^2 \Omega
   {\tilde \varepsilon}_{\ms} {\cal F}_{\msps AB} 
\label{magcharges}
\ee
associated with these transformations
[cf.\ Eq.\ (\ref{Bmatch}) above] can be
derived in exact parallel with the derivation for the electric case
discussed in Sec.\ \ref{sec:gauge-specialization} \cite{Strominger:2015bla}.
However, the solutions generated by the transformations (\ref{notgauge})
are generically singular in the interior of the spacetime.  One
example of such a solution is a static magnetic dipole at the origin
$r=0$.  In this paper we restrict attention to initial data on $\scri_-$ which
evolves into smooth solutions in the interior of the spacetime, which
requires that
\be
\Psi_{\msms}^{\rm m} = 0,
\label{kkk1}
\ee
and disallows the transformations (\ref{notgauge}).  A similar analysis at
$\scri^+$ yields the condition
\be
\Psi_{\psps}^{\rm m} = 0.
\label{kkk2}
\ee
The other limits
$\Psi_{\psms}^{\rm m}$ and
$\Psi_{\msps}^{\rm m}$ of the magnetic potentials at spatial infinity are generally nonzero (see Appendix
\ref{app:freesolns}). However in this paper we restrict attention
to the sector of the theory in which they vanish, as discussed around
Eq.\ (\ref{kkk3}).

\section{Asymptotic edge modes}
\label{app:edge}

\subsection{Equivalence of soft degrees of freedom and edge modes}

In the literature on soft charges it has been conjectured \cite{Strominger:lectures}
that soft modes can be alternatively described in the language
of edge modes used in Refs.\ \cite{Donnelly:2016auv,Speranza:2017gxd,2017NuPhB.924..312G,Harlow:2016vwg}.
In this appendix we demonstrate the equivalence explicitly by giving
an alternative construction of the phase space and symplectic product
of Sec.\ \ref{sec:Dirac} using edge modes.
A general treatment of edge modes in electromagnetism at finite
boundaries has been given by Freidel and Pranzetti \cite{Freidel:2018fsk}.

We slightly generalize the treatment in the body of the paper by considering a region ${\cal R}$ of $\scri^-$ of points with $v_1 \le v \le v_2$.
At the end we will specialize to $v_1 = -\infty$, $v_2 = \infty$, the
case treated in the body of the paper.
The presymplectic product evaluated on this region is not gauge
invariant: the derivation that led to Eq.\ (\ref{pres1}) gives in this
context
\be
\Omega_{{\cal R}}(\delta A_a, \delta \Phi; \delta_\ve A_a, \delta_\ve \Phi) = \frac{1}{e^2} \int d^2 \Omega 
\left[ \delta {\cal F}_{\ms rv}
  \varepsilon_{\ms}
    \right]^{v_2}_{v_1}.
\label{pres111}
\ee
Now the field component at the boundary $v=v_1$ is given by, from
Eq.\ (\ref{c1}) and an equation analogous to Eq.\ (\ref{maxwellexpand1}),
\be
{\cal F}_{\ms rv}(v_1) = \int_{-\infty}^{v_1} dv \left[ e^2 {\cal
    J}_{\ms v} - D^A {\cal F}_{\ms Av} \right].
   \ee
If we now vary just the fields within the region ${\cal R}$, as is
relevant for the presymplectic form (\ref{pres111}), the result
vanishes, since the field ${\cal F}_{\ms rv}$ at $v_1$ depends only on fields
outside the region:
\be
\delta {\cal F}_{\ms rv}(v_1) = 0.
\ee
Hence the one of the two terms in (\ref{pres111}) vanishes.  However
the second term at $v=v_2$ is generally nonzero and so the presymplectic form is
not gauge invariant.

The key idea of edge modes is to use the Stueckelberg trick\footnote{A
  nice review of this technique and its applications in gauge theory 
as well as gravitational theories can be found in Ref.\ \cite{Hinterbichler:2011tt}.}
to restore gauge invariance by introducing new degrees of freedom
\cite{Donnelly:2016auv,Speranza:2017gxd,2017NuPhB.924..312G}.
We introduce a $U(1)$ field $e^{i e^2 \lambda}$ on the boundary at $v =
v_2$, which transforms under gauge transformations as $\lambda \to
\lambda + \ve_{\ms}/e^2$.  We add to the presymplectic form $\Omega_{\cal
  R}$ a term
\be
\int d^2 \Omega \left[ \delta_1 \lambda \delta_2 {\cal
    F}_{\ms vr}(v_2) - \delta_2 \lambda \delta_1 {\cal
    F}_{\ms vr}(v_2) \right],
\label{extra}
\ee
so that the total presymplectic form is now gauge invariant.

We next specialize the gauge by using the gauge function $\ve_{\ms}$
to set ${\cal A}_{\ms v}$ to zero, using the analog of Eq.\ (\ref{gg1}).
There is a residual gauge freedom given by
\be
\ve_{\ms} = \ve_{\ms}(\bftheta).
\label{residual1}
\ee
We now follow the derivation given in the body of the paper that leads
to the form (\ref{presymfinal2}) of the presymplectic form, but
keeping the extra term (\ref{extra}).  The final result is
\begin{eqnarray}
  \label{presymfinal22}
\Omega_{{\cal R}} &=&  - \frac{1}{e^2} \int dv \int d^2 \Omega  \left[
  \partial_v {\tilde \Psi}^{\rm e}_{1\,\ms} D^2 {\tilde \Psi}^{\rm
    e}_{2\,\ms}
+  \partial_v \Psi^{\rm m}_{1\,\ms} D^2 \Psi^{\rm
    m}_{2\,\ms}
- ( 1 \leftrightarrow 2) \right] \nonumber \\
 && - \frac{1}{e^2} \int d^2 \Omega  \left\{
  D^2 \Delta \Psi^{\rm e}_{1\,\ms} \left[ {\bar \Psi}^{\rm
    e}_{2\,\ms} - \lambda_2 + 2 \int dv g^\prime {\tilde \Psi}^{\rm e}_{2 \ms} \right]
         - ( 1 \leftrightarrow 2) \right\} \nonumber \\
&&  + \int dv \int d^2 \Omega  \left[
  \partial_v \delta_1
  \chi_{\ms}^* \delta_2 \chi_{\ms}
+    \partial_v \delta_1
  \chi_{\ms} \delta_2 \chi_{\ms}^*
- (1 \leftrightarrow 2)  \right].
\end{eqnarray}

We may now use the residual gauge freedom (\ref{residual1}) (which was
  not a gauge freedom in the body of the paper) to set to zero the
variable ${\bar \Psi}^{{\rm e}}_{2 \ms}$.  Then the phase space and
presymplectic form (\ref{presymfinal22}) coincide exactly with the
version (\ref{presymfinal2}) from the body of the paper (when we
  specialize to $v_1 = -\infty$, $v_2 = \infty$), except that
the variable ${\bar \Psi}^{{\rm e}}_{\ms}$ has been replaced by $-\lambda$.
Thus we see that of the two canonically conjugate variables $\Delta
\Psi^{\rm e}_{\ms}$ and ${\bar \Psi}^{\rm e}_{\ms}$ that comprise the soft modes,
one can be interpreted as an edge mode at the future boundary at $v
= v_2 = \infty$ of $\scri^-$.

\subsection{Necessity of including edge modes}

As discussed in the introduction and in Sec.\ \ref{sec:Dirac},
there is some disagreement in the literature on the necessity of
including asymptotic edge modes. Ashtekar
\cite{Ashtekar:1987tt,Ashtekar:1981bq}
eliminates these modes by 
imposing $\Psi^{\rm
  e}_{\psms} = \Psi^{\rm e}_{\msps} = 0$ in the definition of the
field configuration space.  Bousso and collaborators argue in the
gravitational case that the edge modes are unobservable
\cite{Bousso:2016wwu,Bousso}.
The issue arises in a different guise in the
algebraic approach to the scattering problem, where one focuses on constructing the algebra
of observables.
The recent analysis of infrared scattering by Prabhu, Satishchandran and Wald \cite{Prabhu:2022mcj} is based on an
algebra obtained by excluding the edge modes.
However that algebra could be modified to incorporate them\footnote{Specifically the algebra of
observables constructed in their Sec.\ 4.2 could be
extended by including the additional quantity
(\ref{hatPsi}) (suitably smeared over angles), which satisfies the
requirement of their Eq.\ (2.5).}.
Finally, the edge modes are included in the treatments by Strominger
and collaborators \cite{He:2014cra,Strominger:lectures}.

One argument for including edge modes is that they are required in
order for there to be an action on the phase space of the large gauge
transformations (\ref{sigmaeven}) associated with the new conserved
charges (\ref{Echarge}) \cite{He:2014cra,Strominger:lectures}.

Another argument is that it is not possible to eliminate edge modes using a gauge
transformation (since the corresponding transformations are not
degeneracy directions of the presymplectic form and are thus not true
gauge transformations).  However, they can be instead eliminated by
specializing the definition of the field configuration space.

The strongest argument for including edge modes is as follows. We will
give the argument for the gravitational case.  The edge modes (or soft
hair) in that case are
observable only relative to a choice of convention, a BMS frame,
equivalent to the choice of zero readings on a set of clocks on an asymptotic two
sphere. This is the reason that the edge modes are sometimes argued to be
unobservable \cite{Bousso:2016wwu,Bousso}.  However, the same
non-observability argument would also apply to the direction in space
of the angular momentum of a stationary gravitational source, which is
observable only relative to a choice of convention.  One could make a
diffeomorphism to fix the direction of angular momentum to be in the
$z$ direction.  However, quantum fluctuations in the direction of
angular momentum have physical consequences, and cannot be
correctly described in a phase space where angular momentum always
points in the $z$ direction.  By analogy, one expects quantum
fluctuations in gravitational edge modes to have physical consequences, and they would not be
correctly described in a phase space in which the edge modes are fixed.

Finally we note that the results of this paper are unchanged if
the edge modes are excluded.  This can be seen as follows.  The general
form of the scattering map (\ref{eq:general2}) transforms
appropriately under the transformation freedom (\ref{tr1}) and
(\ref{tr2}).  Eliminating the edge modes ${\bar \Psi}^{\rm e}_{\ms}$ and
${\bar \Psi}^{\rm e}_{\ps}$ fixes this
transformation freedom and also eliminates Eq.\ (\ref{eq:generalbarpsie2}). 
However the soft-hard coupling discussed in
Sec.\ \ref{sec:higherorders} is still
be present in the second term on the right hand side of
Eq.\ (\ref{eq:generalcheckpsie2}), specifically in the difference
between its $u \to \pm \infty$ limits.  As in
Sec.\ \ref{sec:higherorders} this soft-hard coupling cannot be eliminated.

\section{Derivation of generating functional representation of scattering map}
\label{app:symplectomorphism}

In this appendix we derive some properties of the generating functional $G$ for the scattering map
which were discussed in Sec.\ \ref{sec:gf}.

First, in order for the scattering map ${\cal S}$ to be of
the specific form (\ref{eq:general2}) derived in Sec.\ \ref{sec:generalform}, the generating functional $G$
must be of the form
\be
\label{eq:calGdef}
G\left[
  {\tilde \Psi}^{\rm e}_{\ps},
  \Psi^{\rm m}_{\ps},
  \chi_{\ps},
  \Delta \Psi^{\rm  e}_{\ps},
  {\hat \Psi}^{\rm e}_{\ps}
  \right] = {\cal G}\left[{\breve \Psi}^{\rm
      e}_{\ps},  \Psi^{\rm
      m}_{\ps}, e^{-i {\bar \Psi}^{\rm e}_{\ps}} e^{i {\Delta \Psi}^{\rm e}_{\ps}}
  \chi_{\ps}  \right]
  \ee
  for some functional ${\cal G}$, where 
  ${\breve \Psi}^{\rm e}_{\ps}$ and
  ${\bar \Psi}^{\rm e}_{\ps}$ are defined in terms of the fields on the
  left hand side by Eqs.\ (\ref{hatPsi}) and (\ref{brevePsidef}).  
In deriving the representation (\ref{eq:calGdef}) we used
the decomposition (\ref{eq:scdecompose}) of the scattering map, the generating functional parameterization (\ref{eq:gf}), the Poisson brackets (\ref{pbs}), and the linear order scattering map (\ref{eq:freee}).

We next define variational derivatives of the 
the functional ${\cal G}[{\breve \Psi}^{\rm e}, \Psi^{\rm m}, \chi]$.
These are defined by
\begin{eqnarray}
  \label{eq:varderiv}
\delta {\cal G} &=& \int du \int d^2 \Omega \frac{ \delta {\cal
    G}}{\delta \Psi^{\rm m}(u,\bftheta)} \delta \Psi^{\rm
  m}(u,\bftheta)
+ \int du \int d^2 \Omega \left[\frac{ \delta {\cal
    G}}{\delta \chi(u,\bftheta)} \delta \chi(u,\bftheta) + {\rm c.c.} \right] \nonumber \\
&& + \int du \int d^2 \Omega \frac{ \delta {\cal
    G}}{\delta {\breve \Psi}^{\rm e}(u,\bftheta)} \delta {\breve \Psi}^{\rm
  e}(u,\bftheta) + \int d^2 \Omega
\frac{ \delta {\cal
    G}}{\delta {\Delta \Psi}^{\rm e}(\bftheta)} \delta {\Delta \Psi}^{\rm
  e}(\bftheta).
\end{eqnarray}
The only nontrivial point here is that the field variation 
\be
\label{eq:fv}
   {\breve \Psi}^{\rm e} \to
      {\breve \Psi}^{\rm e} + \delta {\breve \Psi}^{\rm e} = \breve
      \Psi^{\rm e} + \delta {\tilde \Psi}^{\rm e} + g(u) \delta \Delta
      \Psi^{\rm e},
\ee
generates two terms in the variation $\delta {\cal G}$ of the
functional, a bulk term [first term on the second line of Eq.\ (\ref{eq:varderiv})], and a
boundary term (second term on the second line), since the variation
$\delta {\breve \Psi}^{e}$ is generally nonzero at the boundaries $u =
\pm \infty$, with
\be
\label{notind}
\delta {\breve \Psi}^{\rm e}(\pm \infty) = \pm
\delta \Delta \Psi^{\rm e}/2.
\ee
Also, the integral of $\delta {\cal G}
/ \delta {\breve \Psi}^{\rm e}(u,\bftheta)$ against any test function
$f(u,\bftheta)$ is only defined when $f(\infty,\bftheta) = - f(-\infty,\bftheta)$.

One might worry that there is some inconsistency in the
definition (\ref{eq:varderiv}) since the variables ${\breve \Psi}^{\rm e}$ and $\Delta
\Psi^{\rm e}$ are not independent, by Eq.\ (\ref{notind}).  However
there is no inconsistency, as can be seen by formulating the definitions
in terms of the independent variables ${\tilde \Psi}^{\rm e}$ and
$\Delta \Psi^{\rm e}$ and then translating using Eq.\ (\ref{eq:fv}), which yields
\begin{subequations}
\begin{eqnarray}
\frac{\delta {\cal G}}{\delta {\breve \Psi}^{\rm e}(u,\bftheta)} &=&
\left(
\frac{\delta {\cal G}}{\delta {\tilde \Psi}^{\rm e}(u,\bftheta)}
\right)_{\Delta \Psi^{\rm e}}, \\
\frac{\delta {\cal G}}{\delta {\Delta \Psi}^{\rm e}(\bftheta)} &=&
\left(
\frac{\delta {\cal G}}{\delta {\Delta \Psi}^{\rm e}(\bftheta)}
\right)_{{\tilde \Psi}^{\rm e}} - \int du g(u)
\left(
\frac{\delta {\cal G}}{\delta {\tilde \Psi}^{\rm e}(u,\bftheta)}
\right)_{{\Delta \Psi}^{\rm e}}.
\end{eqnarray}
\end{subequations}
We use the variational derivatives on the left hand side rather than
those on the right hand side, since they are independent of the choice
of function $g(u)$.

We can now compute the various functionals 
${\cal H}^{\rm m}$, ${\cal K}$,
  ${\cal H}^{\rm e}$ and ${\cal H}^{\rm e}_\infty$ that appear in the
scattering map (\ref{eq:general2}) in terms of the generating
functional ${\cal G}$.
By using the generating functional parameterization (\ref{eq:gf})
  with $y^a$ taken to be $\Psi^{\rm m}$, $\chi$, ${\tilde \Psi}^{\rm e}$ and $\Delta \Psi^{\rm
    e}$, respectively, the Poisson brackets (\ref{pbs}), the variational derivative definitions (\ref{eq:varderiv}),
  the decomposition (\ref{eq:scdecompose}) of the scattering map, the linear order scattering map (\ref{eq:freee}),
  and comparing with the definitions (\ref{eq:generalpsim2}), 
(\ref{eq:generalchi2}), (\ref{eq:generaltildepsie2n}) and (\ref{eq:generalcheckpsie2soft}),
  we can read off the following expressions:
\begin{subequations}
\label{eq:fnalformulae}
\begin{eqnarray}
\label{eq:fnalformulae1}  
{\cal H}^{\rm m}[ u,\bftheta ; {\breve \Psi}^{\rm e}, \Psi^{\rm m},
  \chi ] &=& \frac{1}{2} e^2 D^{-2} \int du' K(u-u') \frac{ \delta {\cal
    G}}{\delta \Psi^{\rm m}(u',\bftheta)} \left[ {\breve \Psi}^{\rm
    e}, - \Psi^{\rm m}, - \chi \right], \ \ \ \\
\label{eq:fnalformulae2}  
{\cal K}[ u,\bftheta ; {\breve \Psi}^{\rm e}, \Psi^{\rm m},
  \chi ] &=& -\frac{1}{2} \int du' K(u-u') \frac{ \delta {\cal
    G}}{\delta \chi(u',\bftheta)} \left[ {\breve \Psi}^{\rm
    e}, - \Psi^{\rm m}, - \chi \right]^*, \\
\label{eq:fnalformulae3}  
{\cal H}^{\rm e}[ u,\bftheta ; {\breve \Psi}^{\rm e}, \Psi^{\rm m},
  \chi ] &=& \frac{1}{2} e^2 D^{-2} \int du' K(u-u') \frac{ \delta {\cal
    G}}{\delta {\breve \Psi}^{\rm e}(u',\bftheta)} \left[ {\breve \Psi}^{\rm
    e}, - \Psi^{\rm m}, - \chi \right], \ \ \ \\
\label{eq:fnalformulae4}  
{\cal H}^{\rm e}_\infty[ \bftheta ; {\breve \Psi}^{\rm e}, \Psi^{\rm m},
  \chi ] &=& \frac{i}{2} e^2 D^{-2} \int du \, \chi(u,\bftheta) \frac{ \delta {\cal
    G}}{\delta \chi(u,\bftheta)} \left[ {\breve \Psi}^{\rm
    e}, - \Psi^{\rm m}, - \chi \right] + {\rm c.c.},\ \ \ \ 
    \end{eqnarray}
    \end{subequations}
where $K(u) = \Theta(u)-1/2$.  These expressions are accurate up to
corrections of order $O({\cal G}^2)$ or equivalently $O(\alpha^6)$. 

The generating functional ${\cal G}$ must obey two additional constraints.
First, by imposing
consistency of the formulae (\ref{eq:fnalformulae3}) and (\ref{eq:fnalformulae4}) with
the definition (\ref{Hinftydef}), and making the substitutions
$\Psi^{\rm m} \to - \Psi^{\rm m}$ and $\chi \to - \chi$, we obtain 
\be
\label{Gconstraint1}
\int du \frac{ \delta {\cal G}}{\delta {\breve \Psi}^{\rm
    e}(u,\bftheta)}[ {\breve \Psi}^{\rm e}, \Psi^{\rm m}, \chi ] =  -2
i \int du \left\{ \chi(u,\bftheta) \frac{ \delta {\cal G}}{\delta
  {\chi}(u,\bftheta)}[ {\breve \Psi}^{\rm e}, \Psi^{\rm m}, \chi ] \right\}
+ {\rm c.c.} + O({\cal G}^2).
\ee
Second, using the generating functional parameterization
  (\ref{eq:gf}) of the scattering map
  with $y^a = \hat \Psi^{\rm e}$, making use of the Poisson bracket (\ref{pbs3})
  and the variational derivative definitions (\ref{eq:varderiv}),
  and comparing with the scattering map (\ref{eq:generalhatpsie2n})
 and making use of the formulae (\ref{eq:fnalformulae3}) and (\ref{eq:fnalformulae4}) for the
 functionals ${\cal H}^{\rm e}$ and ${\cal H}^{\rm
   e}_\infty$, we obtain the constraint
\be
\label{Gconstraint2}
\frac{ \delta {\cal G}}{\delta {\Delta \Psi}^{\rm
    e}(\bftheta)}[ {\breve \Psi}^{\rm e}, \Psi^{\rm m}, \chi ] =  -
\frac{i}{2} \int du \left\{ \chi(u,\bftheta) \frac{ \delta {\cal G}}{\delta {\chi}(u,\bftheta)}[ {\breve \Psi}^{\rm e}, \Psi^{\rm m}, \chi ]
\right\} + {\rm c.c.}. + O({\cal G}^2).
\ee
By direct substitution one can
show that the general solution of this equation is
\be
\label{mathscrG}
{\cal G} \left[ {\breve \Psi}^{\rm e}, \Psi^{\rm m}, \chi \right] = {\mathscr
  G}\left[ {\breve \Psi}^{\rm e}, \Psi^{\rm m}, e^{-\frac{i}{2} \Delta
    \Psi^{\rm e}} \chi \right] + O(\alpha^6),
\ee
for some functional ${\mathscr G}$, with the restriction that the boundary
variational derivative defined in Eq.\ (\ref{eq:varderiv}) associated
with the first argument of ${\mathscr G}$ vanishes:
\be
\left(\frac{ \delta {\mathscr G}}{\delta \Delta \Psi^{\rm e}}
\right)_{\rm first\ argument} =0.
\label{eq:fa}
\ee
Substituting the
expression (\ref{mathscrG})
into the definition (\ref{eq:calGdef}) of ${\cal G}$ gives that the original generating
functional $G$ can be written as
\be
\label{eq:mathscrGdef}
G\left[
  {\tilde \Psi}^{\rm e}_{\ps},
  \Psi^{\rm m}_{\ps},
  \chi_{\ps},
  \Delta \Psi^{\rm  e}_{\ps},
  {\hat \Psi}^{\rm e}_{\ps}
    \right] = {\mathscr G}\left[{\breve \Psi}^{\rm
      e}_{\ps},  \Psi^{\rm
      m}_{\ps}, e^{-i {\bar \Psi}^{\rm e}_{\ps}} e^{i {\Delta \Psi}^{\rm e}_{\ps}/2}
    \chi_{\ps} \right] + O(\alpha^6).
\ee
The form of the generating functional given by Eqs.\ (\ref{eq:fa}) and
(\ref{eq:mathscrGdef}) is actually valid to all orders  
in $\alpha$ [cf.\ Eq.\ (\ref{eq:mathscrGdef0}) above], even though we have derived it here only to $O(\alpha^5)$.
This is because (i) it guarantees that the quantity $\Psi^{\rm
  e}_{\psms}  = {\hat \Psi}^{\rm e}_{\ps} - 2 \int du g' {\tilde
  \Psi}^{\rm e}_{\ps} - \Delta \Psi^{\rm e}_{\ps}/2$
is preserved under the transformation
(\ref{generatedef}), from the Poisson brackets (\ref{pbs})
and Eqs.\ (\ref{eq:scdecompose}) and (\ref{eq:freee}),
thus ensuring the matching condition (\ref{matching2}); and (ii)
it similarly guarantees that the conservation law
(\ref{cons23}) is satisfied.

\section{Sufficient conditions for decoupling in finite dimensions}
\label{app:theorem}

In this appendix we prove the theorem in symplectic geometry given in Sec.\ \ref{sec:theorem}, which gives
sufficient conditions for two sets of degrees of freedom to decouple in a finite dimensional system.

The derivation is essentially an application of
Frobenius' theorem for vector fields \cite{Wald-book}.  
We define the symmetry generator (Hamiltonian) vector fields ${\vec v}^{(A)}$ associated with the
conserved quantities $s^A$ by
\be
v^{(A)a} = - \Omega^{ab} \nabla_b s^A
\ee
for $1 \le A \le t$.
In order for Frobenius' theorem to be applicable, the Lie brackets of
these vector fields with each other must be expressible as linear
combinations of the vector fields. 
However, the Lie bracket $\left[ {\vec v}^{(A)}, {\vec v}^{(B)} \right]$ is
just the
Hamiltonian vector field associated with the Poisson bracket (\ref{spb}),
which vanishes since $\omega^{AB}$ are assumed to be constants,
$\nabla_a \omega^{AB} =0$.  Thus we obtain $\left[ {\vec v}^{(A)}, {\vec
    v}^{(B)} \right] =0$.

From Frobenius' theorem it now follows that the subbundle of the tangent bundle over $M$
defined by the span of the vector fields ${\vec v}^{(A)}$ is integrable.
That is, locally there exist coordinates $(x^A, h^\Gamma)$ for $1 \le A \le
t$ and $1 \le \Gamma \le n-t$ for which the vector fields ${\vec v}^{(A)}$ are
tangent to the surfaces of constant $h^\Gamma$:
\be
   {\vec v}^{(A)} = \chi^{AB}(x,h) \frac{\partial}{\partial x^B}
   \label{integrablesubbundle}
\ee
for some invertible $\chi^{AB}$.  However this implies that
$\omega^{AB} = {\vec v}^{(A)}(s^B) = \chi^{AC} \partial s^B / \partial
x^C$, and since $\omega^{AB}$ and $\chi^{AC}$ are both invertible it
follows from the inverse function theorem that at fixed $h$ we can 
we can express $x^A$ as a function of $s^A$.  Hence without loss of
generality we can take $x^A = s^A$.

It now follows that in these coordinates $(s^A,h^\Gamma)$ we have $\Omega^{AB} =
\omega^{AB}$ and $\Omega^{A\Gamma} = \left\{ s^A, h^\Gamma \right\} = - v^{(A)\Gamma}
= 0$ from Eq.\ (\ref{integrablesubbundle}).  Thus the symplectic form can be expressed as
\be
\Omega = \omega_{AB} ds^A \wedge ds^B + \Omega_{\Gamma\Sigma}(s,h) dh^\Gamma \wedge
dh^\Sigma,
\label{sff}
\ee
where $\omega^{AB} \omega_{BC} = \delta^A_{\ C}$.
From $d\Omega=0$ it follows that $\Omega_{\Gamma\Sigma}$ is independent
of $s$, and demanding that the symplectic form (\ref{sff}) be preserved
under the pull back ${\cal S}_{2*}$ now shows that the
symplectomorphism ${\cal S}_2$ must have the form (\ref{simpleans}).

\section{Computation of \Lorentz gauge scattering map to second order}
\label{app:secondorder}

In this appendix we derive the second order \Lorentz gauge scattering
map (\ref{eq:freee2l}).

Our starting point is the set of free \Lorentz gauge solutions with
soft charges described in Appendix \ref{app:freesolns}.  Rather than
using the symmetric trace-free tensor expansion used there, it will be
convenient to 
represent those solutions as conventional plane wave expansions, with
the soft degrees of freedom encoded in distributional components of
the mode coefficients at zero frequency.
We write the first order scalar field as
\be
   {\underline \Phi}^{(1)}(t,\bfx) = \sum_{\eta = \pm} \int d^3 k \,
   e^{-i \eta k t} e^{i \bfk \cdot \bfx} f_\eta(\bfk),
   \label{phiexpand0}
\ee
where $k = |\bfk|$ and the functions $f_+(\bfk)$ and $f_-(\bfk)$ are independent.
These functions are related to the initial data
${\underline \chi}^{(1)}_{\ms}(v, \bfn) = \lim_{r \to \infty} r
{\underline \Phi}^{(1)}(v-r, r \bfn)$  on $\scri^-$, with $\bfn =
\bfx/r$, by [cf.\ Eq.\ (9.0.29) of Ref.\ \cite{Strominger:lectures}]
\be
f_\eta(\bfk) = - \frac{i \eta}{2 \pi k} {\tilde {\underline \chi}}^{(1)}_{\ms}(
\eta k, - \eta {\hat \bfk}).
\label{modefnf}
\ee
Here $\eta = \pm 1$, ${\hat \bfk} = \bfk/k$ and
${\tilde {\underline
    \chi}}^{(1)}_{\ms}$ is the Fourier transform
\be
\label{ftdef}
{\tilde {\underline
    \chi}}^{(1)}_{\ms}(\omega,\bfn) = \frac{1}{2 \pi} \int dv e^{i
     \omega v} {\underline \chi}^{(1)}_{\ms}(v, \bfn).
   \ee
Similarly the first order vector potential is given by ${\underline
  A}^{(1)t} =0$ and
\be
   {\underline \bfA}^{(1)}(t,\bfx) = \sum_{\gamma = \pm} \int d^3 p \,
   e^{-i \gamma p t} e^{i \bfp \cdot \bfx} {\bf a}_\gamma(\bfp),
\label{Aexpand0}
   \ee
with $p = | \bfp |$,
\be
   \bfp \cdot {\bf a}_\gamma(\bfp)=0
\label{lorentzformode}
\ee
and ${\bf a}_\gamma(-\bfp)^* = {\bf a}_{-\gamma}(\bfp)$.
The mode coefficients ${\bf a}_\gamma(\bfp)$ are related to the
initial data ${\underline {\cal A}}^{(1)}_{\ms A}(v,\bfn)$ defined in
Sec.\ \ref{sec:found} by
\be
a^i_\gamma(\bfp) = - \frac{i \gamma}{2 \pi p} {\tilde {\underline
    {\cal A}}}^{(1) i}_{\ms}( \gamma p, - \gamma {\hat \bfp}).
\label{modefnf1}
\ee
Here $\gamma = \pm 1$, ${\hat \bfp} = \bfp / p$, ${\tilde {\underline
    {\cal A}}}^{(1) i}_{\ms}( \omega, {\bf n})  = {\tilde {\underline
    {\cal A}}}^{(1)}_{\ms A}( \omega, \bfn) h^{AB} e_B^i$,
$\theta^A = (\theta,\varphi)$ are coordinates
on the two-sphere, $h_{AB}$ is the metric given by $h_{AB} d\theta^A
d\theta^B = d\theta^2 + \sin^2 \theta d\varphi^2$, 
$h^{AB} h_{BC} = \delta^A_{\ C}$, $e_A^i =
\partial n^i/\partial \theta^A$ and
\be
{\tilde {\underline {\cal A}}}^{(1)}_{\ms A}( \omega, \bfn) = 
\frac{1}{2 \pi} \int dv e^{i
  \omega v} {\underline {\cal A}}^{(1)}_{\ms A}(v, \bfn).
\ee

We next discuss how the soft charges are encoded in the mode
coefficients.   For the incoming scalar field, we have excluded by
assumption any
soft degrees of freedom, and the fall off assumption
(\ref{assumedfalloff}) implies that the function ${\tilde {\underline
    \chi}}^{(1)}_{\ms}(\omega,\bfn)$ is bounded.
For the incoming vector vector potential, on the other hand, using the
fall-off assumption (\ref{assumedfalloff0}), the definitions (\ref{transform})
and (\ref{AAdecompose1}), and the vanishing magnetic charges condition
(\ref{kkk3}) we obtain
for the Fourier transform
\be
   {\tilde {\underline {\cal A}}}^{(1)i}_{\ms}(\omega,\bfn) =
\frac{i}{2\pi}      \Delta {\underline {\cal A}}^{(1)i}_{\ms}(\bfn) \, 
          {\rm P.V.}
          \frac{1}{\omega} +
        {\overline {\underline {\cal A}}}^{(1)i}_{\ms}(\bfn) 
   \delta(\omega) + 
   {\tilde {\underline {\cal A}}}^{(1)i\,{\rm rest}}_{\ms}(\omega,\bfn),
   \label{decomposeid}
   \ee
   where P.V. means ``principal value'',
   \be
   \Delta {\underline {\cal A}}^{(1)i}_{\ms}(\bfn) 
=  e^i_A h^{AB} D_B
\Delta {\underline \Psi}^{{\rm e}\,(1)}_{\ms}(\bfn), \ \ \ \ 
        {\overline {\underline {\cal A}}}^{(1)i}_{\ms}(\bfn) =
e^i_A h^{AB} D_B {\bar {\underline \Psi}}^{{\rm
       e}\,(1)}_{\ms}({\bf n}),
   \label{decomposeid1}
   \ee
   and 
\be
   {\tilde {\underline {\cal A}}}^{(1)i\,{\rm rest}}_{\ms}(\omega,\bfn) = O(\omega^{-1 +\epsilon})
   \label{restscaling}
   \ee
as $\omega \to 0$ with $\epsilon >0$.  We will see that the second and third terms in
Eq.\ (\ref{decomposeid}) give a vanishing contribution to the second
order scattering while
the first term gives a nonvanishing contribution.
   
We next write the equation of motion (\ref{eqn-subleading-scalar}) for
the second order scalar field as $\Box {\underline \Phi}^{(2)} = s$, where the
source $s$ is $s = 2 i {\underline A}^{(1)\,a} \nabla_a {\underline
  \Phi}^{(1)}$.
Defining the final data on $\scri^+$ as ${\underline \chi}^{(2)}_{\ps}(u,\bfn) =
\lim_{r\to\infty} r {\underline \Phi}^{(2)}(u+r, r \bfn)$, we make use of the
general identity derived in Sec.\ 2.2 of Ref.\ \cite{Saha:2019tub}
\be
   {\tilde {\underline \chi}}^{(2)}_{\ps}(\omega,\bfn) = - 2 \pi^2
   {\tilde s}(\omega, \omega \bfn),
   \label{Sahaidentity}
   \ee
   where ${\tilde s}$ is the spacetime Fourier transform
   \be
   {\tilde s}(\omega, \bfk) = \frac{1}{(2 \pi)^4} \int dt \int d^3 x
   e^{i \omega t} e^{-i \bfk \cdot \bfx} s(t,\bfx)
   \ee
and the Fourier transform ${\tilde {\underline \chi}}^{(2)}_{\ps}$ is
defined as in Eq.\ (\ref{ftdef}) but with $v$ replaced by $u$.
We now evaluate the source $s(t,\bfx)$ using the 
the mode expansions
(\ref{phiexpand0}) and (\ref{Aexpand0}),
and multiply by a regulator factor $\exp[-t^2 /
  (2 T^2)]$, where we 
will eventually take the limit $T \to \infty$.
We then evaluate the spacetime Fourier transform, make use of the 
the identity (\ref{Sahaidentity}), and 
eliminate the mode functions in favor of the initial
data on $\scri^-$ using Eqs.\ (\ref{modefnf}) and (\ref{modefnf1}).
This gives
\begin{eqnarray}
   {\tilde {\underline \chi}}^{(2)}_{\ps}(\omega,\bfn) &=& - \lim_{T
     \to \infty}
   \sum_{\eta,\gamma = \pm} \int d^3 p \, \frac{ \eta \gamma}{ l p} \delta_T( \nu k - \eta l - \gamma p )
   \nonumber \\   && \times
             {\tilde {\underline \chi}}^{(1)}_{\ms}\left[ \eta l, -
     \eta \frac{(\bfk - \bfp)}{l} \right] \, \bfk \cdot {\tilde {\bf
       {\underline {\cal A}}}}_{\ms}^{(1)}(\gamma p, - \gamma {\hat
               \bfp}),
\label{stage1}
\end{eqnarray}
where we have defined
\be
\bfk = \omega \bfn, \ \ \ k = | \omega |, \ \ \ \nu = {\rm sgn}
(\omega), \ \ \ l = | \bfk - \bfp|,
\label{defs1}
\ee
and $\delta_T$ is the approximate $\delta$-function 
\be
\delta_T(\omega) = \frac{T}{\sqrt{2 \pi}} \exp\left[- \frac{1}{2}
  \omega^2 T^2 \right].
\label{approxdelta}
\ee

To evaluate the momentum space integral (\ref{stage1}) it is convenient to
switch to prolate spheroidal coordinates.  We pick Cartesian
coordinates for which $\bfk = (0,0,k)$, and define coordinates $(\sigma,
\tau, \varphi)$ with $1 \le \sigma < \infty$, $-1 \le \tau \le 1$, $0
\le \varphi \le 2 \pi$ by
\be
\bfp = \frac{1}{2} k ( \sqrt{\sigma^2-1} \sqrt{1-\tau^2} \cos \varphi,
\sqrt{\sigma^2-1} \sqrt{1-\tau^2} \sin \varphi, 1 + \sigma \tau).
\label{prolatedef}
\ee
It follows that $d^3 p =(k^3/8) (\sigma^2 - \tau^2) d\sigma d\tau
d\varphi$ and that
\be
l = |\bfk - \bfp| = \frac{1}{2} k (\sigma-\tau), \ \ \ \ p =
\frac{1}{2} k (\sigma+\tau).
\label{idents22}
\ee
This gives
\begin{eqnarray}
   {\tilde {\underline \chi}}^{(2)}_{\ps}(\omega,\bfn) &=& -
   \frac{1}{2} \lim_{T \to \infty}
   \sum_{\eta,\gamma = \pm} \int_1^\infty d\sigma \int_{-1}^1 d\tau
   \int_0^{2 \pi} d\varphi \, \eta \gamma \, \delta_{k T}\left[ 1 - \frac{\eta (\sigma-\tau)}{2} - \frac{\gamma (\sigma+\tau)}{2} \right]
   \nonumber \\   && \times
             {\tilde {\underline \chi}}^{(1)}_{\ms}\left[ \nu \eta l, -
     \nu \eta \frac{\bfk - \bfp}{l} \right] \, \bfk \cdot {\tilde {\bf
       {\underline {\cal A}}}}_{\ms}^{(1)}(\nu \gamma p, - \nu \gamma {\hat
               \bfp}),
\label{stage2}
\end{eqnarray}
where we have redefined $\eta \to \nu \eta$ and $\gamma \to \nu
\gamma$.

We now discuss in turn the four different terms in the expression
(\ref{stage2}).  Two of these give vanishing contributions, and the
contributions from the remaining two are equal.
For the first term with $\eta = \gamma = -1$, the approximate delta
function reduces in the limit $T \to \infty$ to $\delta(\sigma+1)$,
which gives a vanishing contribution since the range of $\sigma$ is
$\sigma \ge 1$.  For the second term with $\eta = -1$, $\gamma = 1$,
the approximate delta function reduces instead to
$\delta(\tau-1)$,
and the expression reduces to an integral over the half line
$\bfp = (0,0,k + (\sigma-1) k/2)$ for $1 \le \sigma < \infty$.
It follows that the frequency argument $\nu \gamma p$ of ${\tilde {\bf
    {\underline   {\cal A}}}}_{\ms}^{(1)}$ is bounded away from zero, 
and therefore the second line of the integrand is bounded, from
Eq.\ (\ref{decomposeid}); the soft degrees of freedom or
distributional components do not
contribute.  The \Lorentz gauge condition 
(\ref{lorentzformode}) now 
makes the result vanish; this is the essentially the Feynman diagram
argument given in the last paragraph of Sec.\ \ref{sec:secondorder}.

The third term with $\eta = \gamma = 1$ is a little more involved.
Inserting the decomposition (\ref{decomposeid}) of the initial data
for the vector potential, this term can be written as
\begin{eqnarray}
    && - \frac{1}{2} \lim_{T \to \infty}
   \int_1^\infty d\sigma \int_{-1}^1 d\tau
   \int_0^{2 \pi} d\varphi \, \delta_{k T}( 1 - \sigma) \,
             {\tilde {\underline \chi}}^{(1)}_{\ms}\left[ \nu  l, -
               \nu \frac{\bfk - \bfp}{l} \right] 
             \nonumber \\   && \times 
             \bfk \cdot \left[
               \frac{i \nu }{2 \pi}
   {\bf \Delta {\underline {\cal A}}}^{(1)}_{\ms}(-\nu {\hat
     \bfp}) \, {\rm P.V.}
   \frac{1}{p} + 
   {\bf {\overline {\underline {\cal A}}}}^{(1)}_{\ms}(-\nu {\hat \bfp}) 
   \delta(p) + 
   {\tilde {\underline {\cal A}}}^{(1)\,{\rm rest}}_{\ms}(\nu 
   p,- \nu  {\hat \bfp}) \right].
\label{stage3}
\end{eqnarray}
Now the limiting $T \to \infty$ delta function $\delta(\sigma-1)$
would enforce an
integral over the line segment
\be
\bfp = (0,0,k (1+\tau)/2)
\label{lineseg}
\ee
for $-1 \le \tau \le 1$
that
extends from $\bfp = {\bf 0}$ to $\bfp = \bfk$ and that overlaps with 
the location $\bfp={\bf 0}$ of the
distributional contributions to the integrand.
Thus we will need to take the $T \to \infty$ limit carefully.
For the second and third terms in the large square brackets on the
second line we can just
use the limiting delta function $\delta(\sigma-1)$.  
The third term scales as $\sim
p^{-(1-\epsilon)} \sim (1 + \tau)^{-(1-\epsilon)}$ with $\epsilon>0$,
from Eq.\ (\ref{restscaling}).  This yields a diverging integrand but
the integral over $\tau$ converges, and thus the result vanishes by
the \Lorentz gauge condition (\ref{lorentzformode}), since it is
evaluated at ${\hat \bfp} = {\hat \bfk}$ by Eq.\ (\ref{lineseg}).
Similarly, for the second term, the
$\delta(\sigma-1)$ factor enforces evaluation at
${\hat \bfp} = {\hat \bfk}$, and then the $\delta(p)$ factor
evaluates at $\tau=-1$ at which the integrand is finite, giving again a
vanishing result by Eq.\ (\ref{lorentzformode}).

Thus we are left with the first term in the large square brackets in
the expression (\ref{stage3}).  We simplify this in several
stages:
\begin{itemize}
\item First, we argue that we can evaluate the factor ${\tilde
  {\underline \chi}}^{(1)}_{\ms}$
  at $\bfp = {\bf 0}$ and pull it  outside the integral. To see this we
  define the variable
  \be
  \mu = {\hat \bfk} \cdot {\hat \bfp} = \frac{ 1 + \sigma \tau}{\sigma
    + \tau}
  \label{mudef1}
  \ee
  from Eq.\ (\ref{prolatedef}), and we note that by the \Lorentz gauge condition (\ref{lorentzformode})
  that $\bfk \cdot \Delta {\underline {\cal
      A}}^{(1)}_{\ms}(-\nu {\hat \bfp})$ must vanish at $\mu =1$ and at
  $\mu = -1$.  Hence it must be expressible as $1-\mu^2$ times a
  bounded function, and
  including the factor of $1/p$ and using Eqs.\ (\ref{idents22}) and (\ref{mudef1})
  we find that the second line in the expression (\ref{stage3}) can be written as a bounded
  function times the factor
  \be
  \frac{ (\sigma^2-1) (1- \tau^2)}{(\sigma + \tau)^3}.
  \label{goodfactor}
  \ee
  Now the approximate delta function $\delta_{kT}(1-\sigma)$ will
  restrict the value of $\sigma$ to $\sigma - 1 = O(\varepsilon_0)$
  where $\varepsilon_0 = 1/(kT)$ and we are taking the limit
  $\varepsilon_0 \to 0$, from Eq.\ (\ref{approxdelta}).  Thus when
  $\tau + 1 = O(1)$ the factor (\ref{goodfactor}) scales like
  $ \sim \varepsilon_0$ and the integrand is suppressed, whereas when $\tau + 1 =
  O(\varepsilon_0)$ the factor (\ref{goodfactor}) scales like $\sim
  \varepsilon_0^{-1}$.  Hence for small $\varepsilon_0$ we can
  evaluate the factor ${\tilde {\underline \chi}}^{(1)}_{\ms}$ in the expression
  (\ref{stage3}) at $\tau=-1$ and
  pull it outside the integral, where it can be written as ${\tilde
    {\underline \chi}}_{\ms}^{(1)}(\omega, -\bfn)$ by Eq.\ (\ref{defs1}).
  
\item Second, we can drop the integral over $\sigma$ and the
  approximate delta function, evaluate the remaining integral at
  $\sigma = 1 + \varepsilon_0$, take the limit $\varepsilon_0 \to 0$,
  and then multiply by a correction factor of $1/2$.  This is because
we can pull the approximate delta function outside 
  the integral over $\tau$ and $\varphi$, and that integral becomes independent of
  $\sigma$ in the limit $\sigma \to 1$ from above (see below).  Then for small
  $\varepsilon_0$ the integral over $\sigma$ of the approximate delta function
  gives an overall factor of $1/2$, since the range of $\sigma$ is
  restricted to $\sigma \ge 1$.

  \item Third, the remaining integral at a fixed value of $\sigma$ is
    an integral over a prolate spheroid that limits as $\sigma \to 1$
    to the line segment (\ref{lineseg}).  We parameterize this
    integral as an integral over solid angles centered at $\bfp = {\bf
      0}$ by changing one of the variables of integration from $\tau$
    to $\mu$, using
    $$
    \frac{1}{\tau + \sigma} d\tau = \frac{1}{\sigma - \mu} d\mu
    $$
    from Eq.\ (\ref{mudef1}).
\end{itemize}

The result of all these simplifications is the expression
\be
- \frac{i}{4 \pi} {\tilde {\underline \chi}}^{(1)}_{\ms}(\omega, - \bfn) \,
\lim_{\sigma \to 1} \, \int_{-1}^1 d\mu \int_0^{2 \pi} d\varphi \,
\frac{1}{\sigma - \mu} \, \nu {\hat \bfk} \cdot \Delta {\underline
  {\cal A}}^{(1)}_{\ms}(- \nu {\hat \bfp})
\ee
for the $\eta = \gamma =1$ piece of Eq.\ (\ref{stage2}).
Defining the unit vectors ${\hat \bfm}=-\nu {\hat \bfp}$ and
${\hat \bfm}_0 = -\nu {\hat \bfk} = - \bfn$ from the definitions (\ref{defs1}),
and taking the limit $\sigma \to 1$, 
we can  rewrite this as
\be
 \frac{i}{4 \pi} {\tilde {\underline \chi}}^{(1)}_{\ms}(\omega, - \bfn) \,
 \, \int d^2 \Omega_{\hat \bfm} \,
 \frac{ {\hat \bfm}_0 \cdot \Delta {\underline
  {\cal A}}^{(1)}_{\ms}({\hat \bfm})}{1 - \mu}.
\label{stage5}
\ee
Next we define the Green's function
 \be
 G({\hat \bfm},{\hat
   \bfm}_0) = \frac{1}{4 \pi} \log(1 - {\hat \bfm} \cdot {\hat \bfm}_0)
 \label{green1}
 \ee
which satisfies $D^A D_A G = \delta^2({\hat
  \bfm},{\hat \bfm}_0) - 1/(4 \pi)$.
Using Eq.\ (\ref{decomposeid1}) we can rewrite the expression
(\ref{stage5}) as
\be
- i {\tilde {\underline \chi}}^{(1)}_{\ms}(\omega, - \bfn) 
 \, \int d^2 \Omega_{\hat \bfm} \,
D^A  G({\hat \bfm},{\hat
   \bfm}_0) D_A \Delta {\underline \Psi}^{{\rm e}(1)}_{\ms}({\hat \bfm}),
\ee
where we  have used $D_A m^i = e^i_A$.
Finally integrating by parts and using the
fact that $\Delta {\underline \Psi}^{{\rm e}(1)}_{\ms}$ has no $l=0$
component [cf.\ Eq.\ (\ref{AAdecompose1}) above] gives
that the  contribution to ${\tilde {\underline \chi}}^{(2)}_{\ps}(\omega, \bfn)$ is
  \be
   i {\tilde {\underline \chi}}^{(1)}_{\ms}(\omega, - \bfn) \Delta {\underline
     \Psi}^{{\rm e}(1)}_{\ms}(-\bfn).
   \label{final44}
   \ee

We now turn to the fourth and final term $\eta = 1$, $\gamma = -1$ in
Eq.\ (\ref{stage2}).  This can be evaluated using the same methods as
for the $\gamma = \eta = 1$ term just described.  One ends up
integrating  over a hyperboloid of revolution of the form $\tau = -1 +
O(\varepsilon_0)$ which limits to the half line 
$\bfp = (0,0, (1-\sigma) k/2)$ for $1 \le \sigma < \infty$.
   The final result is the same as the  result  (\ref{final44}), and
   adding the two contributions gives
   \be
      {\tilde {\underline \chi}}^{(2)}_{\ps}(\omega, \bfn) = 
   2  i {\tilde {\underline \chi}}^{(1)}_{\ms}(\omega, - \bfn) \Delta {\underline
     \Psi}^{{\rm e}(1)}_{\ms}(-\bfn),
   \label{final444}
   \ee
   which is equivalent to Eq.\ (\ref{nonlinearexplicit}).

   The remaining relations  in Eqs.\ (\ref{eq:freee2l}) can be evaluated using
similar methods; all the second order contributions vanish
since they are quadratic in the incoming first order scalar field
by Eq.\ (\ref{eqn-subleading-vec}), and that scalar field has no soft piece.

\section{Leading order change in electromagnetic memory}
\label{app:memory}

In this  appendix we compute explicitly the  leading order
electromagnetic memory produced when there is an incoming scalar field
but no incoming vector potential.  From the
conservation law (\ref{cons23}) this quantity is directly related to the
change $\Delta Q^{\rm hard}(\bfn)$ between the total hard outgoing charge per solid
angle at $\scri^+$ and that at $\scri^-$.  This quantity is the lowest
order coupling of soft and hard degrees of freedom that cannot be
removed by modifying the definitions of the hard and soft sectors,
as demonstrated in Sec.\ \ref{sec:higherorders}.  We shall show that
it is non-zero in general.

The quantity $\Delta Q^{\rm hard}(\bfn)$ is quartic in the incoming
\Lorentz gauge field ${\underline \chi}^{(1)}_{\ms}(v,\bfn)$.  For
any  function $\varepsilon = \varepsilon(\bfn)$ we define $\Delta
Q^{\rm hard}_\varepsilon = \int d^2 \Omega \, \varepsilon(\bfn) \Delta
Q^{\rm hard}(\bfn)$, and the leading order expression for this quantity in terms of
the initial data is (see below for the derivation)
\begin{eqnarray}
  \label{cm}
  \Delta Q^{\rm hard}_\varepsilon &=& 2 i e^2 \sum_{\gamma,\eta,\nu,\sigma = \pm}
\int \frac{d^3 p}{\gamma p} 
\int \frac{d^3 q}{\eta q} 
\int \frac{d^3 k}{\nu k} 
\int \frac{d^3 l}{\sigma l} 
\delta^{(4)}({\vec p} - {\vec q} + {\vec k} - {\vec l}) \varepsilon(\nu {\hat
  {\bf k}})
\nonumber \\ && \times
\,      {\tilde {\underline \chi}}^{(1)}_{\ms}(\gamma p, - \gamma {\hat {\bf p}})^*
\,      {\tilde {\underline \chi}}^{(1)}_{\ms}(\eta q, - \eta {\hat {\bf q}})
\,      {\tilde {\underline \chi}}^{(1)}_{\ms}(\nu k, - \nu {\hat {\bf k}})^*
\,      {\tilde {\underline \chi}}^{(1)}_{\ms}(\sigma l, - \sigma {\hat {\bf l}})
\nonumber \\ && \times
\frac{ ({\vec p} + {\vec q}) \cdot {\vec l} }{( {\vec p} - {\vec
    q})^2} \ \ + {\rm c.c.}
\end{eqnarray}
Here on the right hand side we integrate over four spatial momenta
$\bfp$, $\bfq$, $\bfk$, and $\bfl$ and sum over four corresponding signs $\gamma$, $\eta$,
$\nu$ and $\sigma$, and we define $p = | \bfp |$, ${\hat \bfp} = \bfp
/ p$ and ${\vec p} = (\gamma p, \bfp)$, with similar formulae for
the other three momenta.

We next discuss some properties of the formula (\ref{cm}).
First, the divergent factor $({\vec p} - {\vec q})^{-2}$ on the third
line is made integrable by the fact that when ${\vec p} = {\vec q}$,
the second line is explicitly real, and thus vanishes by virtue of the
factor of $i$ and the $+ {\rm c.c.}$.
Second, the $l=0$ component of the
expression (\ref{cm}) vanishes, which is necessary for consistency
with the conservation law (\ref{eq:cons-law}) since the memory
variables $\Delta \Psi^{\rm e}_{\ps}$ and $\Delta \Psi^{\rm e}_{\ms}$ have no $l=0$
pieces.  This follows from the fact that when $\varepsilon(\bfn)=1$,
taking the complex conjugate of the integrand is equivalent to making
the change of variables ${\vec p} \leftrightarrow {\vec q}$ and ${\vec
  k} \leftrightarrow {\vec l}$.  
Third, of the sixteen combinations of signs on the right hand side,
ten are identically vanishing.  Squaring the 4-momentum conservation
equation ${\vec p} +  {\vec k} = {\vec q} + {\vec l}$ yields that
$\gamma \nu = \eta \sigma$, which eliminates eight combinations. 
Also when both sides of this equation are timelike, requiring that
they both be either future directed or past directed eliminates the
cases $(\gamma,\nu,\eta,\sigma) = (+,+,-,-)$ and $(-,-,+,+)$.

When the momentum transfer ${\vec p} - {\vec q}$ is spacelike, the factor
on the third line in Eq. (\ref{cm}) is proportional 
to the $t$-channel\footnote{This could also be called $u$-channel,
depending on the identifications chosen of the ingoing and outgoing momenta.} scattering
amplitude for 2 particle to 2 particle 
scattering in scalar QED \cite{2014qfts.book.....S}.
This occurs for the four sign combinations
$(\gamma,\nu,\eta,\sigma) = (+,+,+,+), (-,-,-,-), (+,-,+,-)$ and
$(-,+,-,+)$.  Summing over these sign combinations gives the
$t$-channel and $u$-channel contributions to the change in memory
\begin{eqnarray}
  \label{cm-tchannel}
  \Delta Q^{\rm hard, tu-channels}_\varepsilon &=& 2 i e^2 
\int \frac{d^3 p}{p} 
\int \frac{d^3 q}{ q} 
\int \frac{d^3 k}{ k} 
\int \frac{d^3 l}{ l} 
\delta^{(4)}({\vec p} - {\vec q} + {\vec k} - {\vec l}) \Xi({\vec
  p},{\vec q}, {\vec k}, {\vec l})
\nonumber \\ && \times
\frac{ ({\vec p} + {\vec q}) \cdot {\vec l} }{( {\vec p} - {\vec
    q})^2} \ \ + {\rm c.c.}.
\end{eqnarray}
Here on the right hand side we have ${\vec p} = (p, \bfp)$ etc., and
\begin{eqnarray}
\Xi({\vec
  p},{\vec q}, {\vec k}, {\vec l}) &=& 
      {\tilde {\underline \chi}}^{(1)}_{\ms}(p, - {\hat {\bf p}})^*
\,      {\tilde {\underline \chi}}^{(1)}_{\ms}(q, - {\hat {\bf q}})
\,      {\tilde {\underline \chi}}^{(1)}_{\ms}(k, - {\hat {\bf k}})^*
\,      {\tilde {\underline \chi}}^{(1)}_{\ms}( l, - {\hat {\bf l}})
\, \varepsilon( {\hat
  {\bf k}})
\nonumber \\ &&
+       {\tilde {\underline \chi}}^{(1)}_{\ms}(-p,  {\hat {\bf p}})^*
\,      {\tilde {\underline \chi}}^{(1)}_{\ms}(-q,  {\hat {\bf q}})
\,      {\tilde {\underline \chi}}^{(1)}_{\ms}(-k,  {\hat {\bf k}})^*
\,      {\tilde {\underline \chi}}^{(1)}_{\ms}( -l,  {\hat {\bf l}})
\, \varepsilon( -{\hat
  {\bf k}})
\nonumber \\ &&
-      {\tilde {\underline \chi}}^{(1)}_{\ms}(p, - {\hat {\bf p}})^*
\,      {\tilde {\underline \chi}}^{(1)}_{\ms}(q, - {\hat {\bf q}})
\,      {\tilde {\underline \chi}}^{(1)}_{\ms}(-l, - {\hat {\bf l}})^*
\,      {\tilde {\underline \chi}}^{(1)}_{\ms}( -k, - {\hat {\bf k}})
\, \varepsilon( {\hat
  {\bf l}})
\nonumber \\ &&
-      {\tilde {\underline \chi}}^{(1)}_{\ms}(-p,  {\hat {\bf p}})^*
\,      {\tilde {\underline \chi}}^{(1)}_{\ms}(-q,  {\hat {\bf q}})
\,      {\tilde {\underline \chi}}^{(1)}_{\ms}(l,  {\hat {\bf l}})^*
\,      {\tilde {\underline \chi}}^{(1)}_{\ms}( k,  {\hat {\bf k}})
\, \varepsilon( -{\hat
  {\bf l}}),
\end{eqnarray}
where to obtain the third and fourth lines we have made the change of
variables in the integral ${\vec k} \leftrightarrow - {\vec l}$.

To eliminate the $\delta$-function in Eq.\ (\ref{cm-tchannel}) it is
convenient to parameterize the integral in terms of a spatial momentum
${\bf P}$ and spatial momentum transfer ${\bf \Delta}$ for which the
momenta in the center of momentum frame are ${\vec p}' = (P, {\bf P} +
{\bf \Delta})$, ${\vec q}' = (P, {\bf P})$, ${\vec k}' = (P, -{\bf P} -
{\bf \Delta})$ and ${\vec l}' = (P, - {\bf P})$, and in terms of the boost with
rapidity parameter $\psi$ in the direction of the unit vector ${\bfbeta}$
that takes the center of momentum frame to the frame of
Eq.\ (\ref{cm-tchannel}).  We also parameterize ${\bf \Delta}$ in terms of
the unit vector ${\bf w}$ via ${\bf \Delta} = P {\bf w} - {\bf P}$.
The result is 
\begin{eqnarray}
  \label{cm-tchannel1}
  \Delta Q^{\rm hard, tu-channels}_\varepsilon &=& \frac{i e^2 }{2}
  \int_{-\infty}^\infty d\psi \int d^3 P
  \int d^2 \Omega_\beta \int d^2 \Omega_w \,
\Xi({\vec
  p},{\vec q}, {\vec k}, {\vec l})  \,
P \frac{ g(\Theta,\psi)}{\cosh \psi + \cos {\bar \Theta} \sinh \psi}
\nonumber \\ && \times
\left[\frac{\Delta^2 - 8 P^2}{\Delta^2}\right] \ \ + {\rm c.c.},
\end{eqnarray}
where we have defined 
\be
g(\Theta,\psi) = \frac{\sin \Theta \sinh \psi \tanh \psi \left\{ 6 + 2 \cos(2
  \Theta) + 2 \left[1 - \cos(2 \Theta)\right] \cosh (2 \psi)
  \right\}}{\cosh^2 \psi - \cos^2 \Theta \sinh^2 \psi}
\ee
and $\cos \Theta = {\bfbeta} \cdot {\bf w}$ and $\cos {\bar \Theta}
= \bfbeta \cdot {\hat {\bf P}}$.  This form of the integral eliminates
all the divergences in the integrand of Eq.\ (\ref{cm-tchannel})
except for the divergence at ${\bf \Delta}=0$ which was discussed
after Eq.\ (\ref{cm}) above.

The remaining two sign combinations in Eq.\ (\ref{cm}) yield
a timelike momentum transfer ${\vec p} - {\vec q}$ and the $s$-channel
contribution to the change in memory.  These sign combinations are
$(\gamma,\nu,\eta,\sigma) = (+,-,-,+)$ and $(-,+,+,-)$.
In this case we parameterize the integral in terms of a spatial
momentum ${\bf P}$ and a momentum transfer ${\bf \Delta}$ in the frame
where ${\vec p} - {\vec q}$ has no spatial component, for which
${\vec p}' = (P, {\bf P})$, ${\vec q}' = (-P, {\bf P})$, ${\vec k}' =
(-P, {\bf P} + {\bf \Delta})$, ${\vec l}' = (P, {\bf P} + {\bf \Delta})$.
The result for the $s$-channel contribution to the change in memory is
\begin{eqnarray}
  \label{cm-schannel1}
  \Delta Q^{\rm hard, s-channel}_\varepsilon &=& \frac{i e^2 }{2}
  \int_{-\infty}^\infty d\psi \int d^3 P
  \int d^2 \Omega_\beta \int d^2 \Omega_w \,
{\tilde \Xi}({\vec
  p},{\vec q}, {\vec k}, {\vec l})  \,
P \frac{ g(\Theta,\psi)}{\cosh \psi + \cos {\bar \Theta} \sinh \psi}
\nonumber \\ && \times
\left[\frac{\Delta^2 - 2 P^2}{2 P^2} \right] \ \ + {\rm c.c.},
\end{eqnarray}
where
\begin{eqnarray}
{\tilde \Xi}({\vec
  p},{\vec q}, {\vec k}, {\vec l}) &=& 
      {\tilde {\underline \chi}}^{(1)}_{\ms}(p, - {\hat {\bf p}})^*
\,      {\tilde {\underline \chi}}^{(1)}_{\ms}(-q,  {\hat {\bf q}})
\,      {\tilde {\underline \chi}}^{(1)}_{\ms}(-k,  {\hat {\bf k}})^*
\,      {\tilde {\underline \chi}}^{(1)}_{\ms}( l, - {\hat {\bf l}})
\, \varepsilon( -{\hat
  {\bf k}})
\nonumber \\ &&
+       {\tilde {\underline \chi}}^{(1)}_{\ms}(-p,  {\hat {\bf p}})^*
\,      {\tilde {\underline \chi}}^{(1)}_{\ms}(q,  -{\hat {\bf q}})
\,      {\tilde {\underline \chi}}^{(1)}_{\ms}(k,  -{\hat {\bf k}})^*
\,      {\tilde {\underline \chi}}^{(1)}_{\ms}( -l,  {\hat {\bf l}})
\, \varepsilon( {\hat
  {\bf k}}).
\end{eqnarray}
The contributions (\ref{cm-tchannel1}) and (\ref{cm-schannel1}) to the change in memory are
generically nonzero, as can be shown for example by numerical
integrations with specific choices of initial data ${\tilde
  {\underline \chi}}^{(1)}(\omega,\bfn)$.

Finally we turn to the derivation of the formula (\ref{cm}).  We are
assuming that ${\underline {\cal A}}_{\ms A}=0$, and it follows from
the equations of motion (\ref{eqns-lg}) that
${\underline \chi}_{\ps}$
will be non-zero at orders $O(\tepsilon)$, $O(\tepsilon^{3})$ and
at higher orders, while $\underline{\mathcal{A}}_{\ps A}$ will be non-zero
at $O(\tepsilon^{4})$ and at higher orders (its $O(\tepsilon^{2})$ piece
vanishes as argued in Sec.\ \ref{sec:secondorder}).
Combining the definitions (\ref{deltaqhard}) and (\ref{currents})
of $\Delta Q^{\rm hard}(\bfn)$, 
the expansion (\ref{calJu}) of the asymptotic current ${\cal J}_{\ps
  u}$ and a similar expansion for ${\cal J}_{\ms v}$, the expansion
(\ref{pert-ansatz2a}) of the scalar field ${\underline \chi}_{\ps}$,
the definition (\ref{ftdef}) of the Fourier transform and the free
field scattering map (\ref{eq:freee-a}) yields
\be
\label{f33}
\Delta Q^{\rm hard}(\bfn) = -4 \pi \int_{-\infty}^\infty d\omega \omega
\left[ {\tilde {\underline \chi}}^{(3)}_{\ps}(\omega,\bfn) {\tilde
    {\underline \chi}}^{(1)}_{\ms}(\omega,-\bfn)^* + {\rm c.c.}
  \right] + O(\alpha^5).
\ee
To evaluate the third order scattered field ${\tilde {\underline \chi}}^{(3)}_{\ps}$ in this expression we use the same techniques
as in Appendix \ref{app:secondorder}.
We solve the equation of motion (\ref{eqn-subleading-vec}) for
${\underline A}^{(2)}$ using the mode expansion (\ref{phiexpand0}) to
expand the source and using the momentum space retarded Green's
function.  We then insert the result into the equation of motion 
(\ref{eqn-subsubleading-scalar}) for ${\underline \Phi}^{(3)}$, extract
the asymptotic form of the solution ${\underline \chi}^{(3)}_{\ps}$
near $\scri^+$ using the general identity (\ref{Sahaidentity}),
and eliminate the mode functions in favor of the initial data using
Eq.\ (\ref{modefnf}).  Finally inserting the resulting expression for
${\tilde {\underline \chi}}^{(3)}_{\ps}$ into the expression (\ref{f33})
yields the result (\ref{cm}).

\bibliographystyle{JHEP}
\bibliography{softhair}
\end{document}